\documentclass{ieeeaccess}
\usepackage{cite}
\usepackage{amsmath,amssymb,amsfonts}
\usepackage{balance}
\usepackage{tipa}
\usepackage{subfig}
\usepackage{float}
\usepackage{algorithm}
\usepackage{algpseudocode}
\usepackage{graphicx}
\usepackage{textcomp}
\usepackage{placeins}
\usepackage{letltxmacro}

\LetLtxMacro{\originaleqref}{\eqref}
\renewcommand{\eqref}{Eq.~\originaleqref}

\def\BibTeX{{\rm B\kern-.05em{\sc i\kern-.025em b}\kern-.08em
    T\kern-.1667em\lower.7ex\hbox{E}\kern-.125emX}}
\begin{document}
\history{Date of publication xxxx 00, 0000, date of current version xxxx 00, 0000.}
\doi{10.1109/ACCESS.2017.DOI}

\title{TVOR: Finding Discrete Total Variation Outliers among Histograms}
\author{\uppercase{Nikola Bani{\'{c}}}\authorrefmark{1} and \uppercase{Neven Elezovi{\'{c}}}\authorrefmark{2}}
\address[1]{Gideon Brothers, Radni{\v{c}}ka 177, 10000 Zagreb, Croatia (e-mail: nbanic@gmail.com)}
\address[2]{Faculty of Electrical Engineering and Computing, University of Zagreb, Unska 3, 10000 Zagreb, Croatia (e-mail: neven.elezovic@fer.hr)}

\markboth
{Bani{\'{c}} and Elezovi{\'{c}}: TVOR: Finding Discrete Total Variation Outliers among Histograms}
{Bani{\'{c}} and Elezovi{\'{c}}: TVOR: Finding Discrete Total Variation Outliers among Histograms}

\corresp{Corresponding author: Nikola Bani{\'{c}} (e-mail: nbanic@gmail.com).}

\begin{abstract}
Pearson's chi-squared test can detect outliers in the data distribution of a given set of histograms. However, in fields such as demographics (for e.g. birth years), outliers may be more easily found in terms of the histogram smoothness where techniques such as Whipple's or Myers' indices handle successfully only specific anomalies. This paper proposes smoothness outliers detection among histograms by using the relation between their discrete total variations~(DTV) and their respective sample sizes. This relation is mathematically derived to be applicable in all cases and simplified by an accurate linear model. The deviation of the histogram's DTV from the value predicted by the model is used as the outlier score and the proposed method is named Total Variation Outlier Recognizer~(TVOR). TVOR requires no prior assumptions about the histograms' samples' distribution, it has no hyperparameters that require tuning, it is not limited to only specific patterns, and it is applicable to histograms with the same bins. Each bin can have an arbitrary interval that can also be unbounded. TVOR finds DTV outliers easier than Pearson's chi-squared test. In case of distribution outliers, the opposite holds. TVOR is tested on real census data and it successfully finds suspicious histograms. The source code is given at https://github.com/DiscreteTotalVariation/TVOR.
\end{abstract}

\begin{keywords}
Age heaping, anomaly detection, discrete total variation, expected value, fitting, histogram, Myers' index, outlier detection, Pearson's chi-squared test, total variation, Whipple's index.
\end{keywords}

\titlepgskip=-15pt

\maketitle

\section{Introduction}
\label{sec:intro}

\newtheorem{theorem}{Theorem}
\newtheorem{hypothesis}{Hypothesis}
\newcommand{\E}[1]{\mathbb{E}\left[ #1 \right]}
\newcommand{\Var}[1]{\mathrm{Var}\left[ #1 \right]}
\newcommand{\V}[1]{\left\lVert #1 \right\rVert_{V}}
\newcommand{\Norm}[2]{\left\lVert #1 \right\rVert_{#2}}
\newcommand{\N}[2]{\mathcal{N}\left(#1,#2\right)}

\PARstart{O}{utliers} can be defined as data patterns that do not conform to an expected normal data behavior~\cite{chandola2007outlier}. Since identifying outliers or anomalies can often be useful, performing outlier, i.e. anomaly, detection has an important role in many data related areas. For example, with the ever growing application of machine learning in various fields, having clean training sets, free of any unwanted outliers, can often significantly benefit the final production accuracy. On the other hand, in real-time applications such as network traffic or health monitoring, it is usually highly important to detect anomalies that could represent any form of unwanted behavior to prevent their potentially detrimental effects. Alternatively, it may be required to see which samples differ the most from the rest of the data and study them in more detail.

Since there is a relatively high demand for anomaly and outlier detection methods in fields dealing with some form of data, numerous methods have been proposed for various applications, as can be seen in several review papers~\cite{hodge2004survey,chandola2007outlier,goldstein2016comparative}.

A particular kind of data are histograms. First introduced by Pearson~\cite{pearson1895x}, histograms are by definition estimates of the probability distribution of a continuous variable. If there is a sample of real numbers drawn from the same distribution and all inside a given interval, then histograms can be used as their simple representation, and are also suitable for visual presentation. For histograms to be useful, the bin size should be adjusted accordingly to the data being described~\cite{sturges1926choice,scott1979optimal,freedman1981histogram,shimazaki2007method}. In certain cases for a group of such histograms it may be interesting to know whether some of them are outliers. This may include histograms describing samples drawn from another distribution different from the one of the majority of the samples, but it may also include histograms just describing some less likely samples from the same distribution. To be clear, in such a case, histograms are not used as tools for outlier detection like in e.g.~\cite{goldstein2012histogram}, but they are the data representations to be analyzed for the presence of outliers.

In the simple case when only a single histogram is given, instead of multiple histograms, a straightforward approach to check whether it represents a sample that differs from a given distribution would be to use the Pearson's chi-squared test~\cite{pearson1900x}. It tests how likely it is that any observed difference between the bins counts of the given histogram and the expected bin counts occurred by chance. However, for this to work, it is required to know the expected bin counts. 

On the other hand, if multiple histograms are given for samples that are assumed to have been drawn from the same distribution, then it is possible to find outliers among them by means of the Pearson's chi-squared test even if the distribution is unknown. Namely, under Glivenko-Cantelli theorem~\cite{serfling2009approximation} all the given histograms, except the currently tested one, can be used to get a reliable empirical distribution function, which in turn can be used to get the expected bin counts. Over time, numerous other techniques that can be applied in the described cases have been proposed~\cite{porter2008testing,bityukov2014method,bityukov2016comparison,gagunashvili2017tests}.

While the problem of finding outliers in terms of distribution is common, in some cases it is required to find histogram outliers in terms of some specific histogram property. For example, census data histograms are usually smooth, i.e. the difference between the counts of neighboring bins is relatively low, but in the presence of anomalies such as \textit{age heaping}~\cite{caselli2005demography}, this often stops being the case. One way to measure smoothness is to calculate total variation~\cite{mallat2008wavelet}. This means that by detecting deviations from the expected total variation it could be possible to detect smoothness outliers more easily than by means of some of the previously described techniques. Single-value properties similar to total variation in terms of simplicity, such as skewness, have already been used for outlier detection~\cite{heymann2012outskewer}. As a matter of fact, total variation has also found application in tasks such as classification~\cite{ahmed2017graph} and outlier detection for graph signals~\cite{gopalakrishnan2019identification}.

Therefore, in this paper a new method for outlier detection in terms of discrete total variation~(DTV) among histograms that describe samples drawn from the supposedly same, but unknown distribution is proposed. There are several contributions of this paper. First, it is mathematically proven that in terms of the underlying distribution there are only two possible cases of the relation between the sample size and the expected discrete total variation with the first case only being a special case of the second one. Second, a method is proposed that utilizes this relation to detect outliers that deviate from their expected discrete total variation. Third, it is shown that while the proposed method is not supposed to be used as a general outlier detector in terms of distribution, in some special cases it still performs better in this task than Pearson's chi-squared test. Fourth, the proposed method is shown to be able to detect suspicious histograms on real-life census data. The practical applicability and usefulness of the proposed method are shown on synthetic data and real-life census data. The proposed method is simple to implement and it does not require prior knowledge of any distribution.

The paper is structured as follows: in Section~\ref{sec:tv} the total variation is formally described, in Section~\ref{sec:method} the theoretical derivation of the proposed method and its underlying model are given, in Section~\ref{sec:results} the experimental results obtained on synthetic data and historical real-life census data are presented and discussed, and Section~\ref{sec:conclusions} concludes the paper.

\section{The total variation}
\label{sec:tv}

Total variation of a differentiable function $f$ is defined as~\cite{mallat2008wavelet}
\begin{equation}
    \label{eq:tvd}
    \V{f}=\int_{-\infty}^{+\infty}\lvert f'(t)\rvert dt.
\end{equation}
If $f$ is non-differentiable, its total variation is given as~\cite{mallat2008wavelet}
\begin{equation}
    \label{eq:tvn}
    \V{f}=\lim\limits_{h\to 0}\int_{-\infty}^{+\infty}\frac{\lvert f(t)-f(t-h)\rvert}{\lvert h\rvert}dt.
\end{equation}
If $f_{n}[i]=f\ast \Phi_{n}(i/n)$ is a discrete signal obtained with an averaging filter $\Phi_{n}(t)=1_{[0, N^{-1}]}(t)$ and a uniform sampling at intervals $n^{-1}$, then its discrete total variation~(DTV) is calculated by approximating the signal derivative by a finite difference over the sampling distance $h=n^{-1}$ and replacing the integral in~\eqref{eq:tvn} by a Riemann, which then gives~\cite{mallat2008wavelet}
\begin{equation}
    \label{eq:dtv}
    \V{f_{n}}=\sum\limits_{i}\lvert f_{n}[i]-f_{n}[i-1]\rvert.
\end{equation}

Despite being relatively simple to calculate, total variation is successfully used in areas such as denoising~\cite{rudin1992nonlinear,huang2009new,liu2014generalized,liu2015two}, image restoration~\cite{rudin1994total,osher2005iterative,wang2008new,jia2019color}, image super-resolution~\cite{aly2005image,ng2007total}, image enhancement~\cite{chan2001total,pierre2017variational}, compressive sensing applications~\cite{li2010efficient,jian2016single}, computer graphics~\cite{ihrke2014introduction}, and others.

\section{The proposed method}
\label{sec:method}

In this section, the proposed method for finding discrete total variation outliers among histograms and the method's underlying model are described. In order to try to avoid any misunderstandings, the structure of this section has purposely been slightly extended. Section~\ref{subsec:general} gives the general idea of how to use the discrete total variation for outlier detection, Section~\ref{subsec:background} gives an initial statistical foundation, Sections~\ref{subsec:uniform} and~\ref{subsec:non-uniform} use this foundation to derive the relation between the sample size and its expected discrete total variation for two general cases, Section~\ref{subsec:model} uses this relation to propose the sample models based on the discrete total variation, Section~\ref{subsec:score} describes the score calculation, Section~\ref{subsec:application} explains how to combine all these results into a single method, and, finally, Section~\ref{subsec:name} names this method.

\subsection{The general idea}
\label{subsec:general}

Let there be a sample of $N$ values, $\mathbf{x}_n$ its histogram with $n$ bins, and $x_i$ the number of values that fell in the $i$-th bin with
\begin{equation}
\label{eq:sum}
\sum_{i=1}^{n}x_i=N.
\end{equation}
Each of the $n$ bins has an arbitrary interval that can also be unbounded. The bins are not required to be of the same size. Let $p_i$ be the probability of a value falling in the $i$-th bin and
\begin{equation}
\label{eq:one}
\sum_{i=1}^{n}p_i=1.
\end{equation}
Due to randomness the discrete total variation of $\mathbf{x}_n$, i.e. $\V{\mathbf{x}_n}$ can differ for each sampling, but it should mostly not differ significantly from its expected value $\E{\V{\mathbf{x}_n}}$ for a given $N$ and probabilities $p_i$. For a given $\mathbf{x}_n$ the difference between its $\V{\mathbf{x}_n}$ and $\E{\V{\mathbf{x}_n}}$ can serve as a score of how much the sample differs from the expected behavior. Such a score has several drawbacks as well as advantages.

%Probably the main disadvantage is that such a score can often fail in discriminating samples taken from different distributions. Namely, many different distributions can produce histograms of the same or similar discrete total variation even if the histograms look totally different. 

The main disadvantage is that it is required to know $\E{\V{\mathbf{x}_n}}$ for any given $N$ or at least to know the relation between these two values for proper scaling and comparison.

The main advantage of such a scoring is the simplicity of its calculations due to the very definition of the discrete total variation. Further, because of that it is not necessary to know the desired sample distribution, which significantly widens the application possibilities. Finally, it is not very likely that two samples of the same size have histograms of the same or similar smoothness, i.e. discrete total variation and that their scores differ significantly. That means that if this score is calculated for every sample in a group of samples that are expected to have similar smoothness, then the ones with the highest scores can be considered as outlier candidates.

However, in order for this to be practically usable, first an analytical relation between $N$ and $\E{\V{\mathbf{x}_n}}$ has to be found.
    
\subsection{The statistical background}
\label{subsec:background}

The first step in finding a relation between $N$ and $\E{\V{\mathbf{x}_n}}$ is to examine $\E{\left(x_i-x_j\right)^2}$ in more detail by using the variances of $x_i$ and $x_j$, i.e. $\Var{x_i}$ and $\Var{x_j}$, respectively:
\begin{equation}
\label{eq:exixj2}
\begin{gathered}
\E{\left(x_i-x_j\right)^2}\\
=\E{\left(x_i-\E{x_i}-x_j+\E{x_j}+\E{x_i}-\E{x_j}\right)^2}\\
=\Var{x_i}-2\E{\left(x_i-\E{x_i}\right)\left(x_j-\E{x_j}\right)}\\
+\Var{x_j}+\left(\E{x_i}-\E{x_j}\right)^2.
\end{gathered}
\end{equation}
The value of $x_i$ for a given $i$ has binomial distribution so that
\begin{equation}
\label{eq:bde}
\begin{gathered}
\E{x_i}=Np_i,
\end{gathered}
\end{equation}
\begin{equation}
\label{eq:bde2}
\begin{gathered}
\Var{x_i}=Np_i\left(1-p_i\right).
\end{gathered}
\end{equation}
For the second term of the last form of~\eqref{eq:exixj2} it holds that
\begin{equation}
\label{eq:term2_1}
\begin{gathered}
\E{\left(x_i-\E{x_i}\right)\left(x_j-\E{x_j}\right)}=\E{\left(x_i-\E{x_i}\right)x_j}.
\end{gathered}
\end{equation}
The result of~\eqref{eq:term2_1} can now be further developed as follows:
\begin{equation}
\label{eq:term2_2}
\begin{gathered}
\E{\left(x_i-\E{x_i}\right)x_j}=\E{\E{\left(x_i-\E{x_i}\right)x_j}|x_j}\\
=\E{\left(x_i-\E{x_i}\right)\E{x_j|x_i}}\\
=\E{\left(x_i-\E{x_i}\right)\frac{p_j}{1-p_i}\left(N-x_i\right)}\\
=-\frac{p_j}{1-p_i}\E{\left(x_i-\E{x_i}\right)x_i}\\
=-\frac{p_j}{1-p_i}\left(\E{x_i^2}-\E{x_i}^2\right)\\
=-\frac{p_j}{1-p_i}\Var{x_i}=-Np_ip_j.
\end{gathered}
\end{equation}
Combining~\eqref{eq:bde},~\eqref{eq:bde2}, and~\eqref{eq:term2_2} develops~\eqref{eq:exixj2} to
\begin{equation}
\label{eq:exixj2_all}
\begin{gathered}
\E{\left(x_i-x_j\right)^2}\\
=Np_i\left(1-p_i\right)+2Np_ip_j+Np_j\left(1-p_j\right)+N^2\left(p_i-p_j\right)^2\\
=N^2\left(p_i-p_j\right)^2+N\left(p_i+p_j-\left(p_i-p_j\right)^2\right).
\end{gathered}
\end{equation}
%By using~\eqref{eq:xi2} and~\eqref{eq:exixj} it is possible to further develop~\eqref{eq:exixj2}:
%\begin{equation}
%\label{eq:exixj2_all}
%\begin{gathered}
%\E{\left(x_i-x_j\right)^2}=\\
%\E{x_i^2}+\E{x_j^2}-2\E{x_i x_j}=\\
%Np_i\left(1-p_i+Np_i\right)+\\
%Np_j\left(1-p_j+Np_j\right)-2N\left(N-1\right)p_ip_j=\\
%N^2\left(p_i^2+p_j^2-2p_ip_j\right)+\\
%N\left(p_i\left(1-p_i\right)+p_j\left(1-p_j\right)+2p_ip_j\right)=\\
%N^2\left(p_i-p_j\right)^2+N\left(p_i+p_j-\left(p_i-p_j\right)^2\right).
%\end{gathered}
%\end{equation}
Based on the values of $p_i$ there are two cases of further actions for establishing a relation between $N$ and $\E{\V{\mathbf{x}_n}}$. These two cases are covered in the following subsections.

%\begin{equation}
%\label{eq:e}
%\begin{gathered}
%\E{\sum_{i=1}^{n-1}\left(x_{i+1}-x_i\right)^2}}=\sum_{i=1}^{n-1}\E{\left(x_{i+1}-x_i\right)^2}=\\
%N^2\sum_{i=1}^{n-1}\left(p_{i+1}-p_i\right)^2+N\sum_{i=1}^{n-1}\left(p_i+p_{i+1}-\left(p_{i+1}-p_i\right)^2\right).
%\end{gathered}
%\end{equation}

\subsection{Uniform distribution}
\label{subsec:uniform}

\subsubsection{Upper bound}
\label{subsubsec:uniform_upper_bound}
%\textbf{1. Upper bound.}
The first case is when the distribution of the sample and consequently the distribution of the histogram are uniform so that the probability of a value falling in the $i$-th bin is then
\begin{equation}
\label{eq:uniform_p}
p_1=p_2=\ldots =p_n=\frac{1}{n}.
\end{equation}
When this is applied to~\eqref{eq:exixj2_all}, it eliminates its first term and it simplifies its second term, which then gives the form
%\begin{equation}
%	\label{eq:exixj2_uniform}
%	\begin{gathered}
%	\E{\left(x_i-x_j\right)^2}=\\
%	N^2\left(p_i-p_j\right)^2+N\left(p_i+p_j-\left(p_i-p_j\right)^2\right)=\\
%	N^^2\left(\frac{1}{n}-\frac{1}{n}\right)^2+N\left(\frac{1}{n}+\frac{1}{n}
%	-\left(\frac{1}{n}-\frac{1}{n}\right)^2\right)=\\
%	0+N\left(\frac{2}{n}-0^2\right)=\frac{2N}{n}.
%	\end{gathered}
%\end{equation}
\begin{equation}
	\label{eq:exixj2_uniform}
	\E{\left(x_i-x_j\right)^2}=
	\frac{2N}{n}.
\end{equation}
Taking into account that the square root is a concave function and applying the Jensen's inequality~\cite{garling2007inequalities} to~\eqref{eq:exixj2_uniform} gives
\begin{equation}
\label{eq:jensen1}
\begin{gathered}
\E{\sqrt{\left(x_i-x_j\right)^2}}=\E{\left|x_i-x_j\right|}\leq\sqrt{\E{\left(x_i-x_j\right)^2}}.
\end{gathered}
\end{equation}
This inequality can than be applied to all neighboring bins:
\begin{equation}
\label{eq:jensen2}
\begin{gathered}
\sum_{i=1}^{n-1}\E{\left|x_{i+1}-x_i\right|}\leq\sum_{i=1}^{n-1}\sqrt{\E{\left(x_{i+1}-x_i\right)^2}}.
\end{gathered}
\end{equation}
Due to the basic properties of the expectation, it holds that
\begin{equation}
\label{eq:ee}
\begin{gathered}
\sum_{i=1}^{n-1}\E{\left|x_{i+1}-x_i\right|}=\E{\sum_{i=1}^{n-1}\left|x_{i+1}-x_i\right|}.
\end{gathered}
\end{equation}
Applying~\eqref{eq:dtv},~\eqref{eq:exixj2_uniform}, and~\eqref{eq:ee} to~\eqref{eq:jensen2} gives
\begin{equation}
\label{eq:upper_bound_uniform}
\begin{gathered}
\E{\V{\mathbf{x}_n}}\leq\left(n-1\right)\sqrt{\frac{2N}{n}}.
\end{gathered}
\end{equation}
This gives the upper bound for the expected value of the discrete total variation and thus the first relation between $N$ and $\E{\V{\mathbf{x}_n}}$ if the sample numbers are uniformly distributed.

\subsubsection{Exact values}
\label{subsubsec:uniform_exact_values}
%\textbf{2. Exact values.}
	Let $F(n,N)$ denote the expected value of the discrete total variation as a function of two key parameters $n$ and $N$:
	\begin{equation}
	\label{eq:Function_F}
		F(n,N):=\E{\V{\mathbf{x}_n}}
	\end{equation}

\bigskip
\begin{theorem}
	\label{th:2-dimensional}
	The exact value of $F(2,N)$ in closed form is
	\begin{equation}
	\label{eq:F(2,N)}
	F(2,N)=2^{-N+1} \lfloor(N+1)/2\rfloor {N\choose\lfloor N/2\rfloor}.
	\end{equation}
\end{theorem}

	The proof of Theorem~\ref{th:2-dimensional} is given later in Appendix.
	
	It is relatively easy to show that for each $r$ it holds that
	\begin{equation}
	\label{eq:nekonkavnost}
	F(2,2r)=F(2,2r-1)
	\end{equation}
	and this leads to some unwanted consequences later on in the paper, but there they are mentioned and handled properly.

	The case of uniform distribution means that a histogram 
	is a realization of the multinomial distribution and its bins $x_1,x_2,...,x_n$ are random variables. The distribution of each $x_i$ is ${\cal B}(N,\frac 1n)$, i.e. it is binomially distributed with parameters $N$ and $\frac{1}{n}$. Variables $x_i$ are not independent, since their sum equals $N$. However, because of the symmetry, variables $x_2-x_1$, \dots, $x_n-x_{n-1}$ have the same distribution, which gives
	\begin{align}
	\label{eq:x_2-x_1}
	F(n,N)&=\E{|x_2-x_1|+\dots+|x_n-x_{n-1}|}\nonumber\\
	&=(n-1)\E{|x_2-x_1|}.
	\end{align}

Before continuing, for the sake of convenience, first the notation for the multinomial coefficient has to be given as
	\begin{equation}
	\label{eq:mnc}
		{N\choose k_1,\dots,k_n}=\frac{N!}{k_1!\cdots k_n!}.
	\end{equation}
	
\begin{theorem}
\label{th:n-dimensional}
	The expected value of the total variation of a histogram of uniformly distributed values is calculated as
	\begin{equation}
	\label{eq:n-dimensional}
	\begin{split}
	F(n,N)=2(n-1)\left(\frac{n-2}n\right)^N
	\sum_{\substack{k_1+k_2\le N\\k_1<k_2}}\\
	{N\choose k_1,k_2,N-k_1-k_2}(n-2)^{-(k_1+k_2)}(k_2-k_1).
	\end{split}
	\end{equation}
\end{theorem}
    
    The proof of Theorem~\ref{th:n-dimensional} is given later in Appendix. By using \eqref{eq:n-dimensional} it is possible to calculate the expected total variation for all reasonable values of $n$ and $N$ with some examples being shown in Table~\ref{tab:comparison}. However, if using \eqref{eq:n-dimensional} turns out to be computationally too demanding, the solution is to develop and use some appropriate asymptotic forms.

\subsubsection{Asymptotics}
\label{subsubsec:uniform_asymptotics}
%\textbf{3. Asymptotics.}
	By taking into account the well-known asymptotic form of the central binomial coefficients that is commonly given as
	\begin{equation}
		{2r\choose r}\approx \frac{4^r}{\sqrt{\pi r}}\qquad \text{as }r\to\infty,
	\end{equation}
	it follows that the asymptotic form of $F(2,N)$ is given as
	\begin{align}
	\label{eq:asymtotic-2}
		F(2,N)=2^{-2r+1}r{2r\choose r}
		\approx \sqrt{\frac{2}{\pi}}\,\sqrt{N}.
	\end{align}

	The experimental calculations suggest that the following hypothesis can be stipulated for the uniform  distribution:

\medskip

\begin{hypothesis}
	\label{hp:1-f(n,N)}
	For $N$ sufficiently large, we have
	\begin{equation}
	\label{eq:hypothesis}
		F(n,N)\approx(n-1)F\left(2,\frac{2N}{n}\right).
	\end{equation}
\end{hypothesis}

\medskip

	The right side of this equation represents the sum of the discrete total variations of two-binned histograms of the uniform distribution with sample size being equal to the expected number of values. If this hypothesis is accepted, then the following asymptotic is true for the uniform distribution:
	\begin{equation}
	\label{eq:asymptotic}
		F(n,N)\approx \frac{2(n-1)}{\sqrt{n\pi}} \sqrt{N}.
	\end{equation}
In Table~\ref{tab:comparison} the values obtained by~\eqref{eq:asymptotic} are compared to the exact values of $F(n, N)$ for some chosen $n$ and $N$.

\begin{table}[ht]
\normalsize
%\scriptsize
%\tiny
\caption{The comparison of the exact values of $F(n,N)$ with the values obtained by~\eqref{eq:asymptotic}
for some $n$ and $N$.}
\label{tab:comparison}
\centering
	\begin{tabular}{|l|l|l|l|l|}
	\hline
	$n$ & \multicolumn{2}{|c|}{$N=100$} & \multicolumn{2}{|c|}{$N=1000$}\\
	\hline
	& \eqref{eq:asymptotic} & exact value & \eqref{eq:asymptotic} & exact value\\
	\hline
    2  & 7.97885 & 7.95892 &        25.2313 & 25.2250\\
    3  & 13.0294 & 13.0213 &        41.2026 & 41.2000\\
    4  & 16.9257 & 16.9045 &        53.5237 & 53.5170\\
    5  & 20.1851 & 20.1472 &        63.8308 & 63.8188\\
    6  & 23.0329 & 22.9752 &        72.8366 & 72.8183\\
    7  & 25.5892 & 25.5090 &        80.9203 & 80.8950\\
    8  & 27.9260 & 27.8207 &        88.3096 & 88.2765\\
    9  & 30.0901 & 29.9577 &        95.1533 & 95.1116\\
    10 & 32.1142 & 31.9525 &        101.554 & 101.503\\
    20 & 47.9395 & 47.3907 &        151.598 & 151.427\\
    30 & 59.7437 & 58.6681 &        188.926 & 188.595\\
    40 & 69.5808 & 67.8604 &        220.034 & 219.509\\
    50 & 78.1927 & 75.7182 &        247.267 & 246.522\\
	\hline
	\end{tabular}
\end{table}

\begin{figure}[htb]
    \centering
    
	\includegraphics[width=\linewidth]{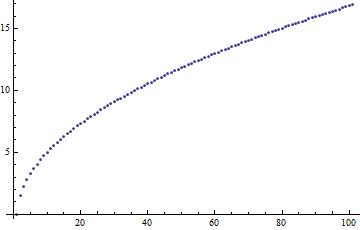}
	
    \caption{The values of $F(4,N)$ for $1\le N\le 100$.}
	\label{fig:F(4,N)}
    
\end{figure}	

	Hypothesis~\ref{hp:1-f(n,N)} and the results of the numerical calculation furthermore suggest that the following hypothesis is true:

\medskip

\begin{hypothesis}
	For each $n\ge 3$, the function
	$N\mapsto F(n,N)$ is increasing and strictly concave, hence, for each $0\le k\le N$
	\begin{equation}
	\label{eq:concave}
		F(n,k)+F(n,N-k)< 2F(n,\frac N2).
	\end{equation}
\end{hypothesis}

\medskip

\begin{figure*}[htb]
    \centering
    
	\subfloat[]{
	\includegraphics[width=0.95\linewidth]{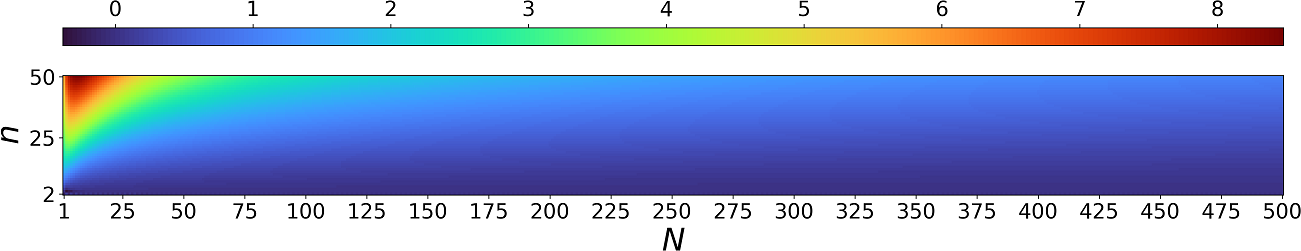}
	\label{fig:variation_absolute}
	}\\
	\subfloat[]{
	\includegraphics[width=0.97\linewidth]{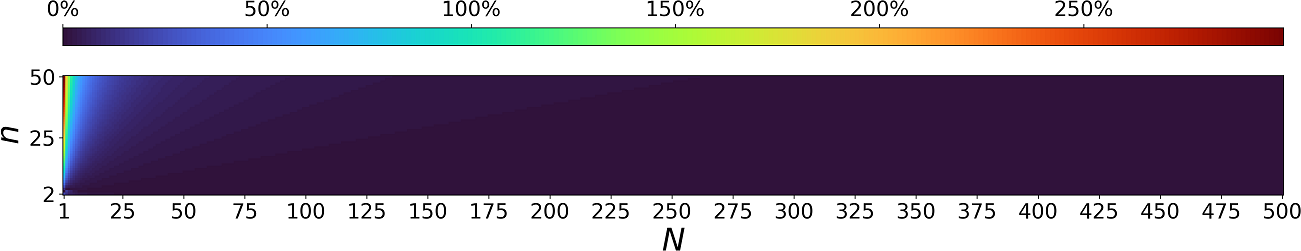}
	\label{fig:variation_relative}
	}\\
	\subfloat[]{
	\includegraphics[width=0.97\linewidth]{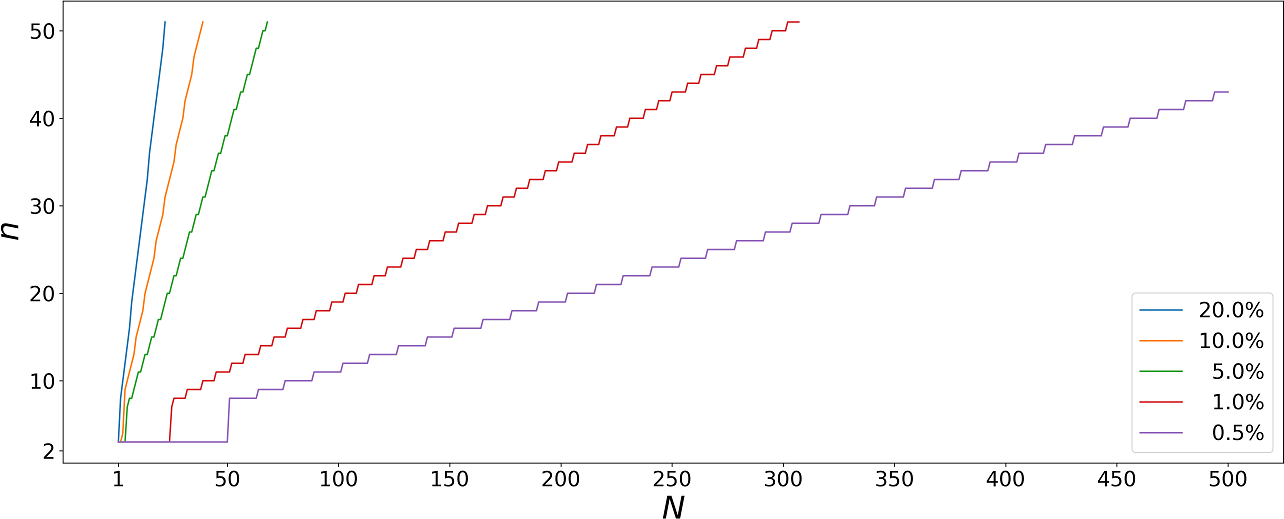}
	\label{fig:variation_lines}
    }
    
    \caption{The difference between the results of \eqref{eq:n-dimensional} and \eqref{eq:asymptotic}, which represent the exact and approximated values of $F(n, N)$, respectively: a)~the absolute error, b)~the relative error, and c)~the dependance of certain relative errors on $n$ and $N$.}
	\label{fig:variation}
    
\end{figure*}

	Function $N\mapsto F(2,N)$ is nondecreasing, but it is not strictly concave, because as demonstrated by \eqref{eq:nekonkavnost} its neighboring values can be equal.
	The proof of these two hypotheses may be very difficult, but they are not essential for the conclusions that are drawn later in the paper. 
	The diagram in Fig.~\ref{fig:F(4,N)} shows the situation for $n=4$ and $1\le N\le100$.

	Let $F_c(n,N)$ denote the expected value of the \textbf{the circular variation}, which unlike the usual variation has an additional term $\left|x_1-x_n\right|$ for the absolute value of the difference between the first and the last bin. $F_c(n,N)$ is then defined as
	\begin{equation}
	\label{eq:circular}
		F_c(n,N)=\E{|x_2-x_1|+\dots+|x_{n}-x_{n-1}|+|x_1-x_n|}.
	\end{equation}
	By taking into account~\eqref{eq:x_2-x_1}, it follows from~\eqref{eq:circular} that
	\begin{equation}
	\label{eq:connection}
		F_c(n,N)=\frac{n}{n-1}\,F(n,N).
	\end{equation}
    Applying \eqref{eq:x_2-x_1} and adjusting the result for later use gives
	\begin{align}
	\label{eq:connection2}
		F_c(n,N)&=\frac{n}{n-1}(n-1)\E{|x_2-x_1|}\nonumber\\
		&=n\,\E{|x_2-x_1|}\nonumber\\
		&=\frac{n}{2}\left(\E{|x_2-x_1|}+\E{|x_n-x_{n-1}|}\right).
	\end{align}
	
\begin{figure*}[htb]
    \centering
    
	\subfloat[]{
	\includegraphics[width=0.97\linewidth]{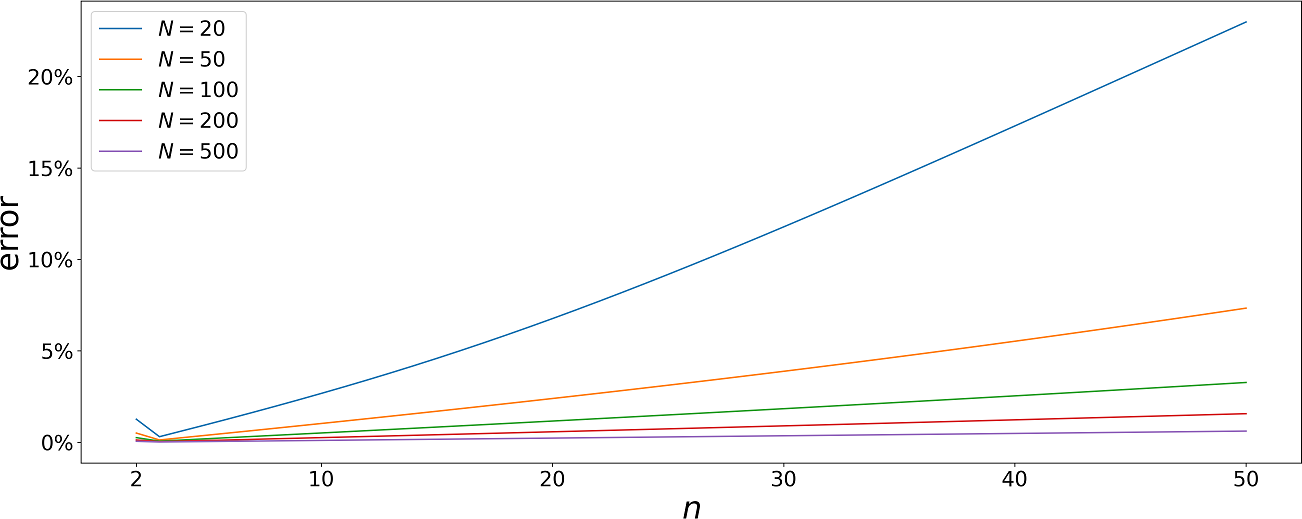}
	\label{fig:error_N}
	}\\
	\subfloat[]{
	\includegraphics[width=0.97\linewidth]{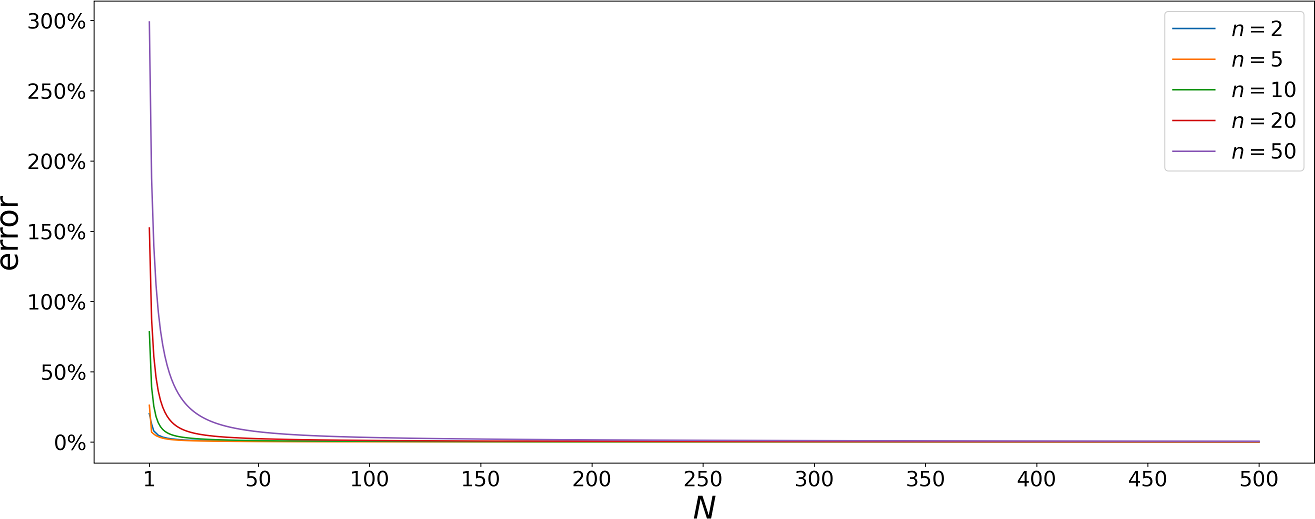}
	\label{fig:error_bins}
    }

	\caption{The relation between the error when using \eqref{eq:asymptotic} and the values of: a)~sample size $N$ and b)~number of bins $n$.}
	\label{fig:error}
    
\end{figure*}

	All possible histograms $\mathbf{x}_n$ can be split into disjoint groups, according to the number of realizations which fall into the first $n/2$ bins. Let $q_k$ be the probability that these bins contain exactly $k$ realizations. Because of the symmetry, $q_k=q_{N-k}$ for each $k$. Since other $n/2$ bins contain exactly $N-k$ realizations, the conditional distribution of the realizations in the first $n/2$ bins is again uniform. Having all this in mind and applying the partition theorem to \eqref{eq:connection2} gives
    \begin{equation}
        \label{eq:before_split}
        \begin{gathered}
        F_c(n,N)\\=\frac{n}{2}\sum_{k=0}^Nq_k\left(\E{|x_2-x_1|\mid k}+\E{|x_n-x_{n-1}|\mid N-k}\right)\\
        =\sum_{k=0}^N q_k\left[F_c\left(\frac{n}{2},k\right)+F_c\left(\frac{n}{2},N-k\right)\right].
        \end{gathered}
    \end{equation}
    Applying \eqref{eq:concave} and the equality $\left(\sum_{k=0}^{N}q_k\right)=1$ leads to the following inequality that holds for each even $n>4$:
    \begin{equation}
        \label{eq:split}
        F_c(n,N)<2 F_c\left(\frac{n}{2},\frac{N}{2}\right).
    \end{equation}
    Here $n$ has to be greater than $4$ because having $n=4$ effectively leads to use of the function $F(2, N)$ on the right side of the inequality, and as explained earlier, this is inappropriate for \eqref{eq:concave}. If $n=k2^r$ where $k\ge 3$ and $r\geq0$ are integers, then taking the inequality above recursively leads further to
	\begin{equation}
	\label{eq:recursion}
		F_c(k2^r,N)<2^{r}F_c\left(k,\frac{N}{2^{r}}\right)
	\end{equation}
	wherefrom for all suitable $N$ and $n$ it then follows that
	\begin{equation}
	\label{eq:lower_bound}
		F_c(n,N)<\frac{n}k\,F_c\left(k,\frac{kN}{n}\right).
	\end{equation}
    If $k=2$ is taken, then the inequality is no longer necessarily valid because of the involvement of $F(2, N)$. However, the obtained form yields a better approximation of $F_c(n, N)$ as
	\begin{equation*}
	F_c(n,N)\approx\frac{n}2\,F_c\left(2,\frac{2N}{n}\right)
	\end{equation*}
	wherefrom after applying \eqref{eq:connection} it then further follows that
	\begin{equation*}
	F(n,N)\approx (n-1)\,F\left(2,\frac{2N}{n}\right),
	\end{equation*}
	which in turn is an approximation stipulated in Hypothesis~\ref{hp:1-f(n,N)}.

\subsubsection{Approximation error}
\label{subsubsec:approximation_error}

Fig.~\ref{fig:variation} shows the difference between the results of \eqref{eq:n-dimensional} and \eqref{eq:asymptotic}, which represent the exact and approximated values of $F(n, N)$, respectively. It can be seen that in cases where $N$ is several times greater than $n$, the approximation error becomes relatively insignificant for practical purposes. The error only becomes significant when the value of $N$ is relatively close to the value of $n$ or below it, but it must be additionally stressed that this rarely occurs in practice since having such values of $n$ and $N$ is not too useful. The plots in Fig.~\ref{fig:error} further suggests that if required, the approximation error could be modelled accurately. However, for the later use here it is enough to conclude that having a sufficiently large value of $N$ renders the approximation error insignificant.

\subsection{Non-uniform distribution}
\label{subsec:non-uniform}

The second case is when the distribution of the sample and consequently the distribution of the histogram are not uniform. In other words this is the case where~\eqref{eq:uniform_p} does not hold, i.e. when $p_i\neq p_j$ for at least one pair of $i$ and $j$. Applying to~\eqref{eq:exixj2_all} all steps that have led to~\eqref{eq:upper_bound_uniform} gives
\begin{equation}
\label{eq:non-uniform}
\begin{aligned}
&\sum_{i=1}^{n-1}\E{\left|x_i-x_j\right|}\\
&\leq\sum_{i=1}^{n-1}\sqrt{N^2\left(p_{i+1}{-}p_{i}\right)^2{+}N\left(p_{i+1}{+}p_{i}
{-}\left(p_{i+1}{-}p_{i}\right)^2\right)}\\
&\leq
\sum_{i=1}^{n-1}\left(\sqrt{N^2\left(p_{i+1}-p_{i}\right)^2}\right.\\
&\qquad\qquad\left.+\sqrt{N\left(p_{i+1}+p_{i}-\left(p_{i+1}-p_{i}\right)^2\right)}\right)\\
&=\sum_{i=1}^{n-1}\left(N\left|p_{i+1}{-}p_{i}\right|+\sqrt{N\left(p_{i+1}{+}p_{i}{-}
\left(p_{i+1}{-}p_{i}\right)^2\right)}\right)\\
&=N\sum_{i=1}^{n-1}\left|p_{i+1}{-}p_{i}\right|+\sqrt{N}\sum_{i=1}^{n-1}
\sqrt{\left(p_{i+1}{+}p_{i}{-}\left(p_{i+1}{-}p_{i}\right)^2\right)}.
\end{aligned}
\end{equation}
	If ${\cal D}$ is the sample's theoretical distribution, then the first term of~\eqref{eq:non-uniform} is the discrete total variation of ${\cal D}$ that is given as
	\begin{equation}
	\label{eq:theoretical_variation}
		\V{{\cal D}}=\sum_{i=1}^{n-1}|p_{i+1}-p_i|.
	\end{equation}
\begin{figure*}[htb]
    \centering
    
    \subfloat[]{
    \includegraphics[width=0.33\linewidth]{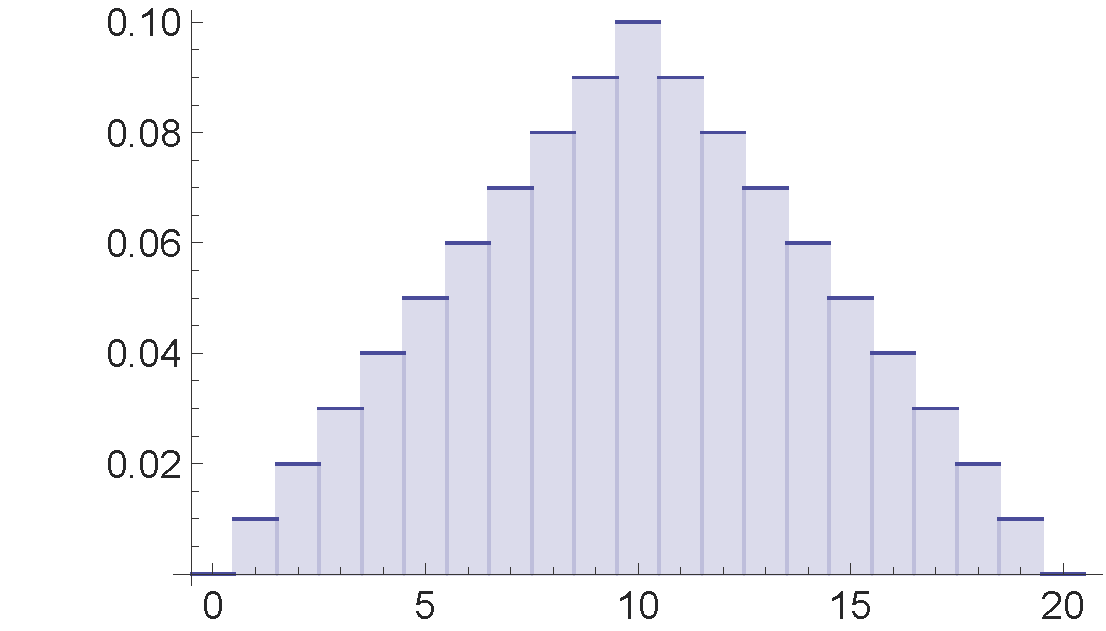}
    \label{fig:triangular}
    }%
    \subfloat[]{
    \includegraphics[width=0.33\linewidth]{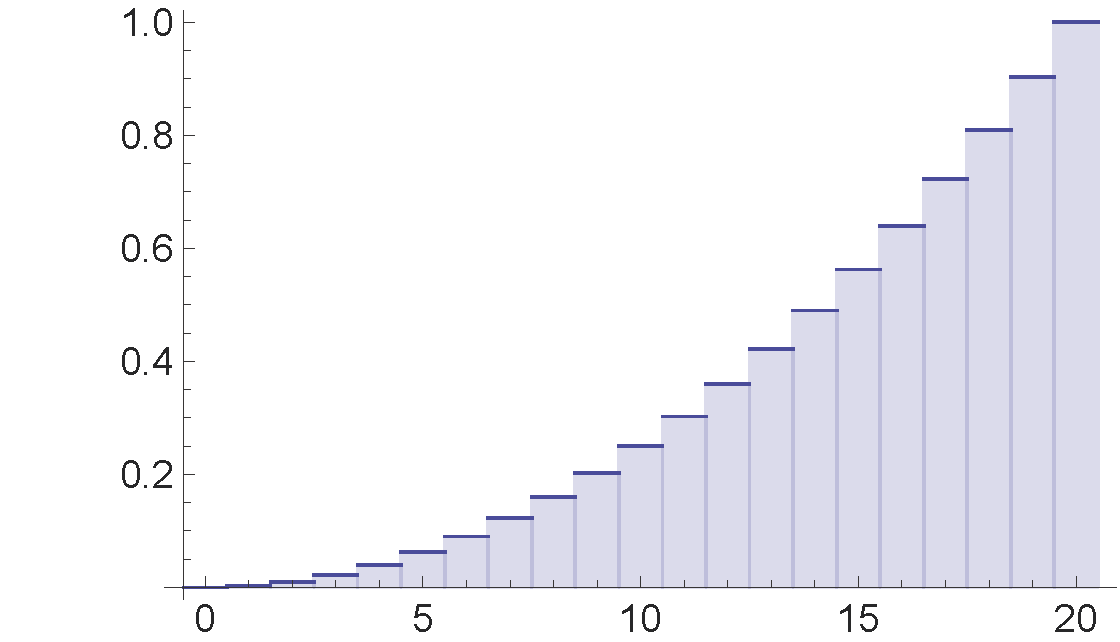}
    \label{fig:quadratic}
    }%
    \subfloat[]{
    \includegraphics[width=0.33\linewidth]{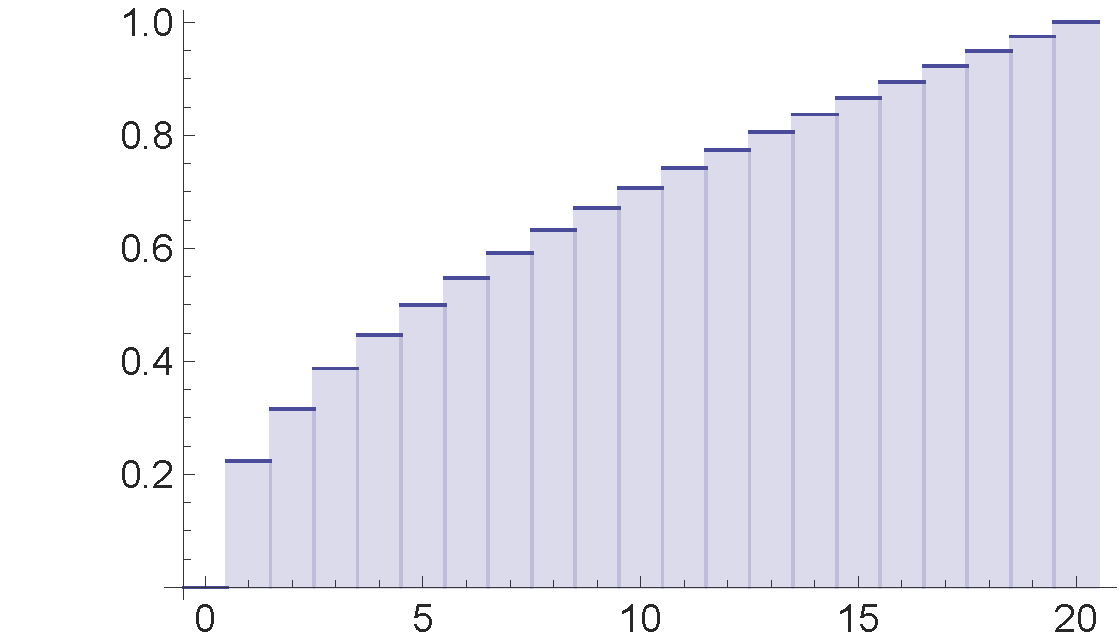}
    \label{fig:squareroot}
    }
    \\
    \subfloat[]{
    \includegraphics[width=0.33\linewidth]{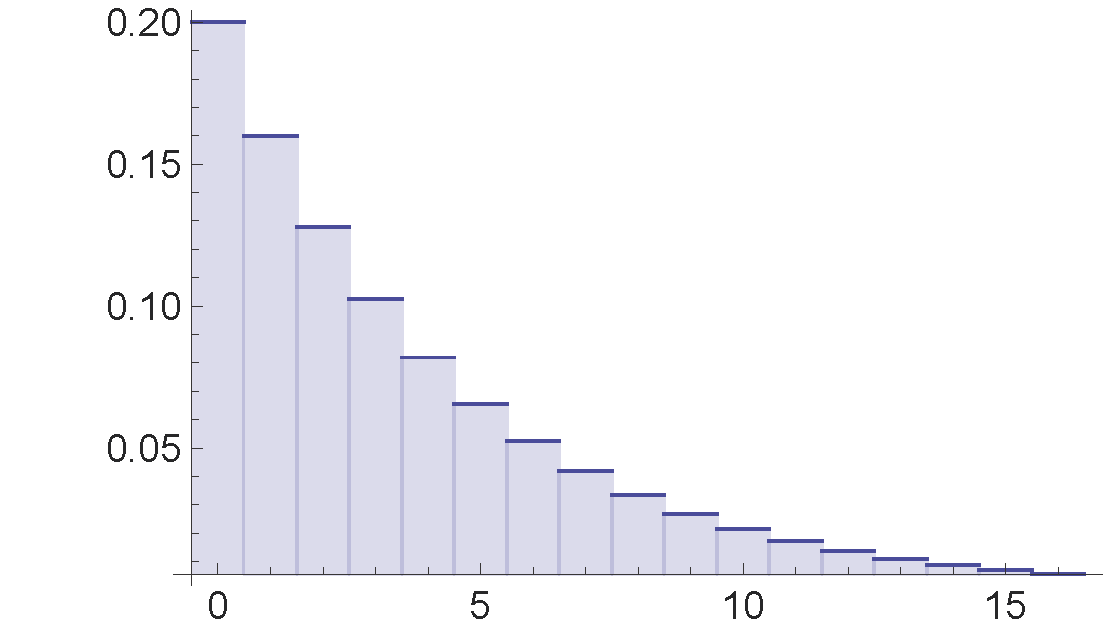}
    \label{fig:geometric}
    }%
    \subfloat[]{
    \includegraphics[width=0.33\linewidth]{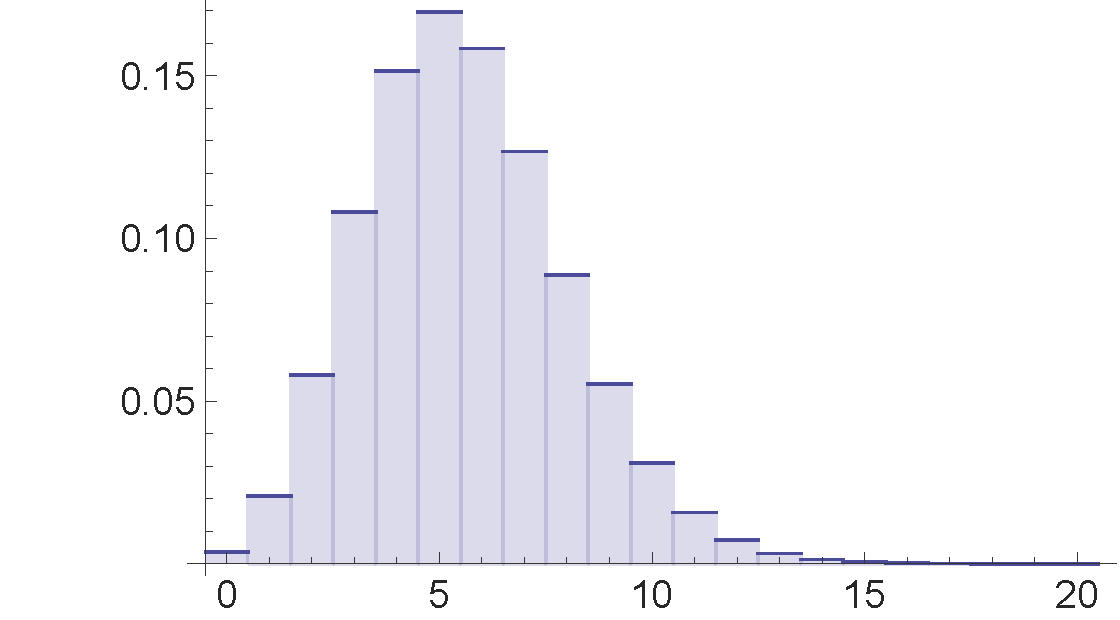}
    \label{fig:poisson}
    }%
    \subfloat[]{
    \includegraphics[width=0.33\linewidth]{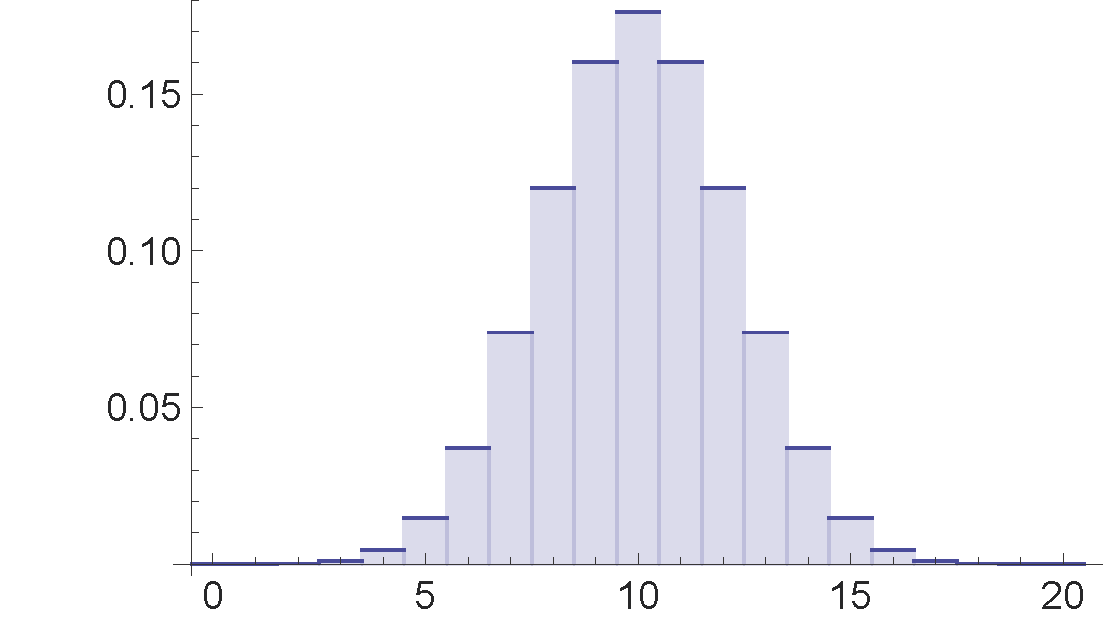}
    \label{fig:binomial}
    }

    \caption{Histograms for the a)~triangular, b)~quadratic, c)~square root, d)~geometric, e)~Poisson, and f) binomial distribution. The shown histograms are merely for the sake of illustration and the $x$-axes do not strictly follow the equations in Section~\ref{subsec:non-uniform}.}
  \label{fig:distributions}
    
\end{figure*}
\begin{figure*}[htb]
    \centering
    \subfloat[]{
    \includegraphics[width=0.33\linewidth]{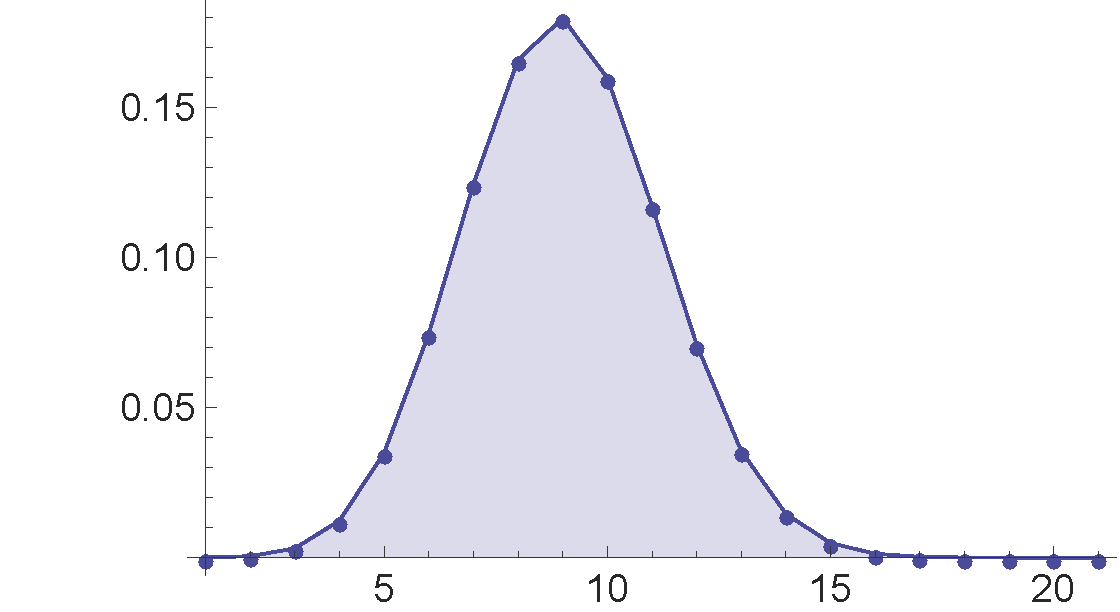}
    \label{fig:distribution1}
    }%
    \subfloat[]{
    \includegraphics[width=0.33\linewidth]{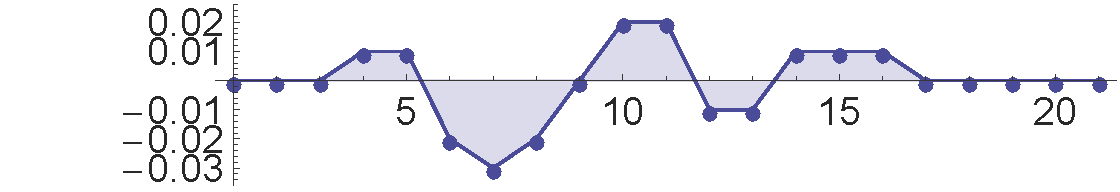}
    \label{fig:distribution2}
    }%
    \subfloat[]{
    \includegraphics[width=0.33\linewidth]{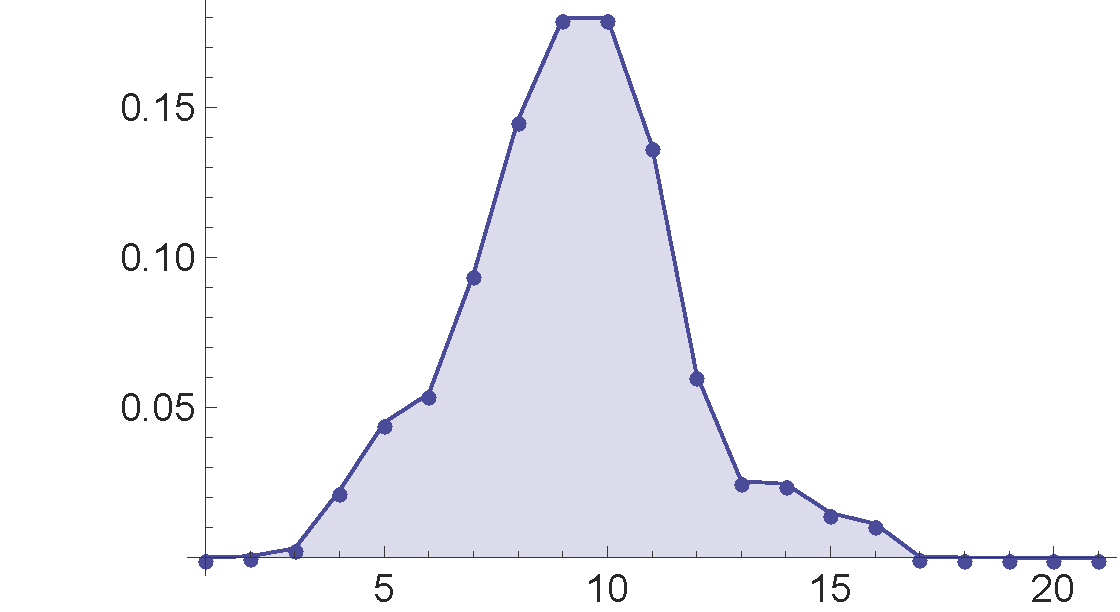}
    \label{fig:distribution3}
    }
    \caption{From a)~the theoretical distribution by adding b)~the deviation due to randomness to c)~the final sample distribution.}
    \label{fig:partition}
\end{figure*}
	The second term is a bound for expectation of the deviation of this given sample from its theoretical distribution.
	A rough estimate for this second term is the value $2\sqrt{n-1}\sqrt{N}$. It is obtained by first removing the subtracting part and applying the inequality $\sqrt{u+v}\leq\sqrt{u}+\sqrt{v}$ for $u, v>0$, which gives
	\begin{equation}
    \label{eq:remove_subtraction}
    \begin{gathered}
	\sum_{i=1}^{n-1}\sqrt{\left(p_{i+1}+p_{i}-\left(p_{i+1}-p_{i}\right)^2\right)}\\
	\leq\sum_{i=1}^{n-1}\sqrt{\left(p_{i+1}+p_{i}\right)}\le\sum_{i=1}^{n-1}\sqrt{p_i}+ \sum_{i=1}^{n-1}\sqrt{p_{i+1}}.
	\end{gathered}
	\end{equation}
	Since $\sqrt{p_i}$ and $\sqrt{p_{i+1}}$ are non-negative, the sums in \eqref{eq:remove_subtraction} can effectively be seen as $L_1$-norms of $(n-1)$-dimensional vectors. Applying the inequality $\Norm{\mathbf{v}}{1}\leq\sqrt{d}\Norm{\mathbf{v}}{2}$ where $d$ is the dimension of the vector $\mathbf{v}$~\cite{datta2004numerical} to these sums gives
	\begin{equation}
    \label{eq:l1_vs_l2}
    \begin{gathered}
	\sum_{i=1}^{n-1}\sqrt{p_i}+\sum_{i=1}^{n-1}\sqrt{p_{i+1}}\\
	\le\sqrt{n-1}\left[\left( \sum_{i=1}^{n-1} p_i\right)^{1/2}+ 
	\left(\sum_{i=i}^{n-1}p_{i+1}\right)^{1/2}\right]\\
	\le \sqrt{n-1}\left(1+1\right)=2\sqrt{n-1}.
	\end{gathered}
	\end{equation}
	
	It is useful to know the discrete total variation of some important distributions. Examples of their histograms are shown in Fig.~\ref{fig:distributions}. The uniform distribution has a zero total variation. For the triangular distribution ${\cal T}$ with $n$ bins this is
	\begin{equation}
	    \label{eq:triangular_even}
		\V{{\cal T}}=\frac{4n-8}{n^2}\approx \frac{4}{n+2}
	\end{equation}
	for an even $n$, while in the case of an odd $n$ this is given as
	\begin{equation}
	    \label{eq:triangular_odd}
		\V{{\cal T}}=\frac{4n-6}{n^2}\approx \frac{4}{n+2}.
	\end{equation}
	The square distribution $\cal Q$ for which $p_i=Ci^2$ with $n$ bins has a discrete total variation that can be approximated as
	\begin{equation}
	    \label{eq:square}
		\V{{\cal Q}}\approx\frac{3}n.
	\end{equation}
	Next, in the case of the square root distribution $\cal S$ for which $p_i=C\sqrt{i}$ and with $n$ bins the approximation is given as
	\begin{equation}
	    \label{eq:square_root}
		\V{{\cal S}}\approx\frac{3}{2n}.
	\end{equation}
    For the geometric distribution $\cal G$ with parameter $p$ this is
	\begin{equation}
	    \label{eq:geometric}
		\V{{\cal G}}=p,
	\end{equation}
	for the Poisson distribution ${\cal P}$ with parameter $\lambda>1$ it is
	\begin{equation}
	    \label{eq:poisson}
		\V{{\cal P}}\approx\frac{2\lambda^{\lfloor \lambda\rfloor}e^{-\lambda}}{\lfloor\lambda\rfloor!}.
	\end{equation}
	The discrete total variation for a unimodal discrete distribution with mode $M$ is bounded by $2M$. The mode for symmetric binomial distribution ${\cal B}(n,\frac12)$ is $\frac{1}{2^n}\binom{n}{\lfloor n/2\rfloor}$ and
	\begin{equation}
	    \label{eq:unimodal}
		\V{{\cal B}}\approx\sqrt{\frac{8}{\pi n}}.
	\end{equation}
	The normal distribution ${\cal N}(0,\sigma^2)$ is a continuous one with unbounded support and its theoretical DTV depends on rasterization. The total variation of the probability density function is
	$\dfrac{2}{\sigma\sqrt{2\pi}}$. If $[-c,c]$ is essentially the support of the distribution and if $n\ge\dfrac{2c}{\sigma}$, then $\V{{\cal N}}$ can be approximated:
	\begin{equation}
	    \label{eq:normal}
		\V{{\cal N}}\approx\frac{2c}{n\sigma}\sqrt{\frac{2}{\pi}}.
	\end{equation}

%Since what is looked for is the relation between $N$ and $\E{\V{\mathbf{x}_n}}$, the probabilities $p_i$ are fixed and they can be considered as constants, which can then simplify~\eqref{eq:non-uniform} to
	Let ${\cal D}$ be any distribution and $\mathbf{x}_n$ the histogram with $n$ bins of a corresponding sample of $N$ values drawn from the distribution ${\cal D}$. Then similarly to~\eqref{eq:non-uniform} it can be written
	\begin{equation}
	\label{eq:non-uniform2}
		\E{\V{\mathbf{x}_n}}\le\V{{\cal D}}\cdot N+\E{\V{{\cal R}}}\sqrt{N}
	\end{equation}
	where ${\cal R}$ is a deviation from the theoretical distribution. If there was no randomness and all values were distributed exactly as predicted by the probabilities, then $\E{\V{\mathbf{x}_n}}$ would be $\V{{\cal D}}\cdot N$. Therefore, the second term is due to the randomness. A further thing to notice here is that as $N$ grows, randomness plays an ever smaller role in~\eqref{eq:non-uniform2} and as $N$ limits at infinity, the term $C_1N$ gets to fully dominate in~\eqref{eq:non-uniform2}, which is also expected in accordance with the Glivenko-Cantelli theorem. In Fig.~\ref{fig:partition} the total variation of the theoretical distribution and the total variation of a sample are equal. This will be the case for all samples which do not alter order between adjacent bins. Therefore, the alteration from the theoretical distribution means that the corresponding sample is essentially different from theoretical one. Deviation from the theoretical distribution can be approximated as total variation of a sample from uniform distribution and therefore the bounds written before can be applied to any distribution.
	
	With regard to the distribution, the use of the discrete total variation that is somewhat similar to the $L_1$-norm may be reminiscent of the assumption of the Laplace distribution. However, no minimization, regularization, or any similar process that requires such an assumption is being performed here. Therefore, it should be stressed again that the relations obtained here can be applied to samples of any distribution.

\subsection{The proposed model}
\label{subsec:model}

After taking into account the previous subsections' results, it is reasonable to consider the model for $\E{\V{\mathbf{x}_n}}$ to be 
\begin{equation}
\label{eq:non-uniform_model}
\begin{gathered}
m=aN+b\sqrt{N}.
\end{gathered}
\end{equation}
This model can be fitted directly to the sizes and discrete total variations obtained on the given histograms that are to be checked for outliers. If there is not enough given histograms to cover the desired value ranges of $N$, then additional ones can be created by randomly subsampling the given ones. In the case where a larger amount of histogram outliers is suspected, then their detrimental influence on fitting of~\eqref{eq:non-uniform_model} can be reduced by applying methods such as RANSAC~\cite{fischler1981random}.

Alternatively, if the distribution, i.e. the values of $p_i$ for the histograms' bins are known, then $a$ and $b$ can be obtained through Monte Carlo simulation by randomly creating arbitrarily many histograms of various sizes $N$ and then fitting the model~\eqref{eq:non-uniform_model} to their sizes and discrete total variations.

\subsection{Score calculation}
\label{subsec:score}

Once the model described by~\eqref{eq:non-uniform_model} has been fitted to data, the next step is to assign an outlier score to each of the given histograms. The first step is to calculate a histogram's discrete total variation. Next, the discrete total variation expected for the histograms's size is obtained by using~\eqref{eq:non-uniform_model}. Finally, the absolute difference between these two values is
\begin{equation}
\label{eq:d}
\begin{gathered}
d=\left|\V{\mathbf{x}_n}-m\right|.
\end{gathered}
\end{equation}
However, $d$ cannot yet be used as the score because the standard deviation of the discrete total variation for histograms of random samples varies depending on the samples size $N$, which means that the significance of $d$ depends on $N$. This means that first the influence of the sample size on the standard deviation has to be removed. Additionally, the discrete total variation is already a statistic of the sample, which means that its standard deviation is actually the standard error~\cite{everitt2002cambridge}. Many standard errors that do not include division by $N$ are proportional or close to being proportional to $\sqrt{N}$, at least in limit, and in practice this is also the case with the discrete total variation. This can intuitively be seen in the form of the second term of~\eqref{eq:non-uniform2} as discussed earlier. Therefore, for practical purposes the influence of $N$ on $d$ can be approximately removed by calculating the distance $d'$ as
\begin{equation}
\label{eq:d2}
\begin{gathered}
d'=\frac{d}{\sqrt{N}}=\frac{\left|\V{\mathbf{x}_n}-m\right|}{\sqrt{N}}=\frac{\left|\V{\mathbf{x}_n}-aN+b\sqrt{N}\right|}{\sqrt{N}}.
\end{gathered}
\end{equation}
The value of $d'$ can now be used instead of the value $d$ as the outlier score for the histogram that it was calculated for because it is normalized with respect to the standard error.

It must be mentioned that strictly speaking~\eqref{eq:d2} is theoretically not correct because the expected value of the discrete total variation is not always proportional to $N$. However, during the research conducted for this paper it has been empirically shown that for all tested distributions the standard error was proportional to $\sqrt{N}$ and that using~\eqref{eq:d2} is a good practice, even though it may introduce inaccuracies. Since~\eqref{eq:d2} was specifically designed to comply with the statistical properties related to the discrete total variation as discussed here, using some other score calculation would potentially require a major overhaul of the whole framework.

An alternative to using~\eqref{eq:d2} that unlike~\eqref{eq:d2} does not include a explicitly derived formula is to take all data from the given histograms, use it in Monte Carlo simulations to create samples of various desired sized, for each of these sizes calculate the discrete total variations and their standard deviation, and fit a model to these sizes and their respective standard deviations. If enough data is available, this should result in a relation that is very similar to the one in~\eqref{eq:d2}.

Since $d'$ is the normalized distance from the expected discrete total variation and since it resembles the $t$-statistic, it could be further used to also provide a probabilistic interpretation for a given histogram. However, the goal of this paper is not to propose a new statistical test that can be used in hypothesis testing with predetermined significance levels. The main goal of this paper is just to find the most likely outlier candidates based on the discrete total variation and the distance $d'$ also suffices for such ranking. Therefore, probabilistic interpretation calculation is omitted in this paper.

\subsection{Application}
\label{subsec:application}

With all the required background given in the previous subsections, it is possible to propose a new method for detecting histogram outliers in terms of the discrete total variation.

First, multiple histograms for the samples of various sizes are given as input. The histograms are supposed to have the same bins where each of the bins can have an arbitrary interval. It is also supposed that all these samples are drawn from the same distribution and the goal is to check which of them are most likely to be outliers in terms of the discrete total variation. Next, the discrete total variation is calculated for each of these histograms. Then, model~\eqref{eq:non-uniform_model} is fitted to histogram sizes and discrete total variations. Finally, each of the histograms is scored by applying~\eqref{eq:d2}. The histograms for which the highest score values were obtained are the most likely outlier candidates in terms of their discrete total variation. All these steps are summarized in Algorithm~\ref{alg:proposed}.

Here it should be additionally stressed that the proposed method has no hyperparameters whatsoever that would have to be tuned or that would otherwise influence the result. It may seem that the number of histogram bins $n$ is a tunable hyperparameter, but the proposed method is agnostic of the underlying histogram samples - it merely receives already existing histograms as inputs. The histograms are only assumed to have the same bins. It is not even important what the range of the bins is nor is it important whether they are bounded.

\subsection{The proposed method's name}
\label{subsec:name}

Due to the proposed method's model's reliance on the discrete total variation, it was named Total Variation Outlier Recognizer~(TVOR) or for the sake of simplicity just Tvor, which is pronounced /t\textscriptv \^{o}\textlengthmark r/ and it means \textit{skunk} in Croatian.
%\Elzpscrv ʋ

\begin{algorithm}
\caption{The proposed method TVOR}
\label{alg:proposed}
\hspace*{\algorithmicindent}\textbf{Input:} $M$ input histograms $\mathbf{x}_n^{\left(1\right)}, \mathbf{x}_n^{\left(2\right)}, \ldots, \mathbf{x}_n^{\left(M\right)}$\\
\hspace*{\algorithmicindent}\textbf{Output:} scores for input histogram $d'_{1}, d'_{2}, \ldots, d'_{M}$
\begin{algorithmic}[1]
\For{$i\in\{1, 2, \ldots, M\}$}
	\State $s_i=\sum_{j=1}^{n} x_{j}^{\left(i\right)}$ \Comment Calculate sample size
	\State $v_i=\V{\mathbf{x}_{n}^{\left(i\right)}}$ \Comment Calculate discrete total variation
\EndFor
\State $a,b=\text{FitModel}\left(\bigcup_{i=1}^{M}\left(s_i, v_i\right)\right)$ \Comment Fit~\eqref{eq:non-uniform_model} to data
\For{$i\in\{1, 2, \ldots, M\}$}
    \State $d'_i=\frac{\left|v_i-as_i+b\sqrt{s_i}\right|}{\sqrt{s_i}}$ \Comment Calculate the score
\EndFor
\end{algorithmic}
\end{algorithm}

\section{Experimental results}
\label{sec:results}

In order to validate the proposed method, several experiments have been conducted on both synthetic and real-life data. Additionally, it is shown why the proposed method is more appropriate than some other similar methods. To give a clear and descriptive overview of the method's properties, the structure of this section is purposely slightly more extended. First, Section~\ref{subsec:baseline} describes a baseline method for histogram outlier detection based on the Pearson's chi-squared test~\cite{pearson1900x} to compare its results to the ones of the proposed method. In Section~\ref{subsec:synthetic_distribution} the behavior of the proposed method in several scenarios of changing conditions is demonstrated and additionally explained by several experiments for distribution outlier detection among histograms of random samples of different sizes drawn from the normal distribution and the beta distribution with various parameter values. Similar to that, Section~\ref{subsec:synthetic_variation} contains experiments for discrete total variation outlier detection among histograms of random samples of various sizes drawn from the beta distribution. The real-life practical use of the proposed method is demonstrated in Section~\ref{subsec:census} on the histograms of the birth years taken from census data of several populations from the same time frame. Section~\ref{subsec:advantages} shows the advantage of the proposed method over some other methods that can be used for similar purposes. The obtained results are discussed in Section~\ref{subsec:discussion}. The online repository with the source code and the data required to recreate the results is described in Section~\ref{subsec:repository}.

\subsection{The baseline method}
\label{subsec:baseline}

The proposed method's goal is to detect outliers speficically in terms of the expected discrete total variation, which can differ significantly from detecting distribution outliers in general. Therefore, the goal of this section is to show the difference in the performance of the proposed method and the Pearson's chi-squared test~\cite{pearson1900x}. This test can be used to check whether a histogram is an outlier by comparing the values of the histogram's bins, which serve here as the categorical variables, to the values that are expected under a supposed distribution. However, since in the problem that is being analyzed in this paper the supposed distribution is unknown, the expected bin values first have to be estimated.

\begin{algorithm}
\caption{The baseline method}
\label{alg:baseline}
\hspace*{\algorithmicindent}\textbf{Input:} $M$ input histograms $\mathbf{x}_n^{\left(1\right)}, \mathbf{x}_n^{\left(2\right)}, \ldots, \mathbf{x}_n^{\left(M\right)}$\\
\hspace*{\algorithmicindent}\textbf{Output:} scores for input histogram $\chi^2_{1}, \chi^2_{2}, \ldots, \chi^2_{M}$
\begin{algorithmic}[1]
\For{$i\in\{1, 2, \ldots, M\}$}
	\State $s_i=\sum_{j=1}^{M} x_{j}^{\left(i\right)}$ \Comment Calculate sample size
\EndFor
\State $S=\sum_{i=1}^{M}s_i$ \Comment Calculate the sum of all bins
\For{$i\in\{1, 2, \ldots, n\}$}
	\State $b_i=\sum_{j=1}^{M} x_{i}^{\left(j\right)}$ \Comment Calculate individual bin sum
\EndFor
\State $\epsilon=10^{-6}$ \Comment A small positive number
\For{$i\in\{1, 2, \ldots, M\}$}
    \For{$j\in\{1, 2, \ldots, M\}$}
    	\State $O_j^{\left(i\right)}=x_{j}^{\left(i\right)}$ \Comment The observed bin value
    	\State $E_j^{\left(i\right)}=\frac{b_j-x_{j}^{\left(i\right)}}{S-s_i}s_i+\epsilon$ \Comment The expected bin value
    \EndFor
    \State $\chi^2_{i}=\sum_{j=1}^{n}\frac{\left(O_j^{\left(i\right)}-E_j^{\left(i\right)}\right)^2}{E_j^{\left(i\right)}}$ \Comment Calculate the score
\EndFor
\end{algorithmic}
\end{algorithm}

\begin{figure}[htb]
    \centering
    
	\includegraphics[width=\linewidth]{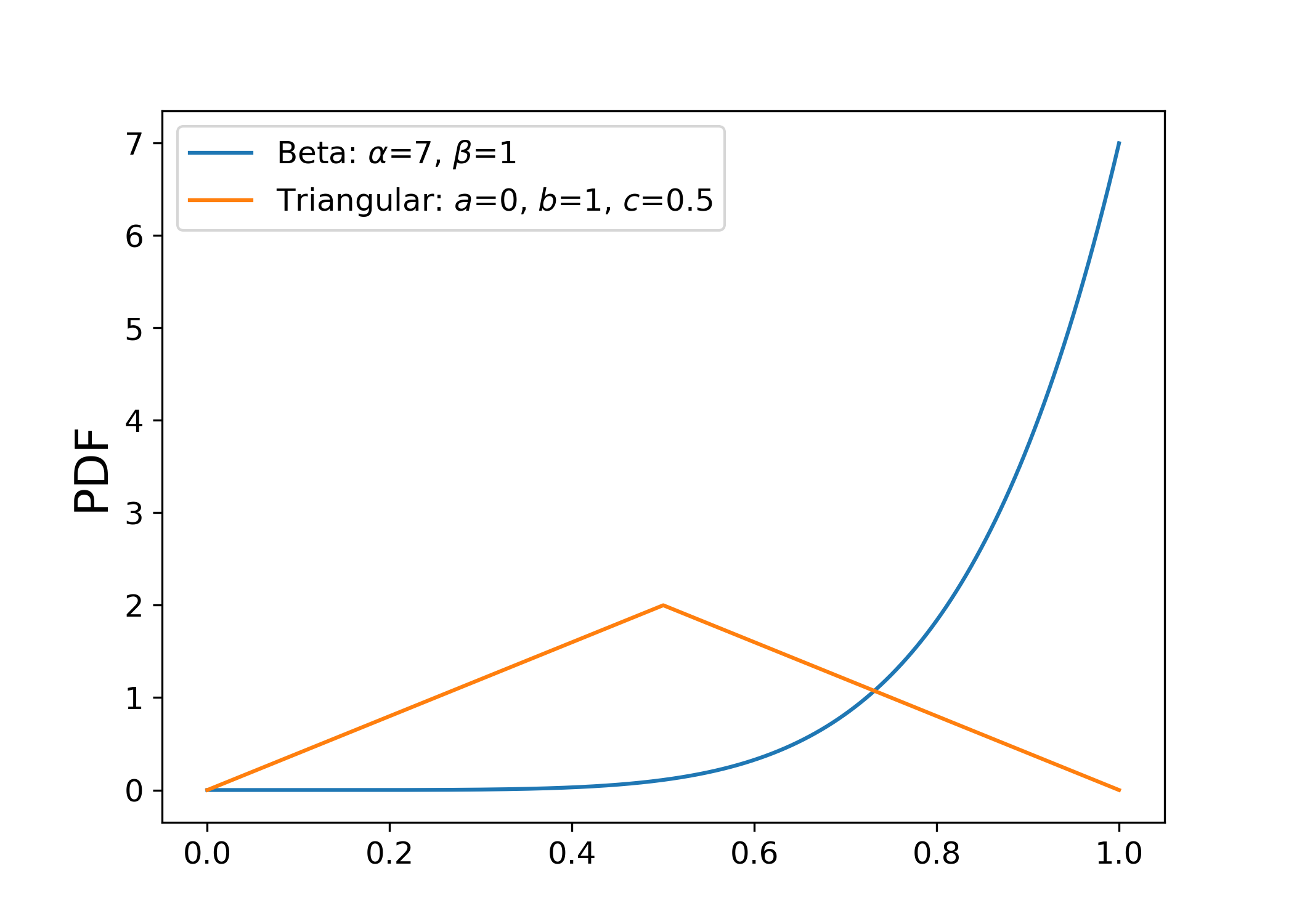}
	
    \caption{The probability density functions of the beta distribution and triangular distribution used in the experiments.}
	\label{fig:beta_triangular_pdf}
    
\end{figure}

\begin{figure*}[htbp]
    \centering
    
  \subfloat[]{
  \includegraphics[width=0.48\linewidth]{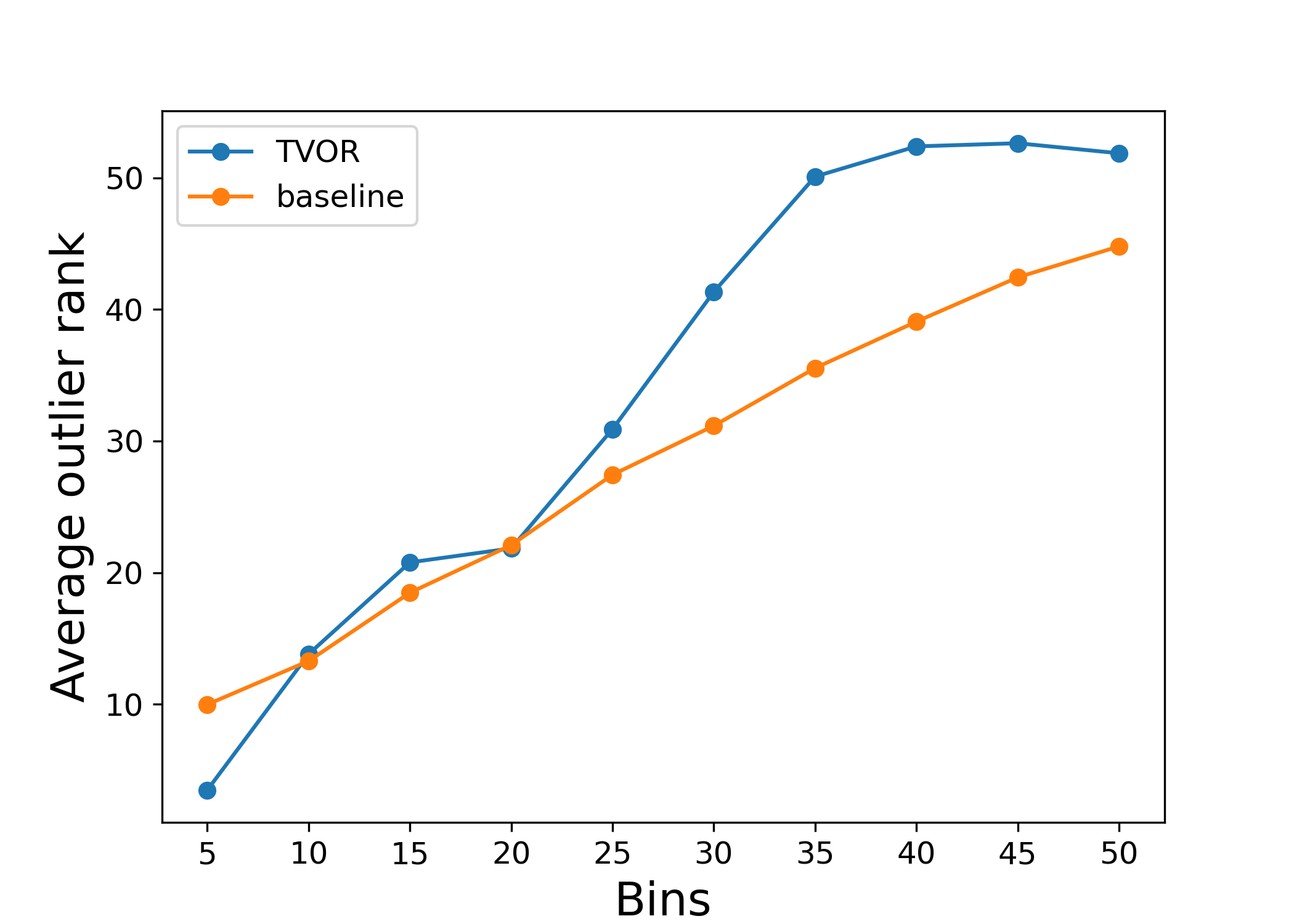}
  \label{fig:sigma_a}
  }%
  \subfloat[]{
  \includegraphics[width=0.48\linewidth]{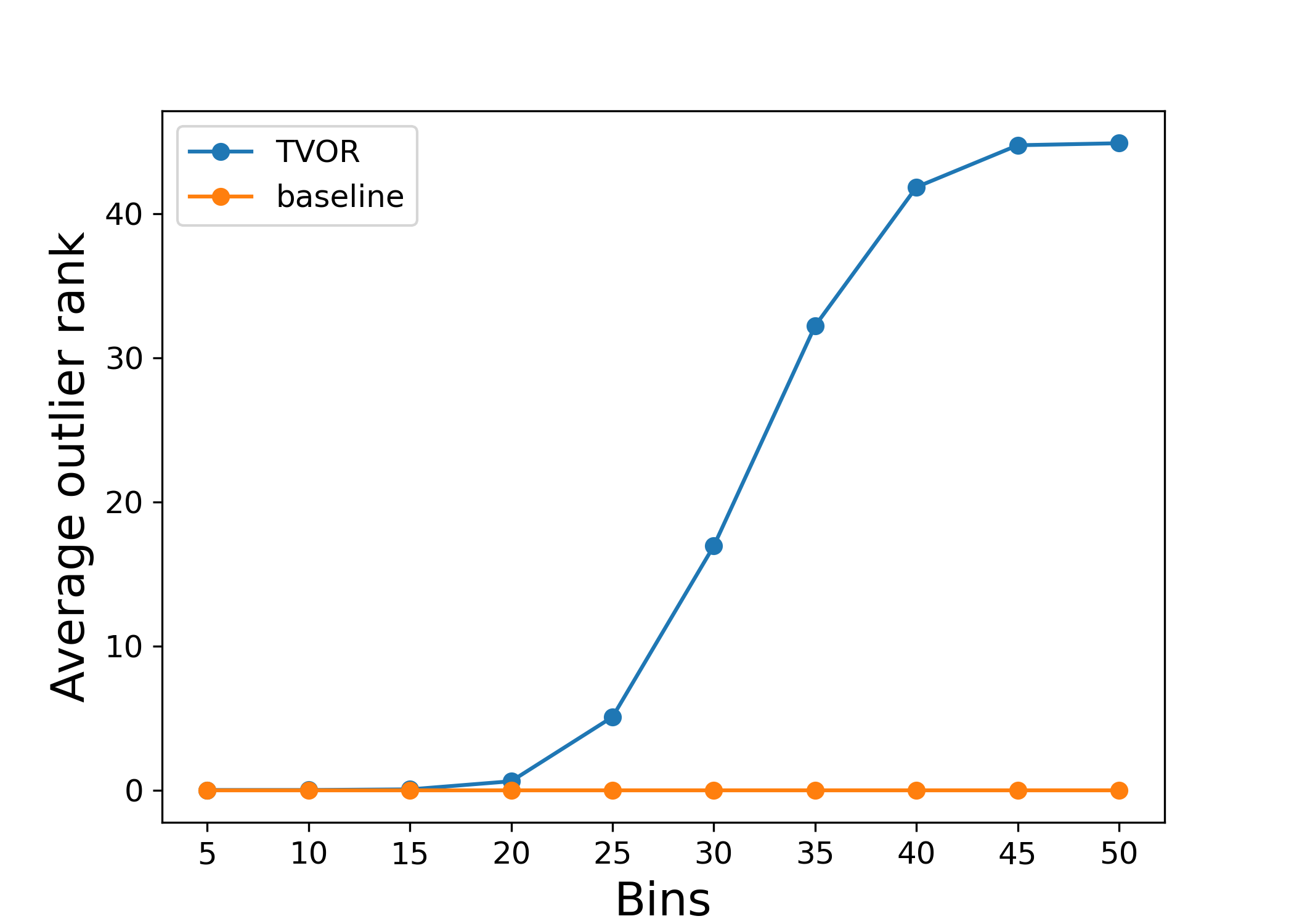}
  \label{fig:sigma_b}
  }
  \\
  \subfloat[]{
  \includegraphics[width=0.48\linewidth]{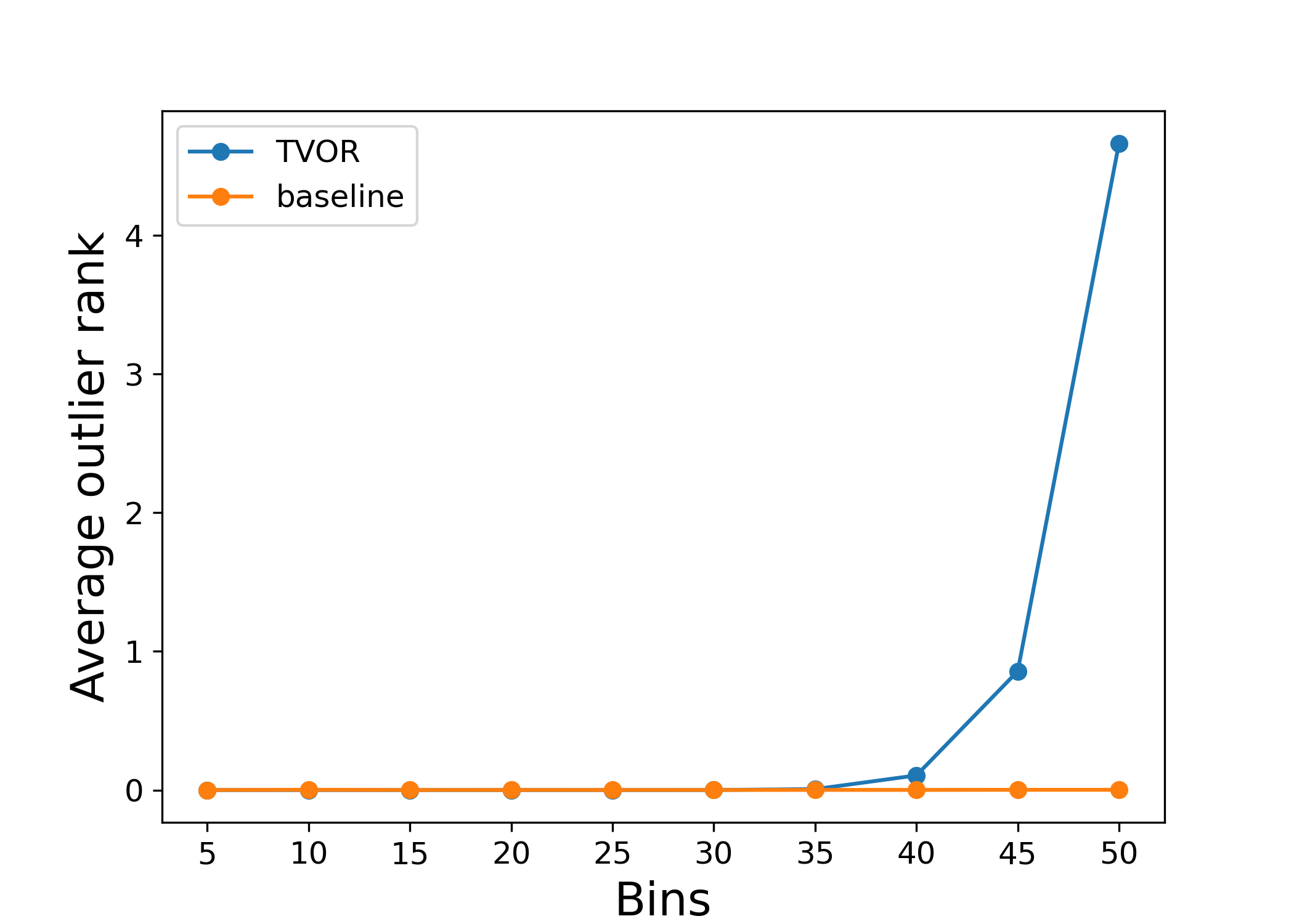}
  \label{fig:b_a}
  }%
  \subfloat[]{
  \includegraphics[width=0.48\linewidth]{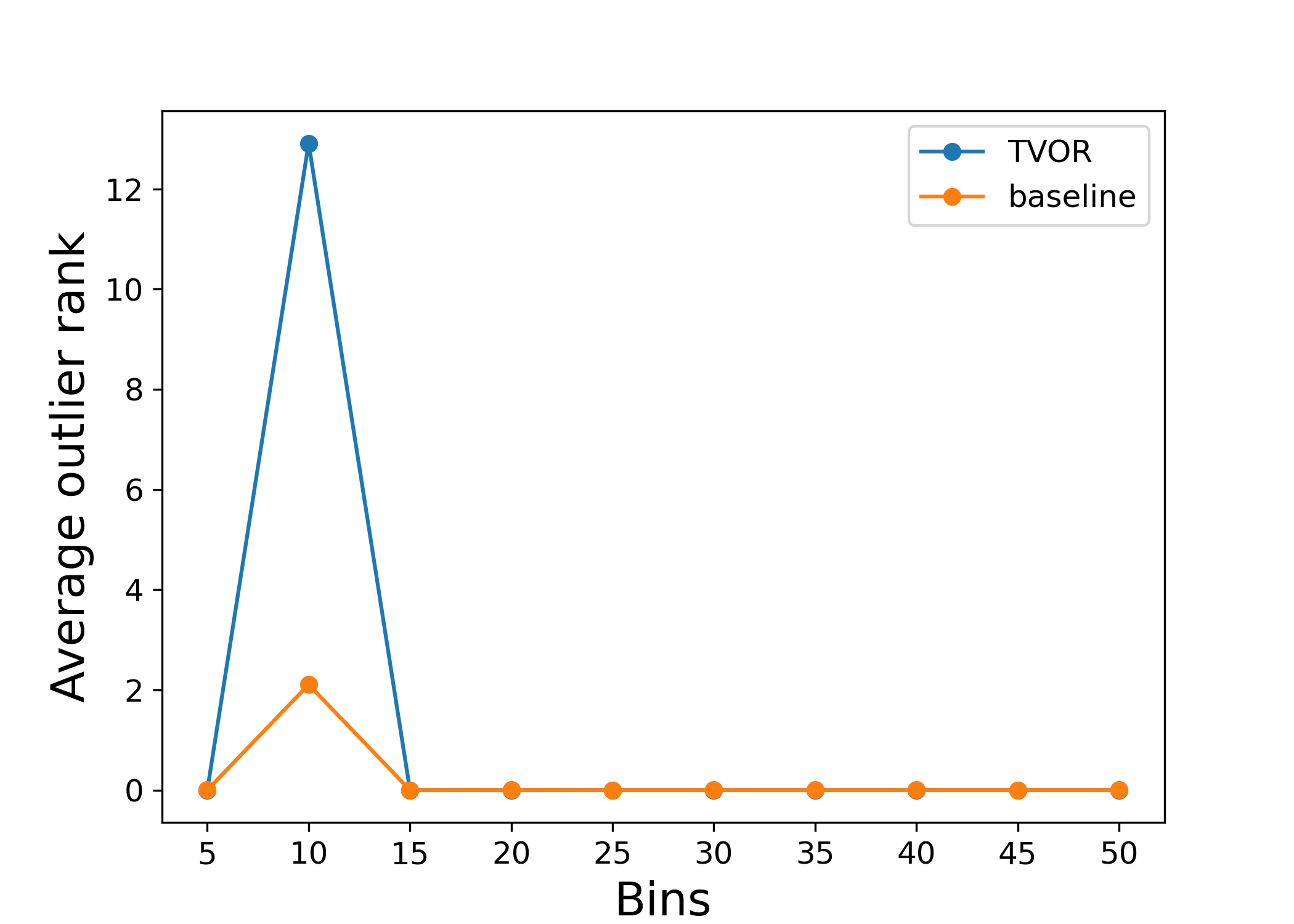}
  \label{fig:b_b}
  }
  \\
  \subfloat[]{
  \includegraphics[width=0.48\linewidth]{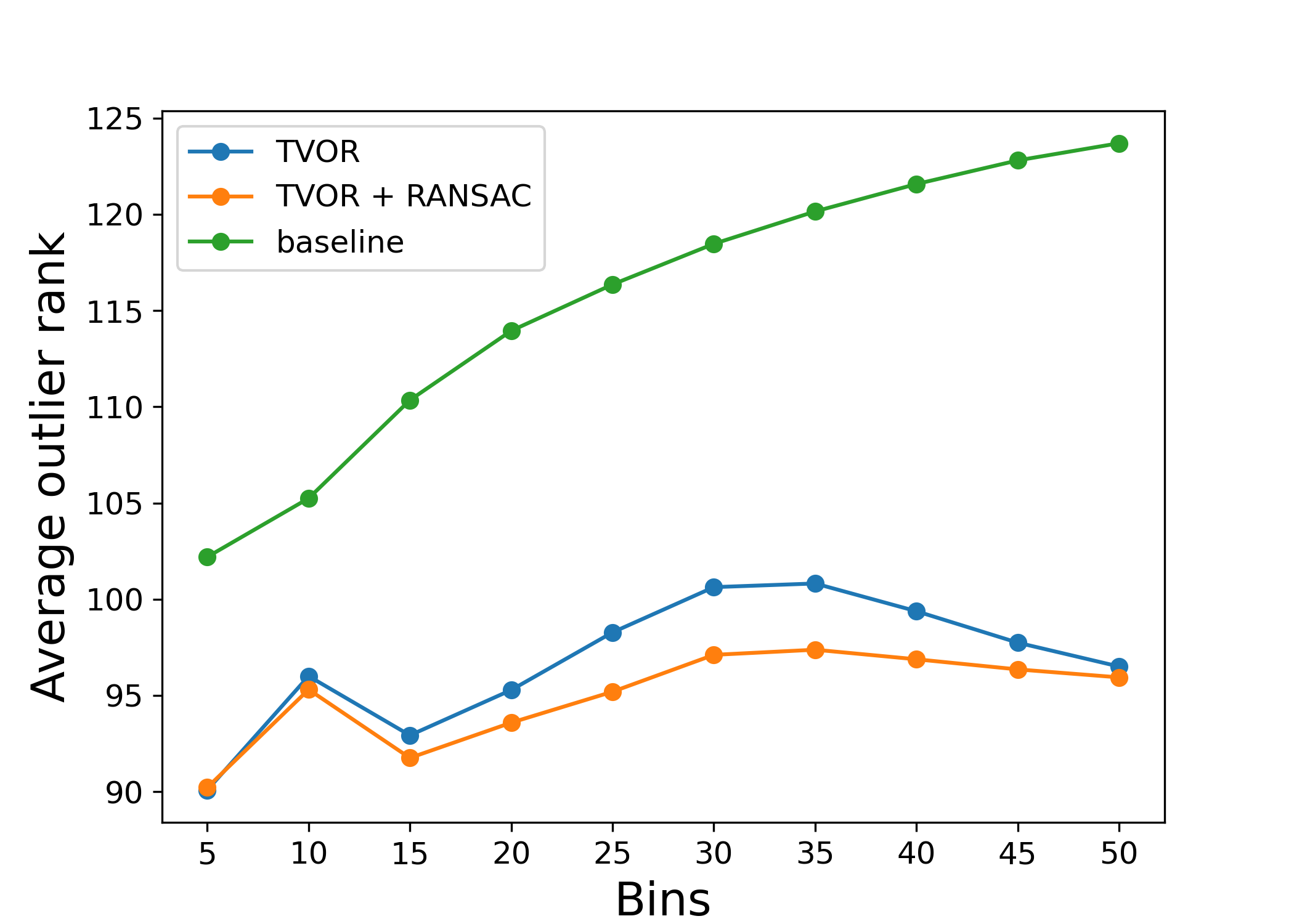}
  \label{fig:osc_a}
  }%
  \subfloat[]{
  \includegraphics[width=0.48\linewidth]{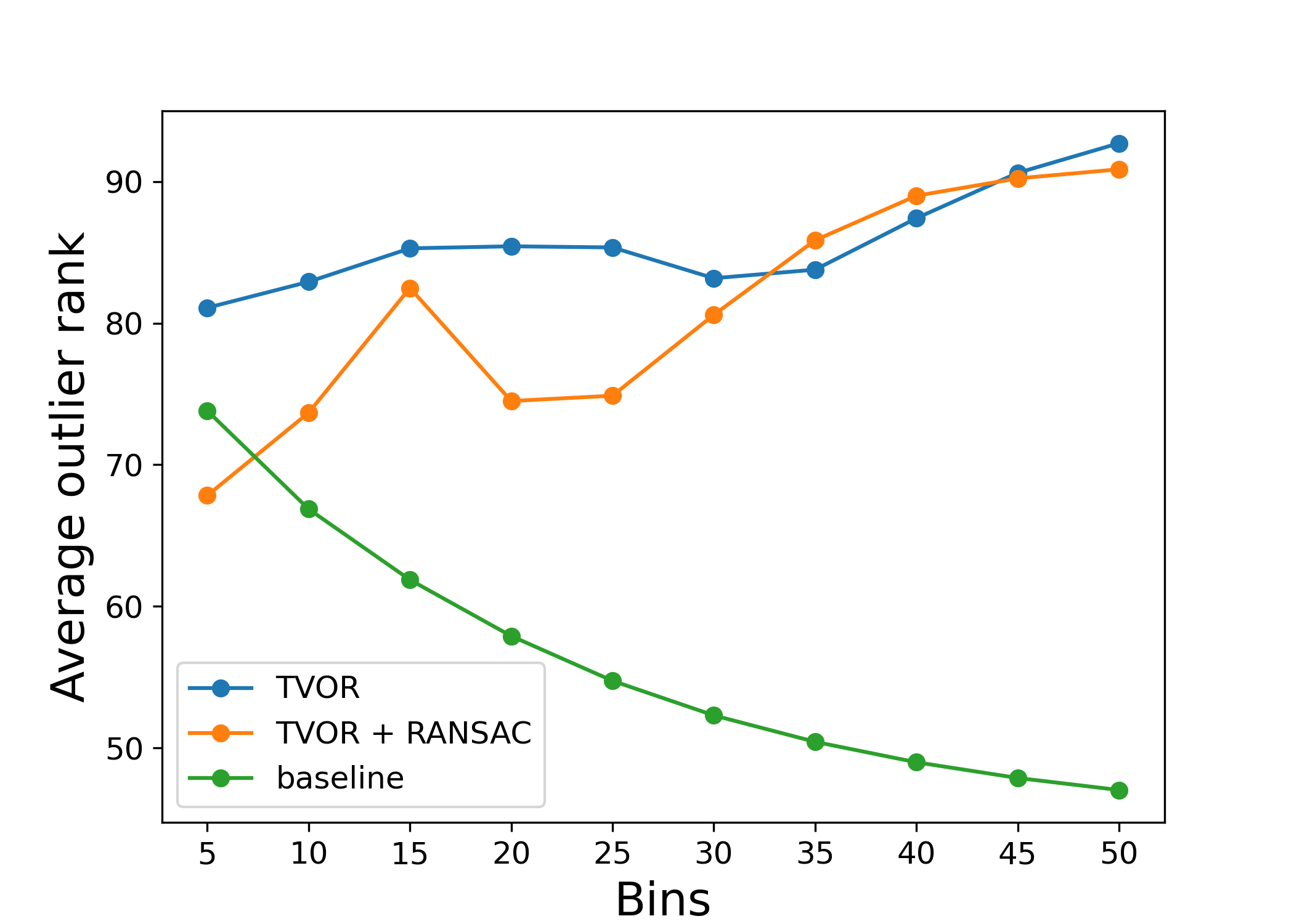}
  \label{fig:osc_b}
  }

    \caption{Comparing the performance of the proposed and baseline methods. \textbf{First row}: performance with $1$ added outlier and $c=5$ for a)~$\sigma=0.9$ and b)~$\sigma=1.5$. \textbf{Second row}: Performance with $1$ added outlier and $\sigma=0.5$ for c)~$c=5$ and d)~$c=10$. \textbf{Third row}: Performance with $90$ added outliers and $c=5$ for e)~$\sigma=0.9$ and f)~$\sigma=1.5$. The results for TVOR~+~RANSAC was added only in the third row because for the results in the first and the second row the difference was not that significant.}
  \label{fig:tvor_vs_baseline}
    
\end{figure*}

\begin{figure*}[htbp]
    \centering
    
	\subfloat[]{
	\includegraphics[width=0.48\linewidth]{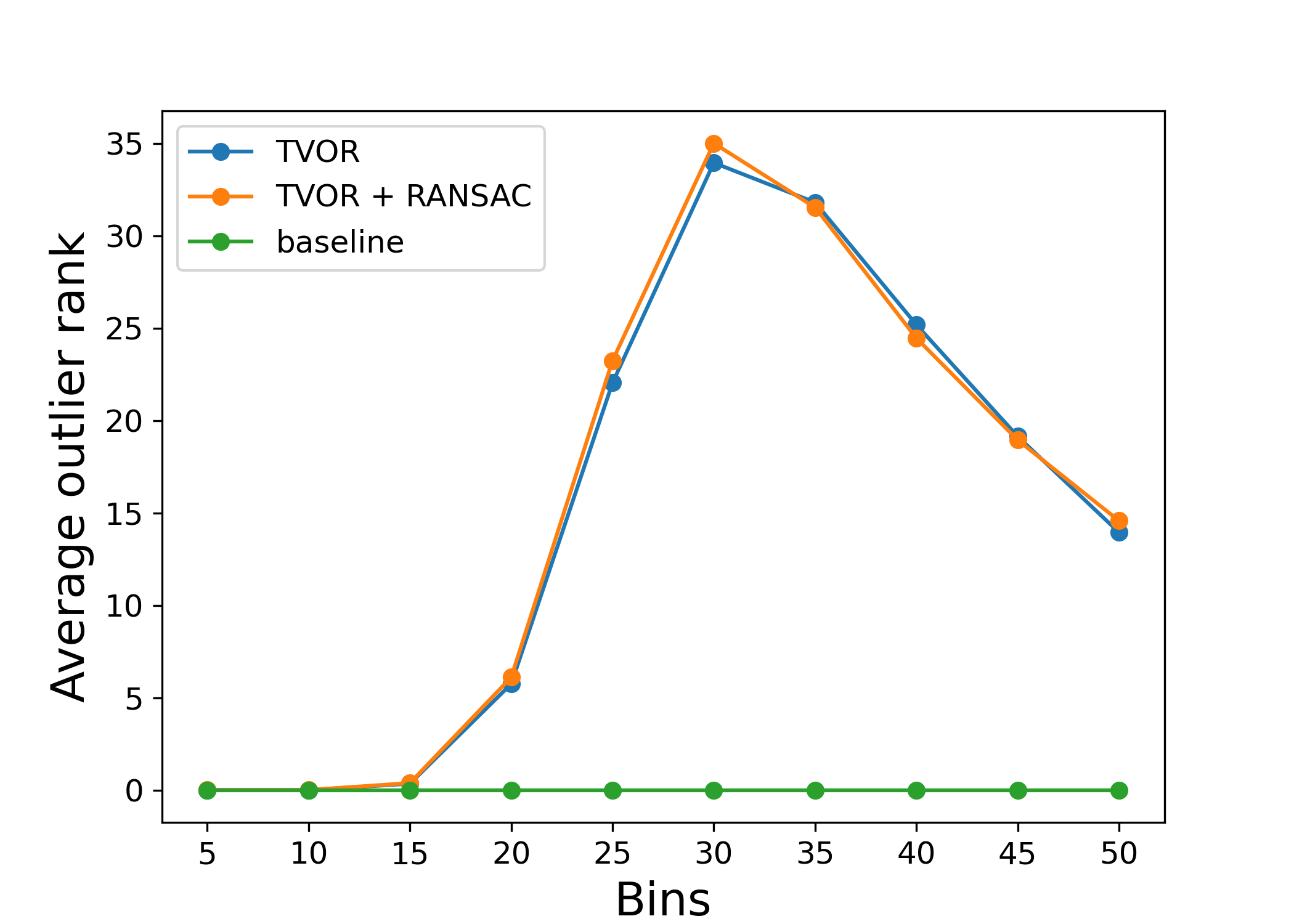}
	\label{fig:beta_triangular_1}
	}
	\subfloat[]{
	\includegraphics[width=0.48\linewidth]{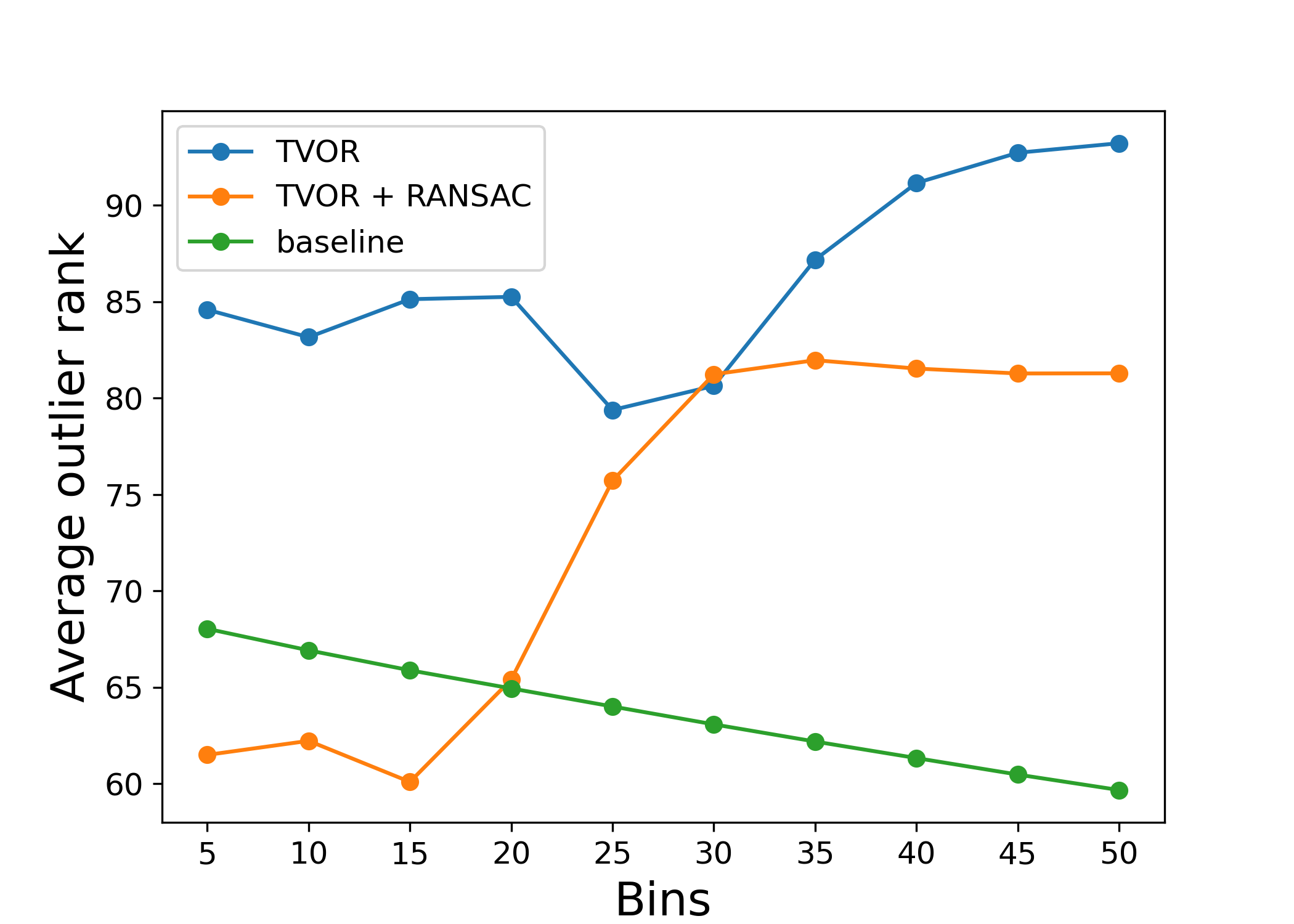}
	\label{fig:beta_triangular_90}
    }

	\caption{Comparing the proposed and the baseline method in terms of distribution outlier detection performance where $100$ inlier random samples are drawn from the beta distribution with $\alpha=7$ and $\beta=1$, while the triangular distribution with $a=0, b=1$, and $c=0.5$ is used to draw the added a)~$1$ outlier sample histogram and b)~$90$ added outlier histograms.}
	\label{fig:beta_comparison_results}
    
\end{figure*}

The first step in calculating the $i$-th expected bin value is to sum the values of the $i$-th bin in all given histograms except the tested one. When this is done for all $n$ bins, all of the obtained bin sums are divided by the sum of values of all bins in all histograms except the tested one. These normalized sums now represent the estimations of the probabilities that a value will fall in each of the histogram bins. The more histogram are given, the better these estimations are under the Glivenko-Cantelli theorem. Next, all these estimated probabilities are then multiplied by the sum of all bin values in the tested histogram. In that way the sum of the bins in the tested histogram and the sum of the estimated expected bin values are the same. Then, a small positive number is added to all scaled bin values in order to avoid division by zero during the calculation of the Pearson's chi-squared test statistics. Finally, the obtained Pearson's chi-squared test statistic is used as the outlier score for the tested histogram. The described procedure is summarized in Algorithm~\ref{alg:baseline}.

\begin{figure*}[htb]
    \centering
    
	\subfloat[]{
	\includegraphics[width=0.48\linewidth]{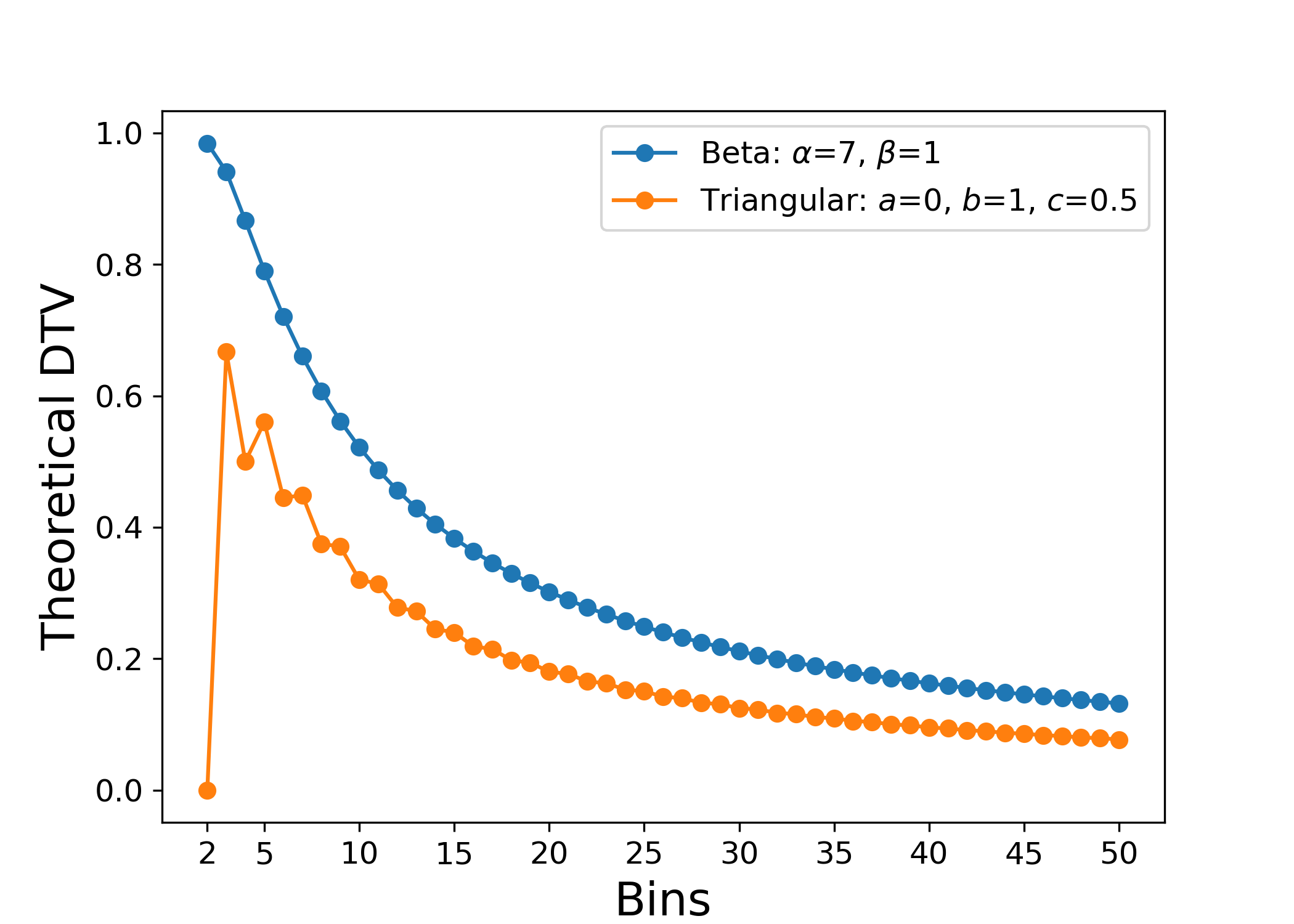}
	\label{fig:beta_triangular_theoretical}
	}
	\subfloat[]{
	\includegraphics[width=0.48\linewidth]{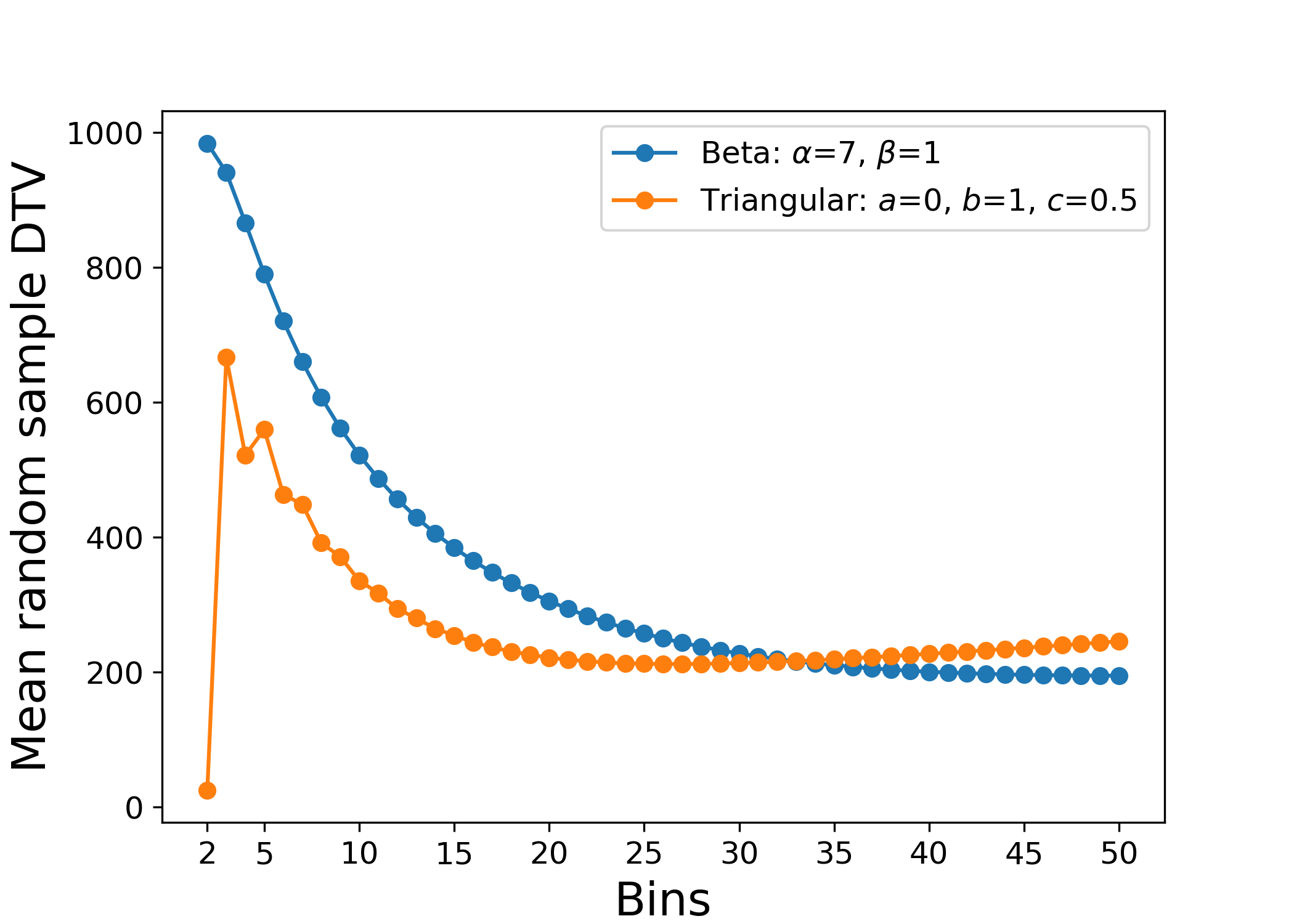}
	\label{fig:beta_triangular_mc}
    }
	
    \caption{The comparison of the used beta and triangular distributions in terms of a) the theoretical discrete total variation $\V{{\cal D}}$ described in \eqref{eq:theoretical_variation} and b) the mean DTV calculated for $10^6$ random samples of size $1000$ for various values of $n$.}
	\label{fig:beta_triangular_dtv}
    
\end{figure*}	

\subsection{Synthetic data for distribution outliers}
\label{subsec:synthetic_distribution}

\subsubsection{The goal}
\label{subsubsec:synthetic_goal}

Since there is much freedom in the overall data generation procedure when using synthetic data and less or no limitations when compared to using real-life data, the goal of this subsection is to demonstrate and explain in more detail the behavior of the proposed method depending on gradual changes of various conditions. The performance is here first measured in terms of distribution outlier detection, even though the proposed method was not designed specifically for that task, while the performance in terms of DTV outlier detection is described in the following subsection. The experiments were performed for cases when the inlier and outlier samples for histograms were from the same distribution with changed parameter values and from different distributions.

\subsubsection{Experimental setup}
\label{subsubsec:synthetic_setup}

The experiments for distribution outlier detection on synthetic data, i.e. histograms of random samples, were conducted by repeatedly first simulating the mixtures of inlier and outlier samples, then trying to recognize the outlier samples by means of applying the baseline method and the proposed method, and finally examining the results of these simulations. The experiments were conducted for two general cases of inlier and outlier random sample distributions by mixing them in $10^4$ simulations. In the first case both the inlier and outlier samples were from the normal distribution.

In each simulation of this first case, the inlier data was prepared by generating $100$ random inlier samples drawn from the normal distribution with mean $0$ and variance $1$, i.e. $\N{0}{1}$. The size of each individual sample was randomly chosen to be between $500$ and $1000$. The histogram bins were set to be evenly spaced on the interval $\left[-c, c\right]$ where $c$ is an arbitrarily chosen value used to check the behavior of various bin arrangements. Each sample value falling outside of the interval $\left[-c, c\right]$ was replaced with the closer one of $c$ and $-c$. Several values of $c$, as well as several values of number of bins $c$, were used to check the effect of changing conditions.

Furthermore, in each simulation, the outlier data was generated by drawing a certain number of random samples from $\N{0}{\sigma^2}$ for various $\sigma\neq 1$. The sample size was randomly determined in the same way as for the inlier samples. For both the inlier and outlier data the values of $c$ and $n$ were set to the same values to assure having histograms with the same bins. Next, the baseline method and the proposed method were applied to the combined inlier and outlier data to score individual histograms. Finally, the mean value of the rank of all outlier examples obtained by each method was calculated as the performance score of each method. A lower mean rank here means a better performance in terms of outlier detection. For the sake of simplicity, zero-based numbering was used for ranks. This means that in the case of a single added outlier sample, the optimal mean rank of a tested method is $0$, while in the case of e.g. $10$ added outlier samples, the optimal mean rank is $4.5$ since this is the average value of the first $10$ zero-based ranks, which should all be assigned to outlier samples' histograms in the case of a method that performs ideally.

In short, every instance of the simulation setup is uniquely determined by the number of histogram bins $n$, the number of added outlier samples, the value $c$ used to determine the interval of the binned values, and the value of $\sigma$ for outlier distribution. Simulations for each instance were repeated $10^4$ times to check the performance of the baseline method and the proposed method in various sampling conditions.

n the second general case, the inlier samples were drawn from the beta distribution with parameter values $\alpha=7$ and $\beta=1$, while the outlier samples were drawn from the triangular distribution with parameter values $a=0$, $b=1$, and $c=0.5$. The probability density functions of these distributions are shown in Fig.~\ref{fig:beta_triangular_pdf}. Similarly to the previous case, several combinations of the number of bins $n$ and the number of outlier sample histograms added to the $100$ inlier sample were checked. For each combination, the results of methods' performance were averaged over $10^4$ simulations.

\begin{figure*}[htb]
    \centering
    
	\subfloat[]{
	\includegraphics[width=0.48\linewidth]{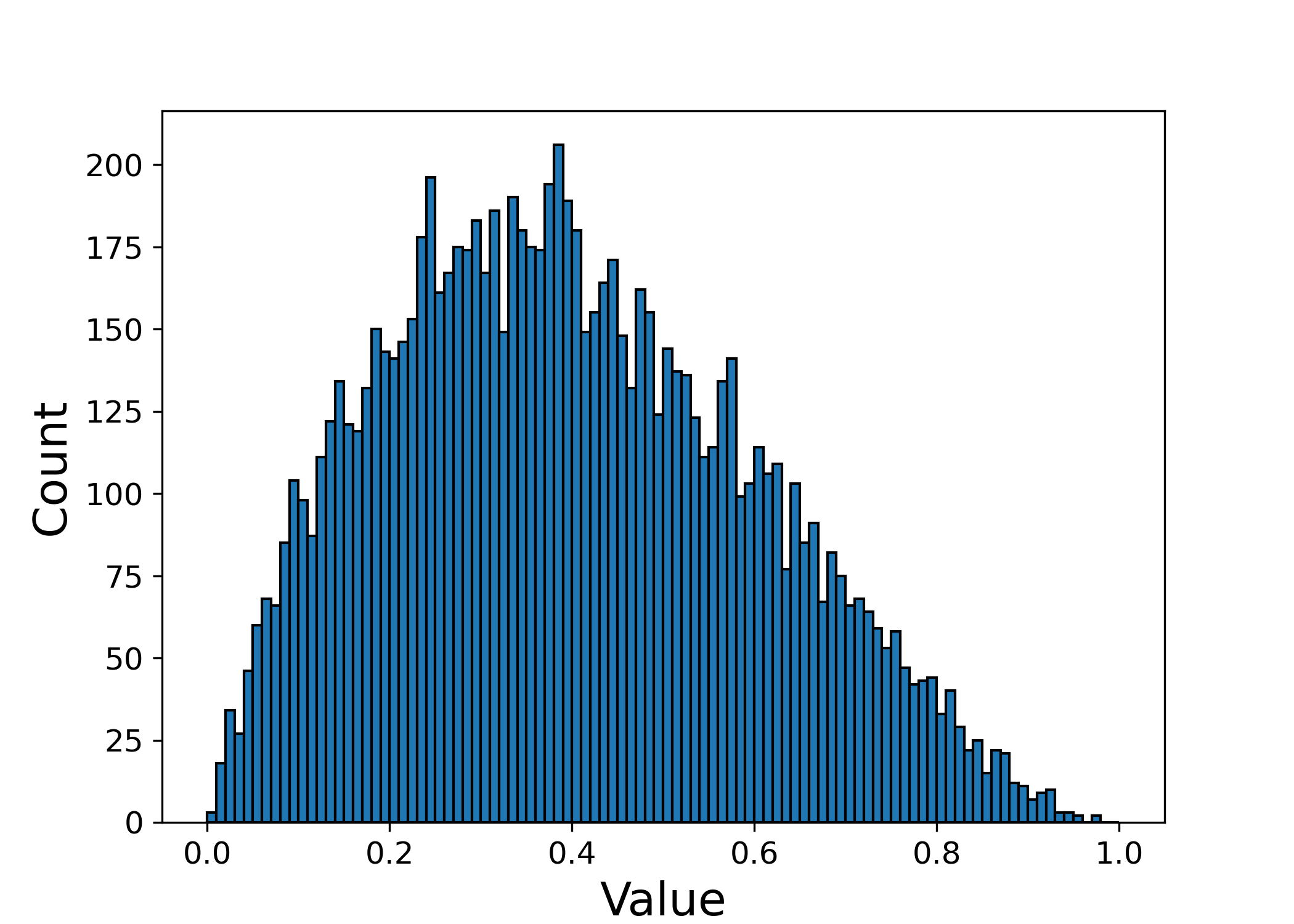}
	\label{fig:beta_sample}
	}
	\subfloat[]{
	\includegraphics[width=0.48\linewidth]{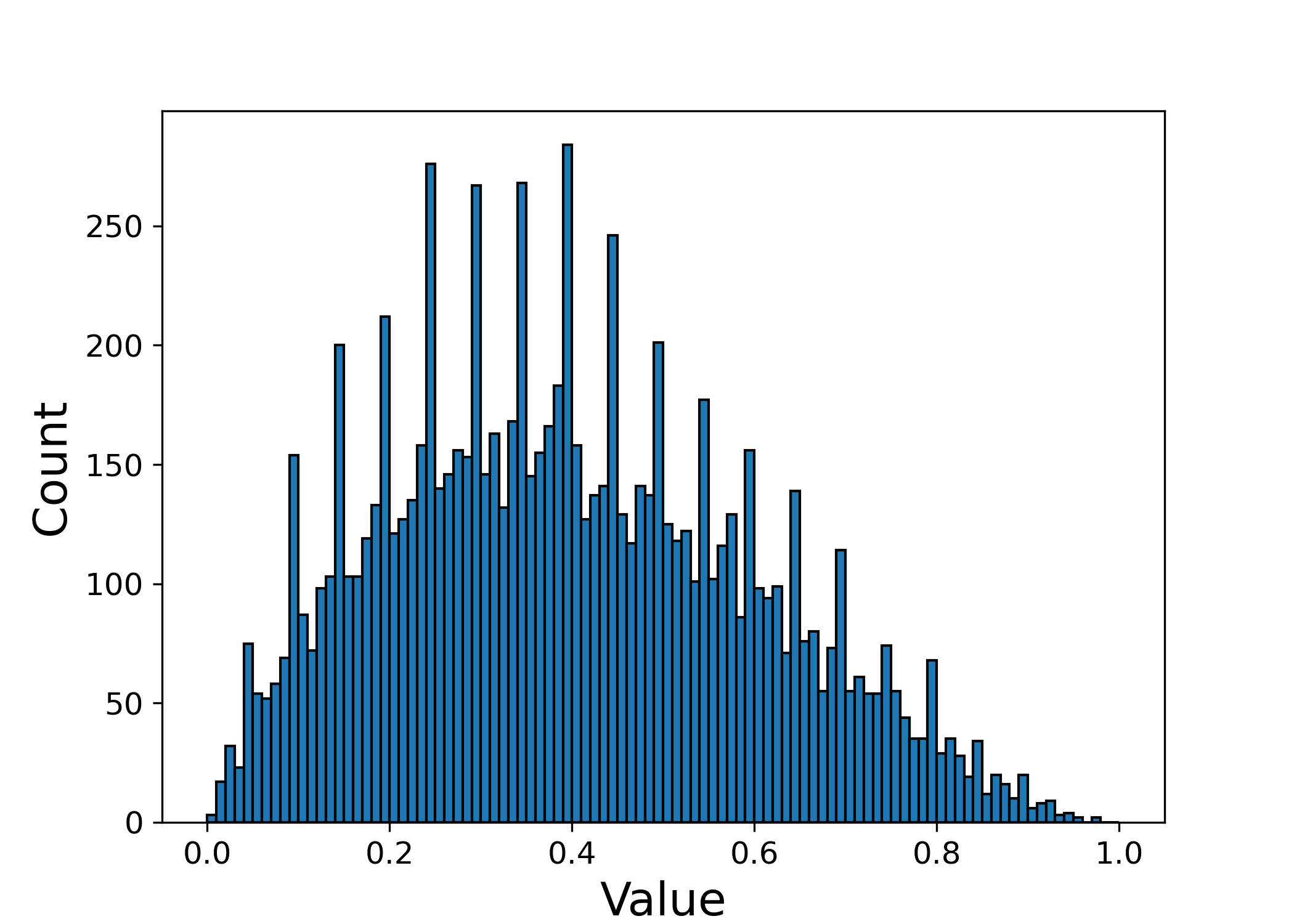}
	\label{fig:beta_sample_ah_10}
    }

	\caption{The histogram of a random sample drawn from the beta distribution with $\alpha=2$ and $\beta=3$ in the case of a)~no heaping and b)~heaping by moving $10$\% of randomly chosen items to bins with ordinal numbers divisible by $5$ closest to them.}
	\label{fig:beta_ah_example}
    
\end{figure*}

\subsubsection{Results}
\label{subsubsec:synthetic_results}

After examining the results of performing simulations for a large number of setups when both the inliers and the outliers are from the normal distribution, due to the similarity of many of the results, it was decided to show only those that can be used to summarize them all. These results are shown in Fig.~\ref{fig:tvor_vs_baseline}. The first thing to observe is that in the majority of the cases the baseline method based on the Pearson's chi-squared test performs better in terms of outlier ranking. This is mainly because the proposed method was not designed to find outliers in general, but to find outliers in terms of the discrete total variation. Interestingly, however, the exception to this are the cases when there is a relatively small number of bins, which can be seen in Figs.~\ref{fig:sigma_a} and~\ref{fig:osc_b}, and cases with a high amount of added outlier sample histograms, which can be seen in Fig.~\ref{fig:osc_a} where the proposed methods outperforms the baseline method for all given numbers of bins. This means that even if the proposed method was not designed for the same task as the baseline method, in some cases it is still able to outperform it, which may be useful should such cases emerge. A more detailed analysis of the performance results shown in Fig.~\ref{fig:tvor_vs_baseline} is given in Appendix, which also explains the sudden drops in the performance such as the one in Fig.~\ref{fig:b_b}.

In short, the proposed method generally performs worse than the baseline method. However, in the cases of smaller values of $n$, i.e. in the cases of a smaller number of bins, as well as in the cases with a high amount of outliers, it may perform better. Similar results can be obtained with some other distributions as well and therefore they have been omitted here. If required, any other experiments with a similar setup can be conducted by using the source code publicly available in the repository that is described later.

Next, Fig.~\ref{fig:beta_comparison_results} shows the results of the experiment where the inlier and the outlier samples were drawn from the beta and the triangular distribution, respectively. As can be expected by viewing Fig.~\ref{fig:beta_triangular_pdf}, the baseline method outperforms the proposed method in most cases since the difference between the used distributions is significant. Nevertheless, Fig.~\ref{fig:beta_triangular_90} again shows that the proposed method may be able to outperform the baseline method in the case of a high amount of outliers.

The performance drop of the proposed method for several values of $n$ shown in Fig.~\ref{fig:beta_triangular_1} deserves some additional comments. As shown in Fig.~\ref{fig:beta_triangular_theoretical}, the theoretical DTVs of both distributions are clearly separated for all shown values of $n$. This means that if the random samples were sufficiently big, then the performance should significantly improve in accordance with \eqref{eq:non-uniform2}. Namely, in that case the influence of the sample size significantly overpowers the influence of the randomness. As a matter of fact, if the whole experiment is repeated with random samples having their sizes increased by several orders of magnitude, then both the proposed method and the baseline method have the same ideal performance. However, as mentioned earlier, the size of each sample used in the experiment whose results are shown in Fig.~\ref{fig:beta_triangular_1} was randomly chosen to be between $500$ and $1000$. For such sizes, the randomness still has a substantial influence on the histograms' DTVs. This is illustrated in Fig.~\ref{fig:beta_triangular_mc}, which shows the mean DTV calculated for $10^6$ random samples of size $1000$ for various values of $n$ created for both the beta and the triangular distribution. It can be clearly seen how this differs from the case of the theoretical DTVs and this can be used to explains the particularly low performance of the proposed method when $n$ is $30$ and $35$ shown in Fig.~\ref{fig:beta_triangular_1}. Namely, for these values of $n$, the mean values of DTVs become so close that, with the influence of randomness included, it becomes difficult to successfully distinguish between the inlier and the outlier histograms based only on their DTVs. The dependence of the proposed method's performance on the size of the samples is further analyzed in more detail in Appendix. Based on all the results shown here and in Appendix, it can be concluded that the proposed method's performance improves as the size of the samples increases.

\begin{figure*}[htbp]
    \centering
    
  \subfloat[]{
  \includegraphics[width=0.48\linewidth]{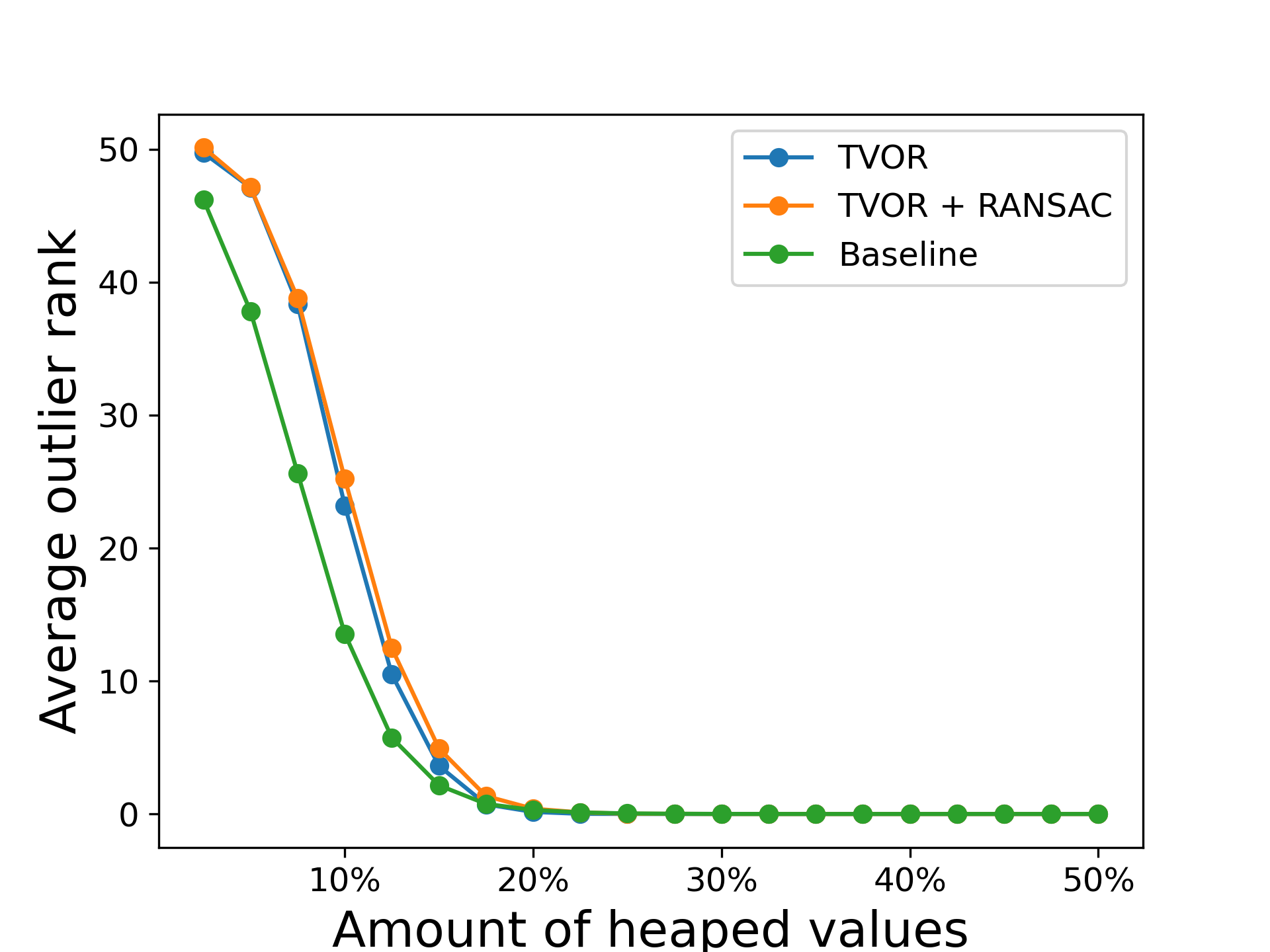}
  \label{fig:osc_1}
  }%
  \subfloat[]{
  \includegraphics[width=0.48\linewidth]{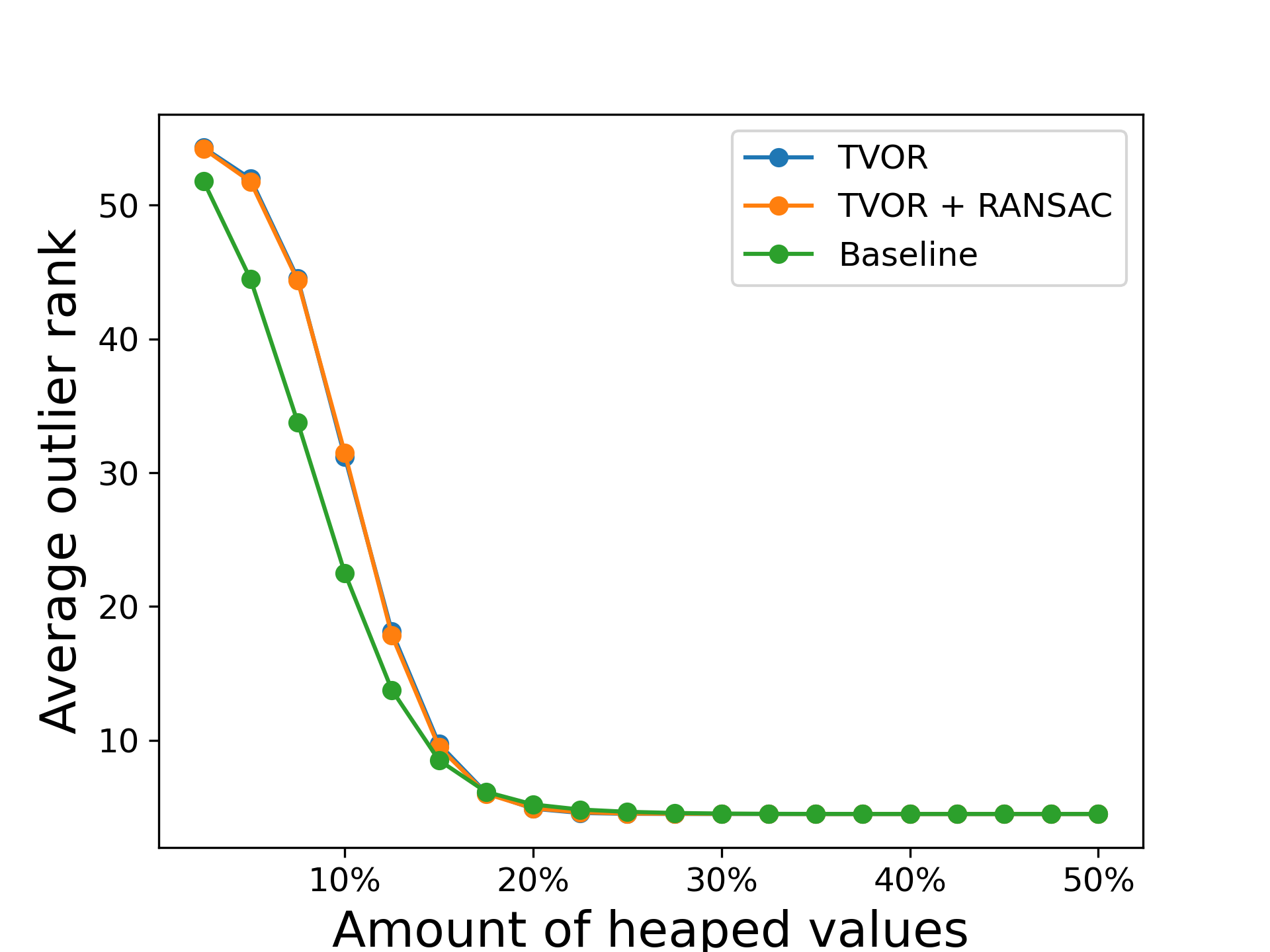}
  \label{fig:osc_10}
  }
  \\
  \subfloat[]{
  \includegraphics[width=0.48\linewidth]{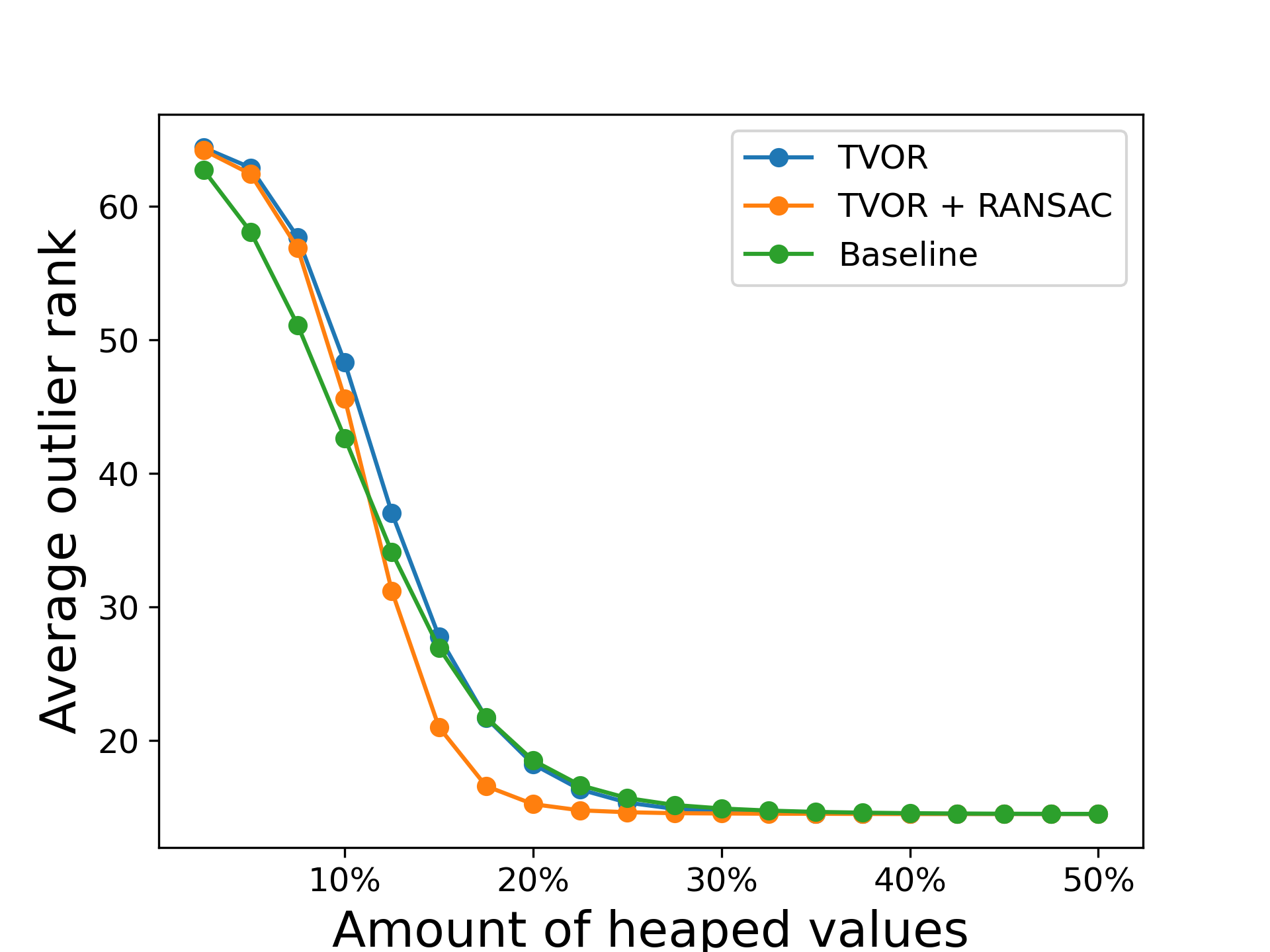}
  \label{fig:osc_30}
  }%
  \subfloat[]{
  \includegraphics[width=0.48\linewidth]{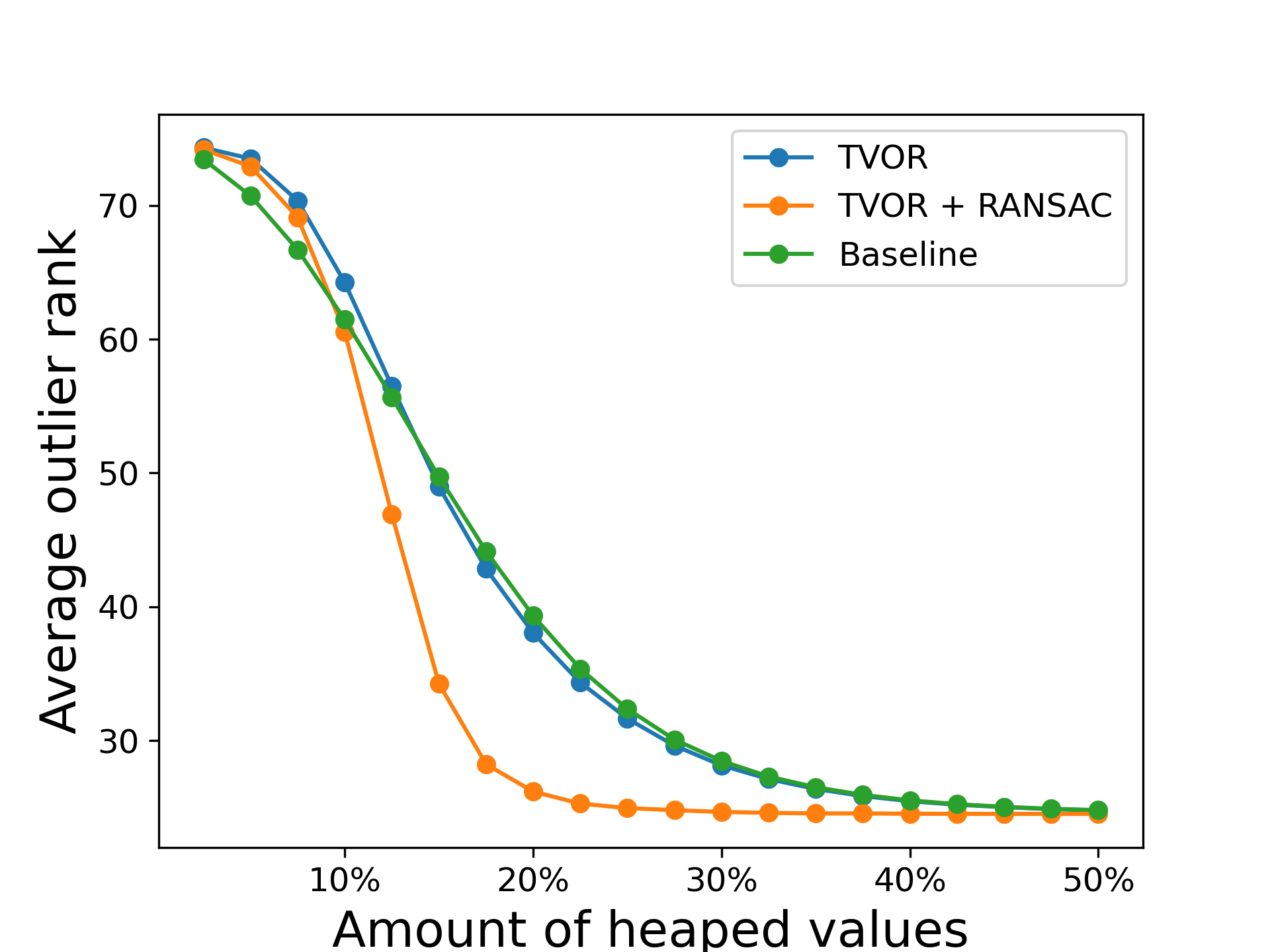}
  \label{fig:osc_50}
  }
  \\
  \subfloat[]{
  \includegraphics[width=0.48\linewidth]{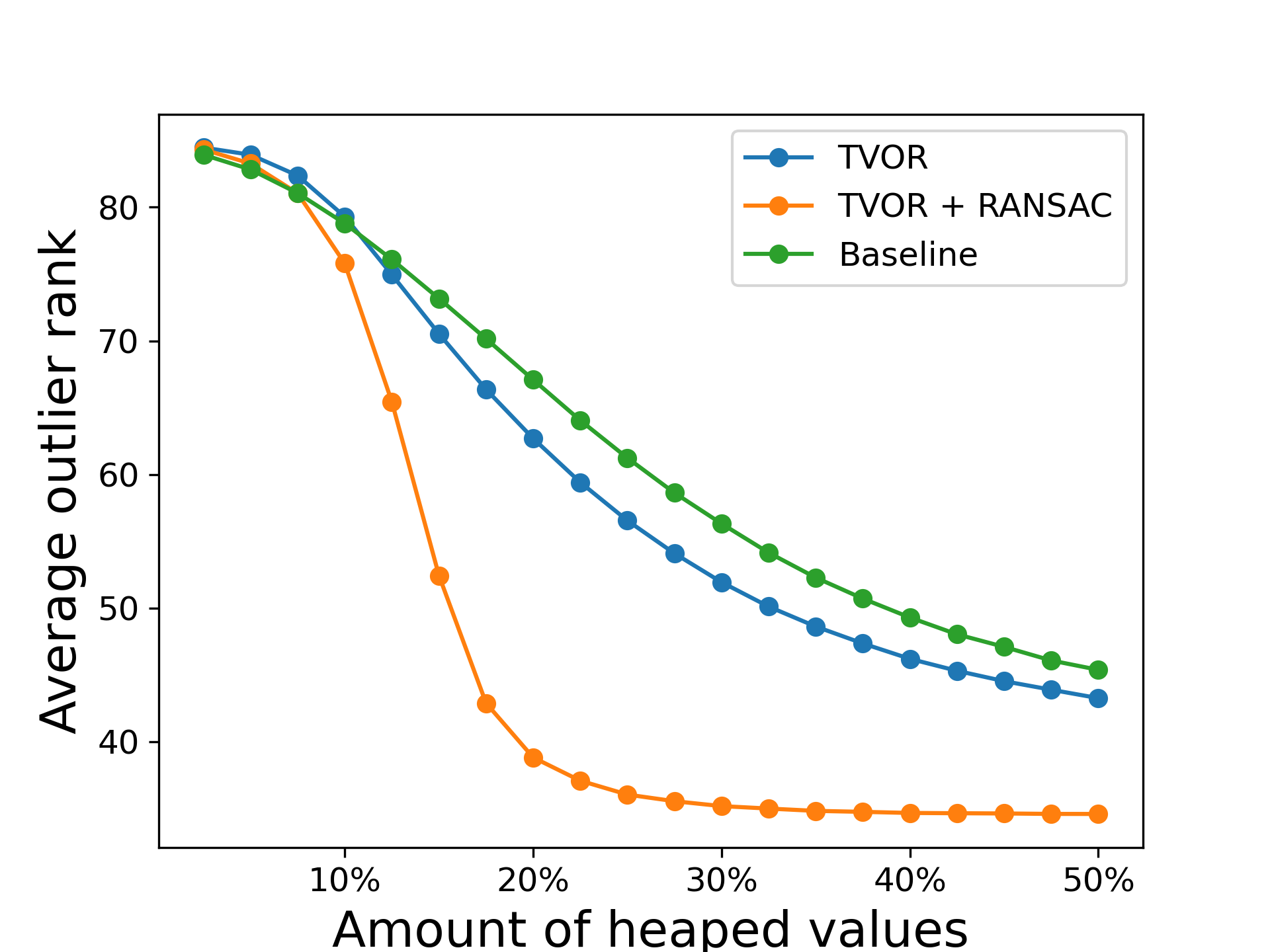}
  \label{fig:osc_70}
  }%
  \subfloat[]{
  \includegraphics[width=0.48\linewidth]{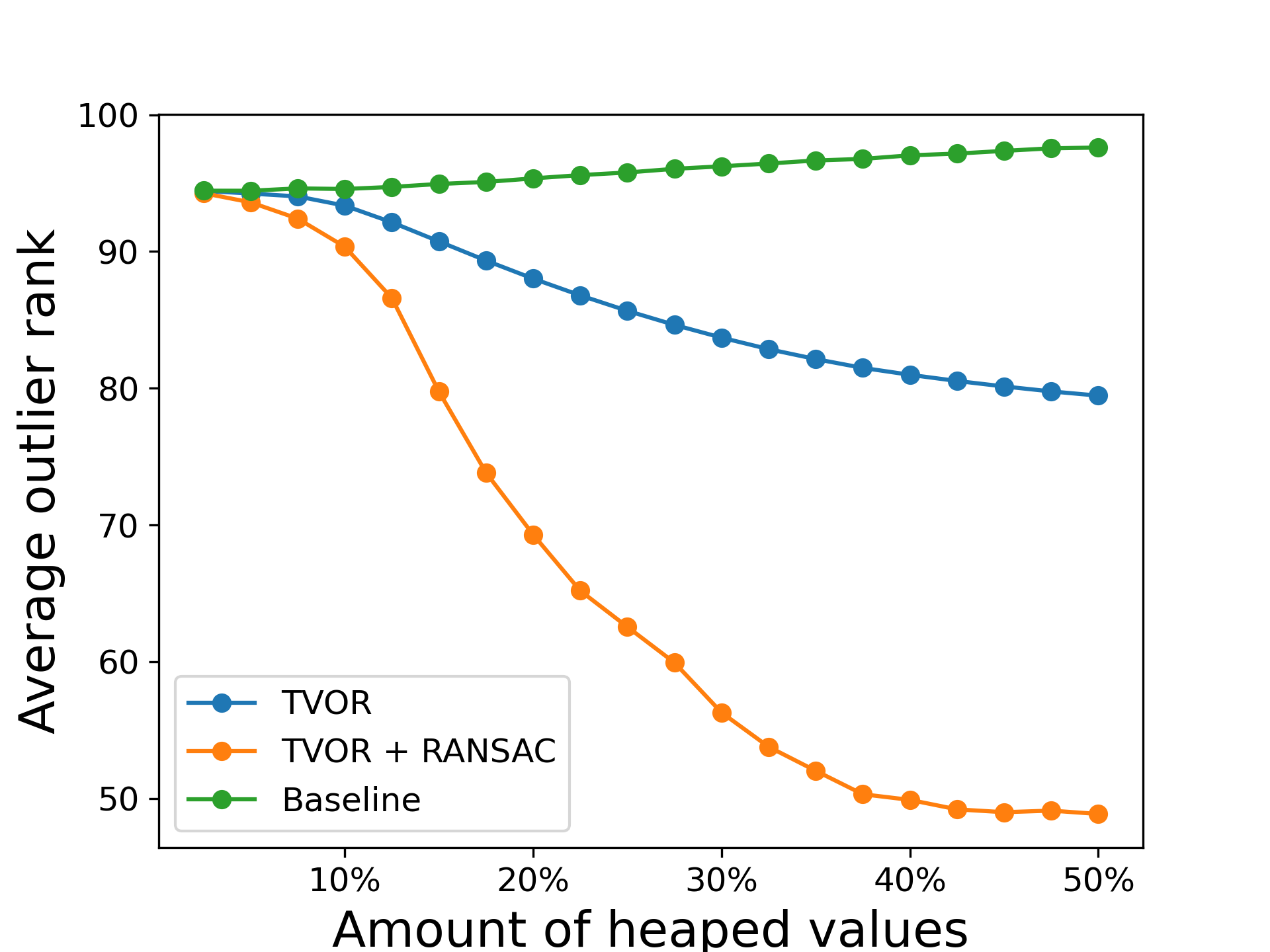}
  \label{fig:osc_90}
  }

    \caption{Comparing the performance of the proposed and baseline methods averaged over $10^4$ random trials in cases where the number of outlier random samples bin values added to the original $100$ inlier random samples was a)~1, b)~10, c)~30, d)~50, e)~70, and f)~90. The inlier and outlier random samples were drawn from the beta distribution with $\alpha=2$ and $\beta=3$, but the outlier samples were additionally changed in order to make their histograms have a prespecified amount of heaped bin values.}
  \label{fig:tvor_vs_baseline2}
    
\end{figure*}

\begin{figure*}[htbp]
    \centering
    
	\subfloat[]{
	\includegraphics[width=0.48\linewidth]{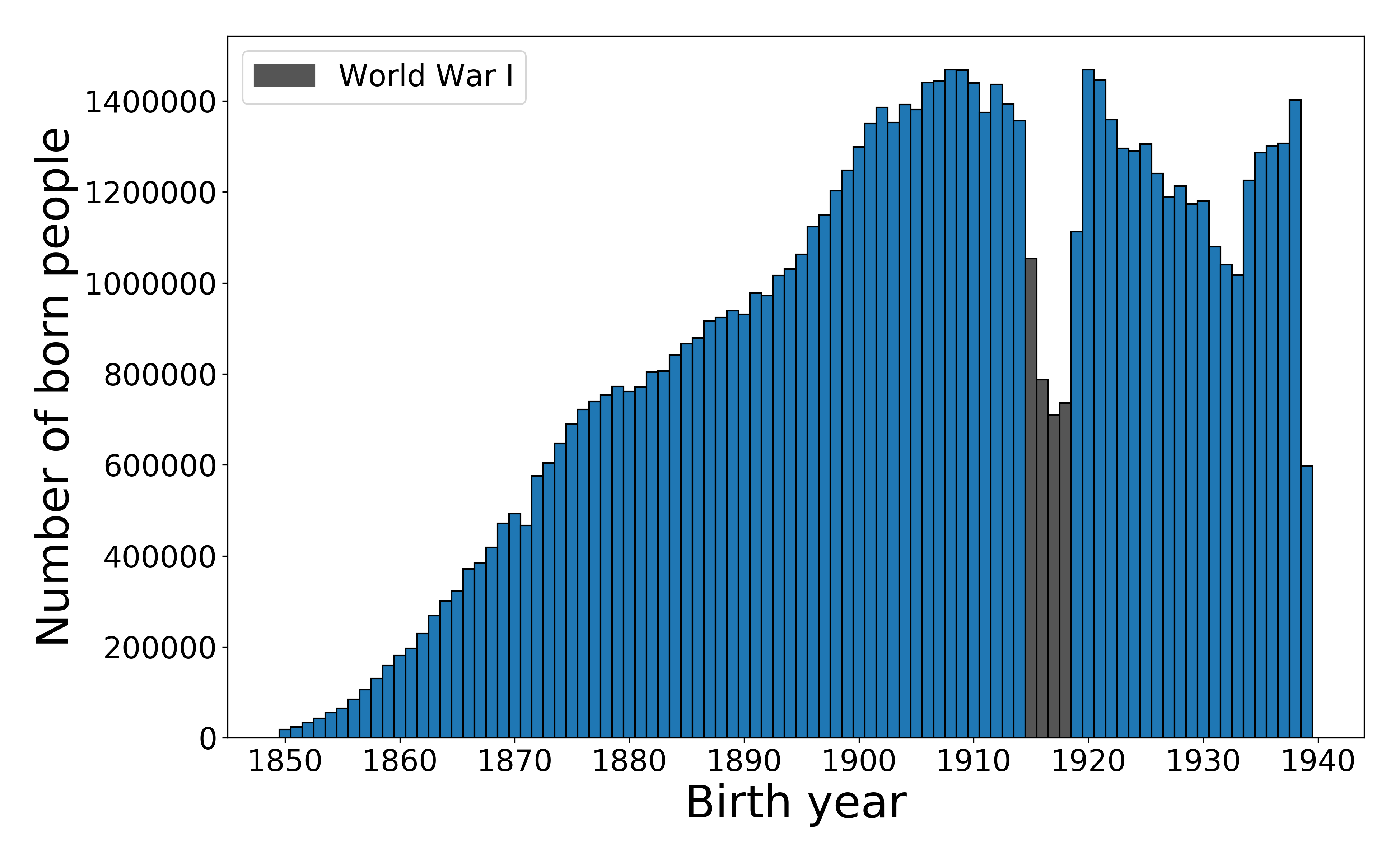}
	\label{fig:germany1939}
	}%
	\subfloat[]{
	\includegraphics[width=0.48\linewidth]{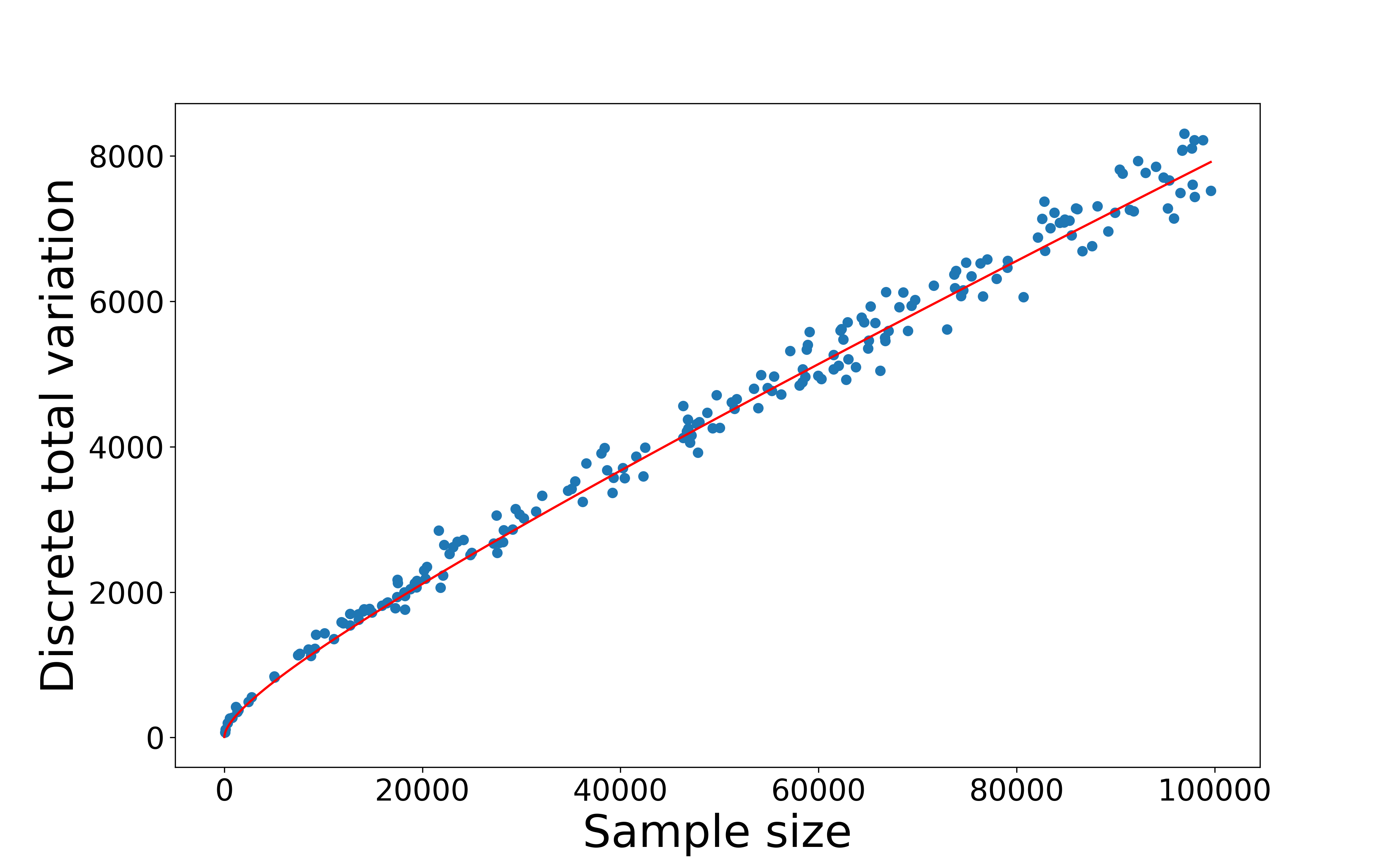}
	\label{fig:simulation}
    }\\
	\subfloat[]{
	\includegraphics[width=0.48\linewidth]{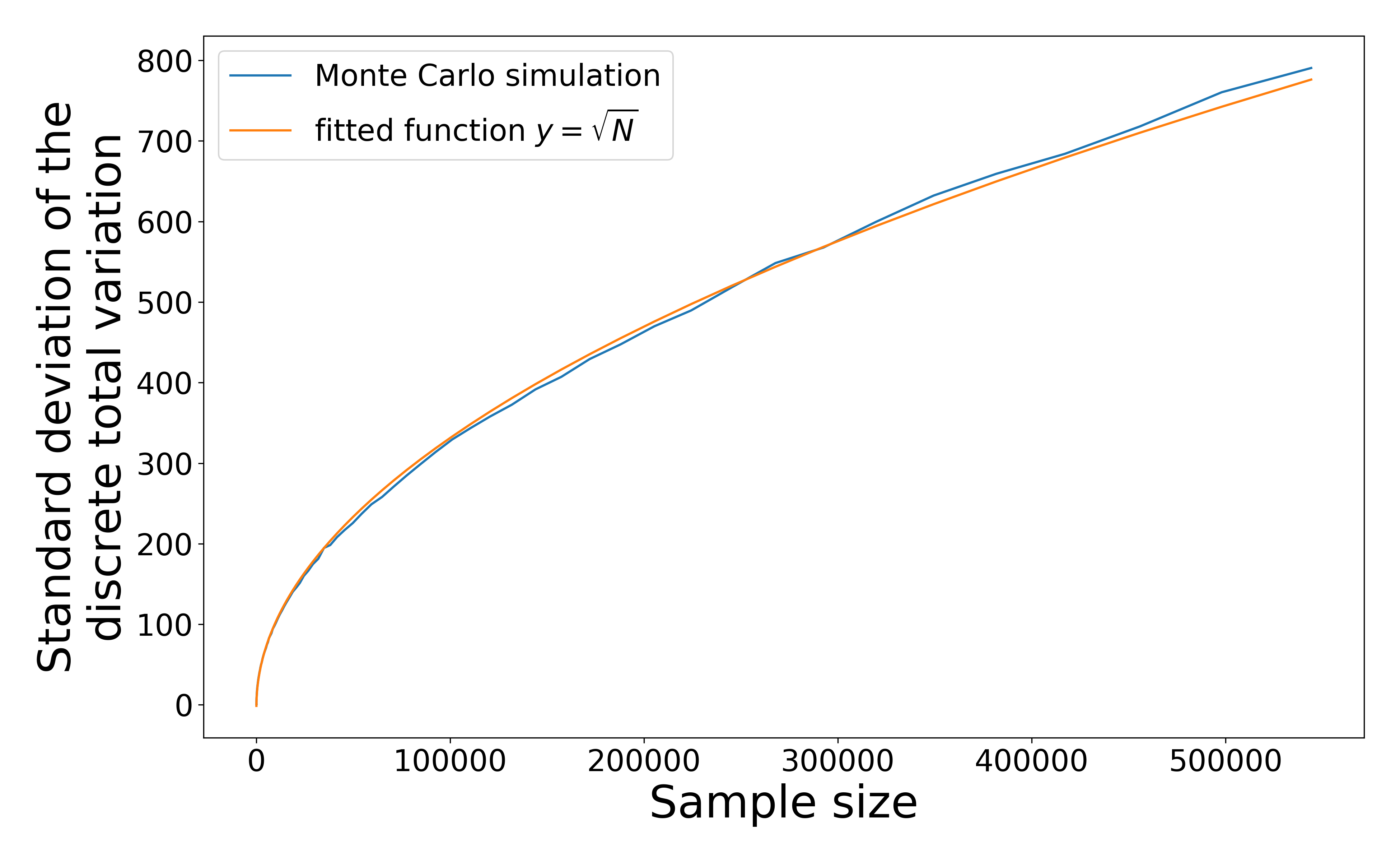}
	\label{fig:sd}
    }%
	\subfloat[]{
	\includegraphics[width=0.48\linewidth]{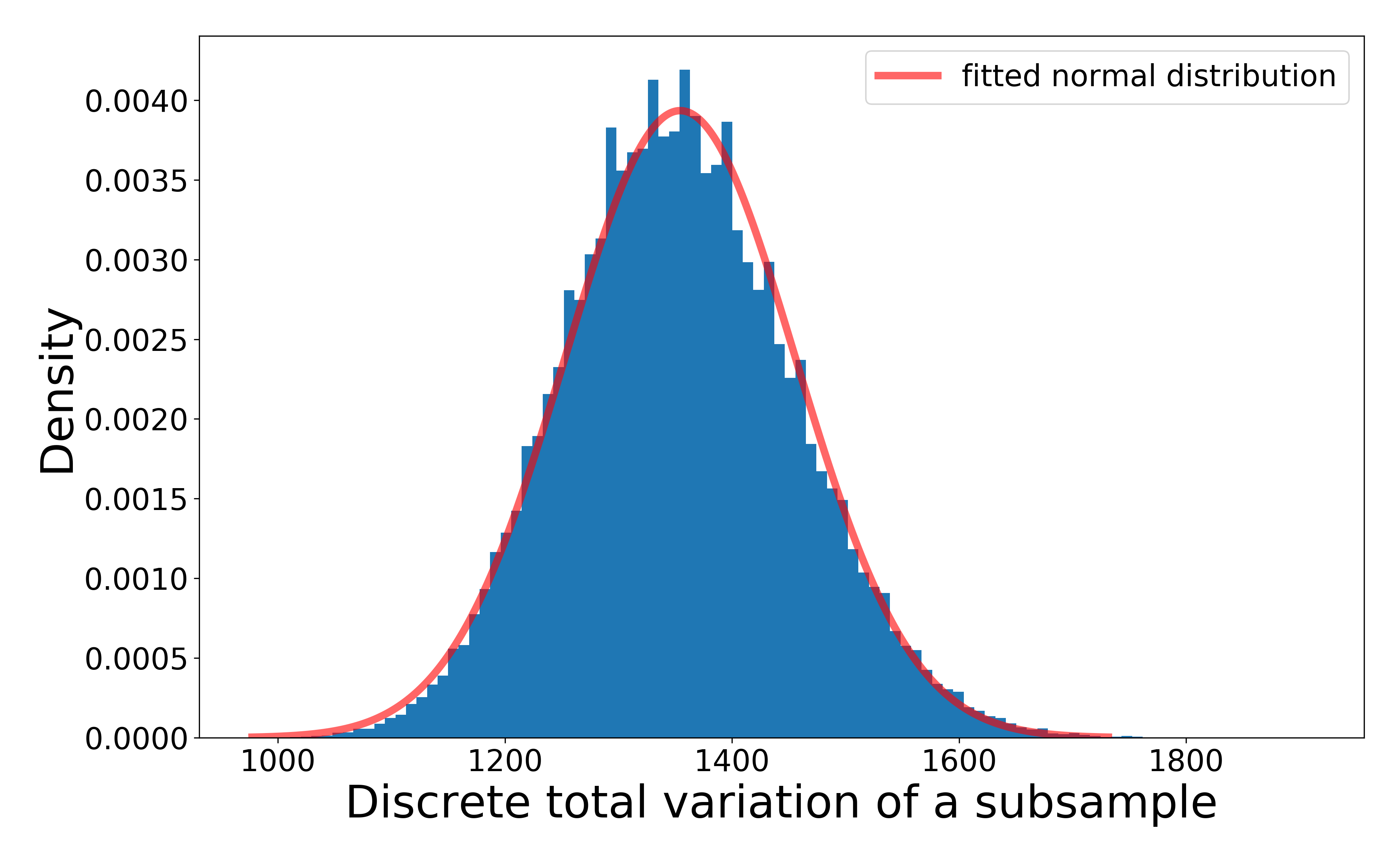}
	\label{fig:sd_distribution}
    }
	
	\caption{The experiments on the German census of 1939: a)~histogram of birth years of the German census of 1939 based on the data from~\cite{population1939germany,jahrbuch1939germany}, starting from year 1850, b)~fitting the proposed method's model in~\eqref{eq:non-uniform_model} to data for subsamples of the German census of 1939, c)~the relation between the sample size and the standard deviation of the discrete total variation obtained through Monte Carlo simulations for the subsamples of the German census of 1939 and a fitted function $y=a\sqrt{N}$, and d)~the distribution of discrete total variations obtained for 100k subsamples of the German census of 1939 of size 10k.}
	
	\label{fig:census}
\end{figure*}

\begin{figure}[htb]
    \centering
    
	\includegraphics[width=\linewidth]{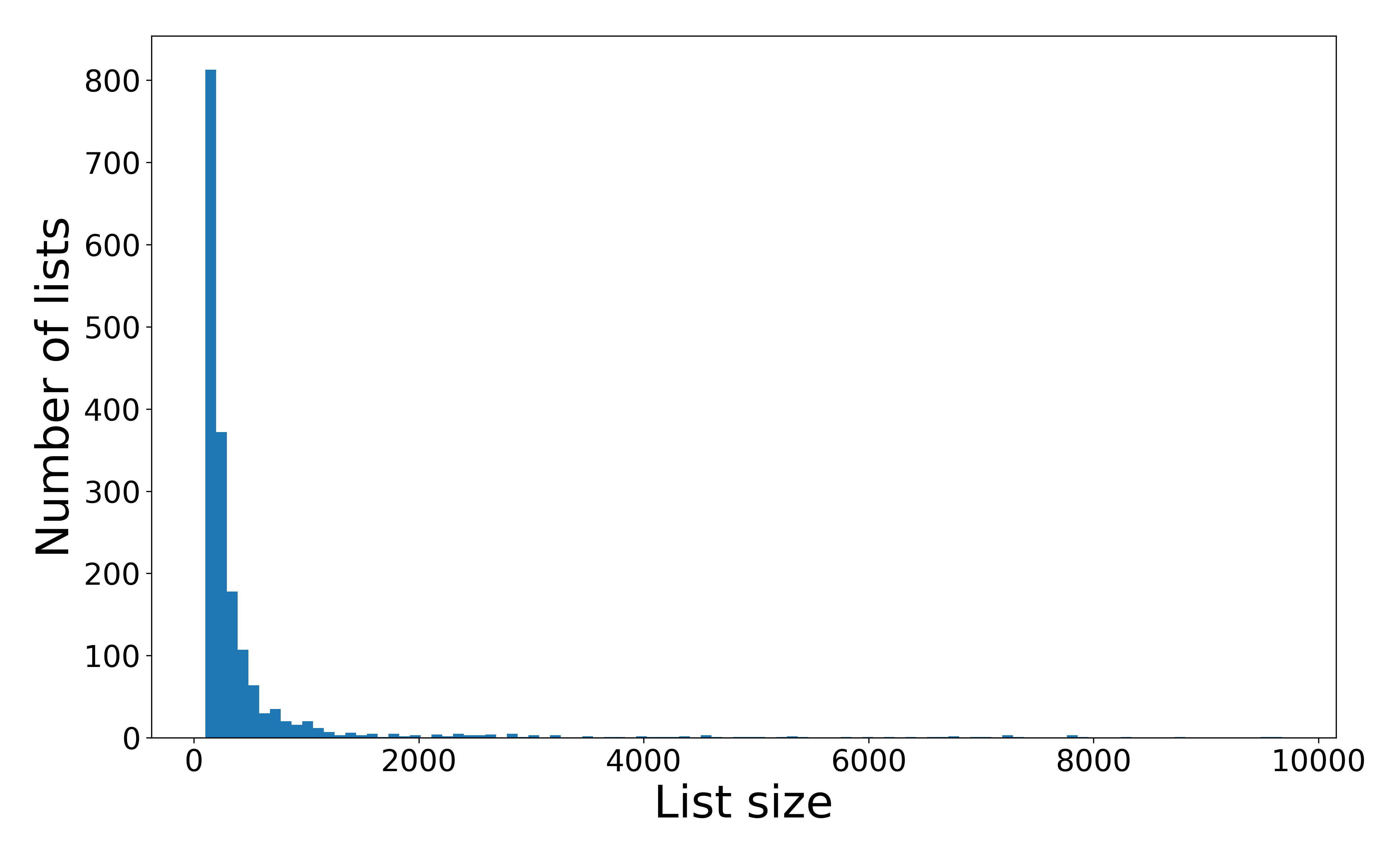}
	
    \caption{The distribution of the sizes of the majority of the $7106$ USHMM lists that are used for the experiments.}
	\label{fig:ushmm_sizes}
    
\end{figure}

Overall, in terms of distribution outlier detection, the performance of the proposed method is only indirectly dependent on the inlier and the outlier distributions. As shown, it is directly dependent on the difference between the theoretical DTVs of these distributions, which is in turn dependent on the chosen histogram bins. This means that, depending on the histogram bins, the proposed method may perform well even when the inlier and the outlier distribution are same, but with slightly different parameters. On the other hand, for significantly different inlier and outlier distributions that have similar theoretical DTVs for the chosen bins, the proposed method may perform poorly. The opposite cases are also possible. Nevertheless, this is not too problematic because the proposed method was not designed for distribution outlier detection, but specifically for the DTV outlier detection.

\subsection{Synthetic data for total variation outliers}
\label{subsec:synthetic_variation}

\subsubsection{The goal}
\label{subsubsec:synthetic_goal2}

The goal of this subsection is to demonstrate the behavior of the proposed method for the case that it was originally designed for, i.e. for discrete total variation outlier detection. Additional emphasis is specifically put on cases where the number of outliers gets very close to the number as the inliers.

\begin{figure*}[htb]
    \centering
    
	\includegraphics[width=0.98\linewidth]{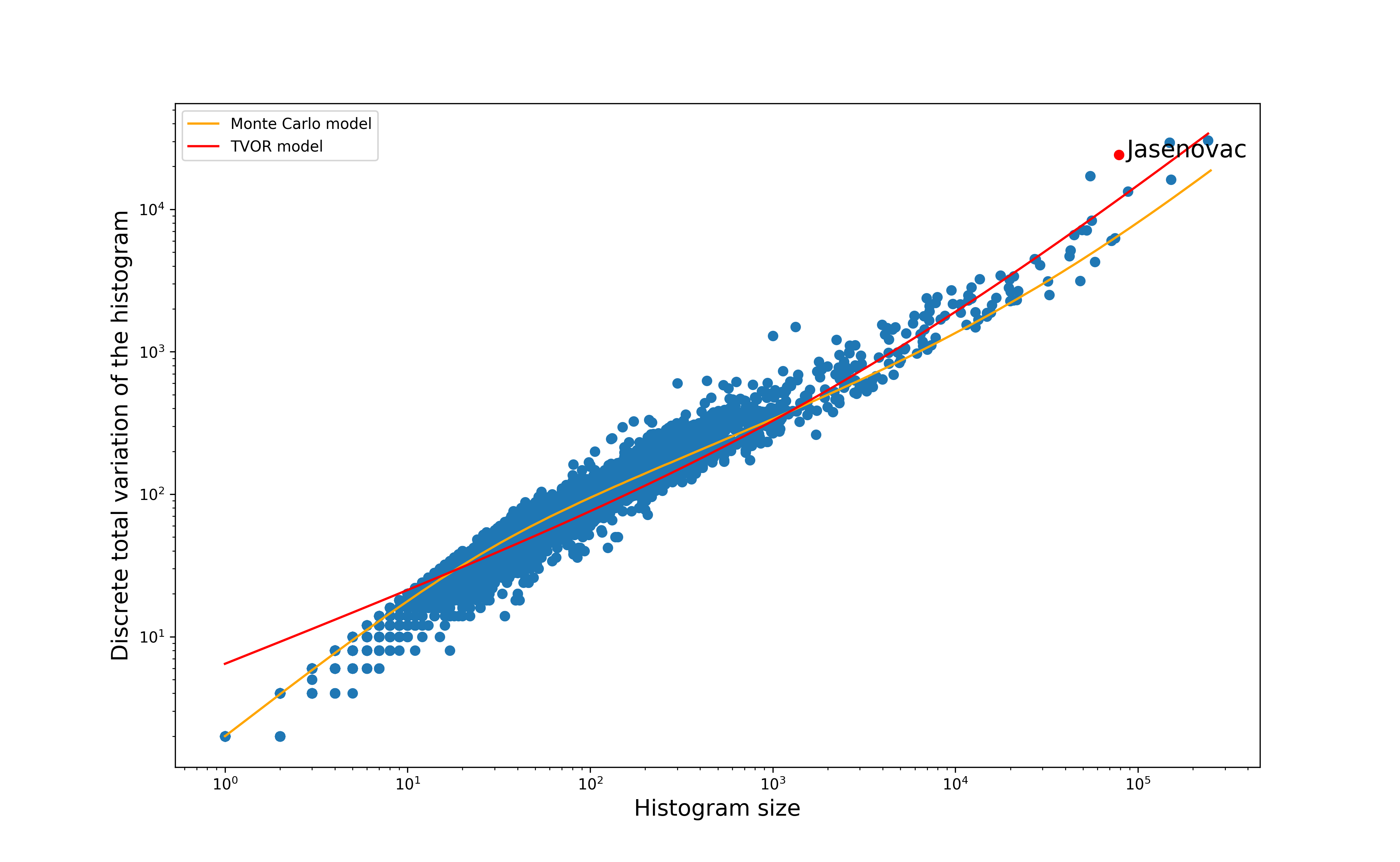}
	
    \caption{Applying the proposed method to $7106$ lists of the USHMM data. The model based on applying the Monte Carlo simulation to the German census of 1939 is shown for comparison. Note that the plot axes use the logarithmic scale.}
	\label{fig:all}
    
\end{figure*}

\begin{figure}[htb]
    \centering
    
	\includegraphics[width=\linewidth]{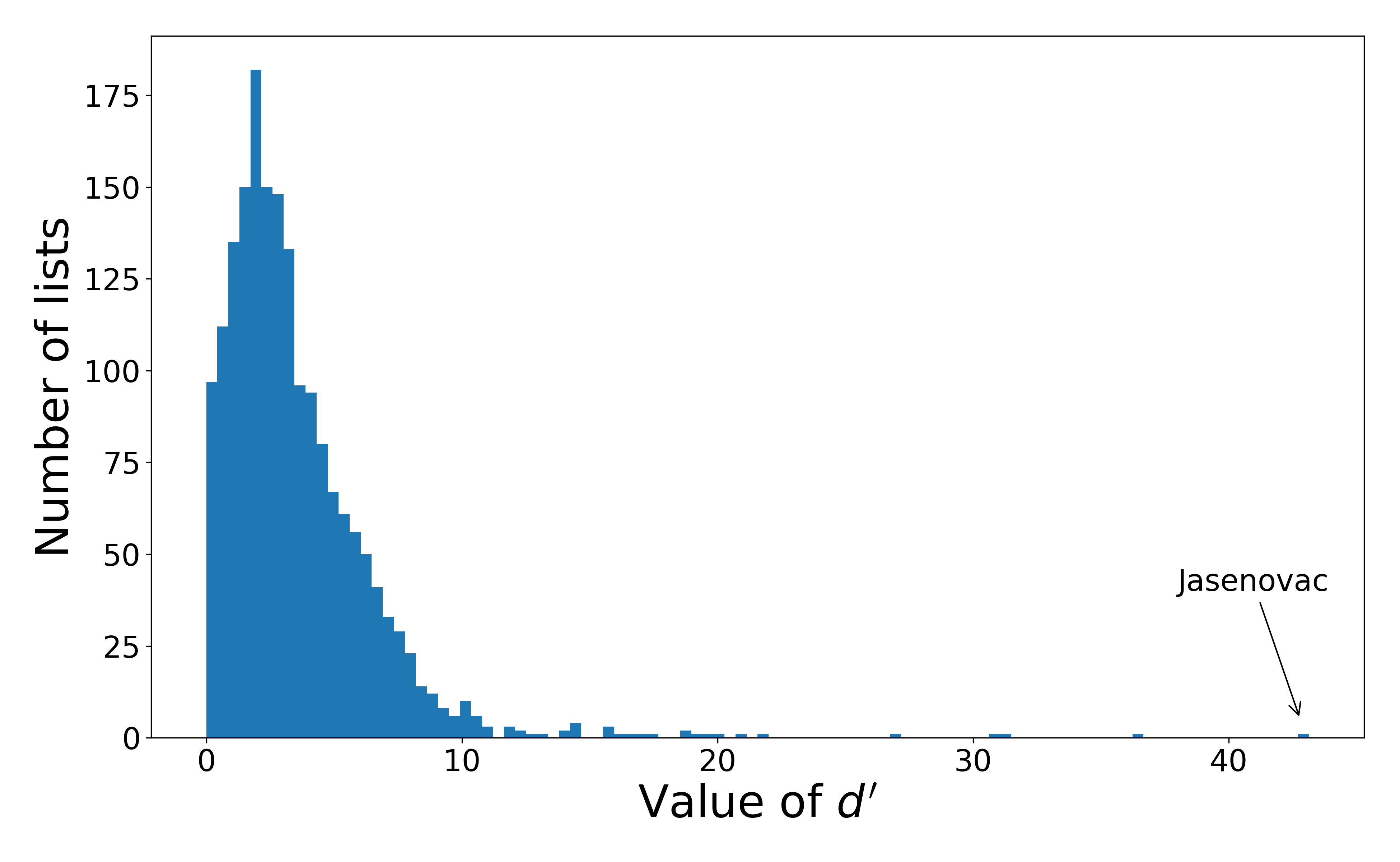}
	
    \caption{The distribution of values $d'$ calculated by the proposed method for birth years from the USHMM lists.}
	\label{fig:tvor_scores}
    
\end{figure}

\subsubsection{Experimental setup}
\label{subsubsec:synthetic_setup2}

Since earlier in the paper it was mentioned that demographics is one of the fields that can benefit from discrete total variation outlier detection, the beta distribution with $\alpha=2$ and $\beta=3$ was chosen for the inlier samples' distribution. The reason is the resemblance of its histograms to the histograms of some population age distributions. For all experiments the number of bins $n$ was fixed to $100$. The outlier samples were initially also drawn from the same beta distribution and their histograms also had $100$ bins. However, the outlier samples' histograms underwent an additional change to simulate the so called age heaping~\cite{caselli2005demography}. Namely, for a certain amount of randomly chosen bins with a count greater than $0$, their count was decreased by $1$ and the count of the closest bin to each of them whose ordinal number was divisible by $5$ was increased by $1$ as can be seen on the example that is shown in Fig.~\ref{fig:beta_ah_example}. This was done for various combinations of the amount of outlier samples and the amount of randomly chosen bins that were changed for these outlier samples' histograms. Finally, the performances of the proposed method and of the baseline method were then compared for all these combinations.

\subsubsection{Results}
\label{subsubsec:synthetic_results2}

The obtained results and comparisons are shown in Fig.~\ref{fig:tvor_vs_baseline2}. It can be seen that if there are only a few outliers, then the proposed and the baseline methods are on par with each other and there are only some smaller differences in performance for various amount of heaped values. However, as the number of outliers increases, the proposed method starts to significantly outperform the baseline method, especially in cases where RANSAC is used as suggested in Section~\ref{subsec:model}. This is especially noticeable in Fig.~\ref{fig:osc_90} where the number of outliers is very close to the number of inliers. There the baseline effectively degrades to a random chooser, while the proposed method used in combination with RANSAC excels at outlier detection. This shows the usefulness of the proposed methods for the task of finding the discrete total variation outliers.

\begin{figure*}[htb]
    \centering
    
	\subfloat[]{
	\includegraphics[width=0.48\linewidth]{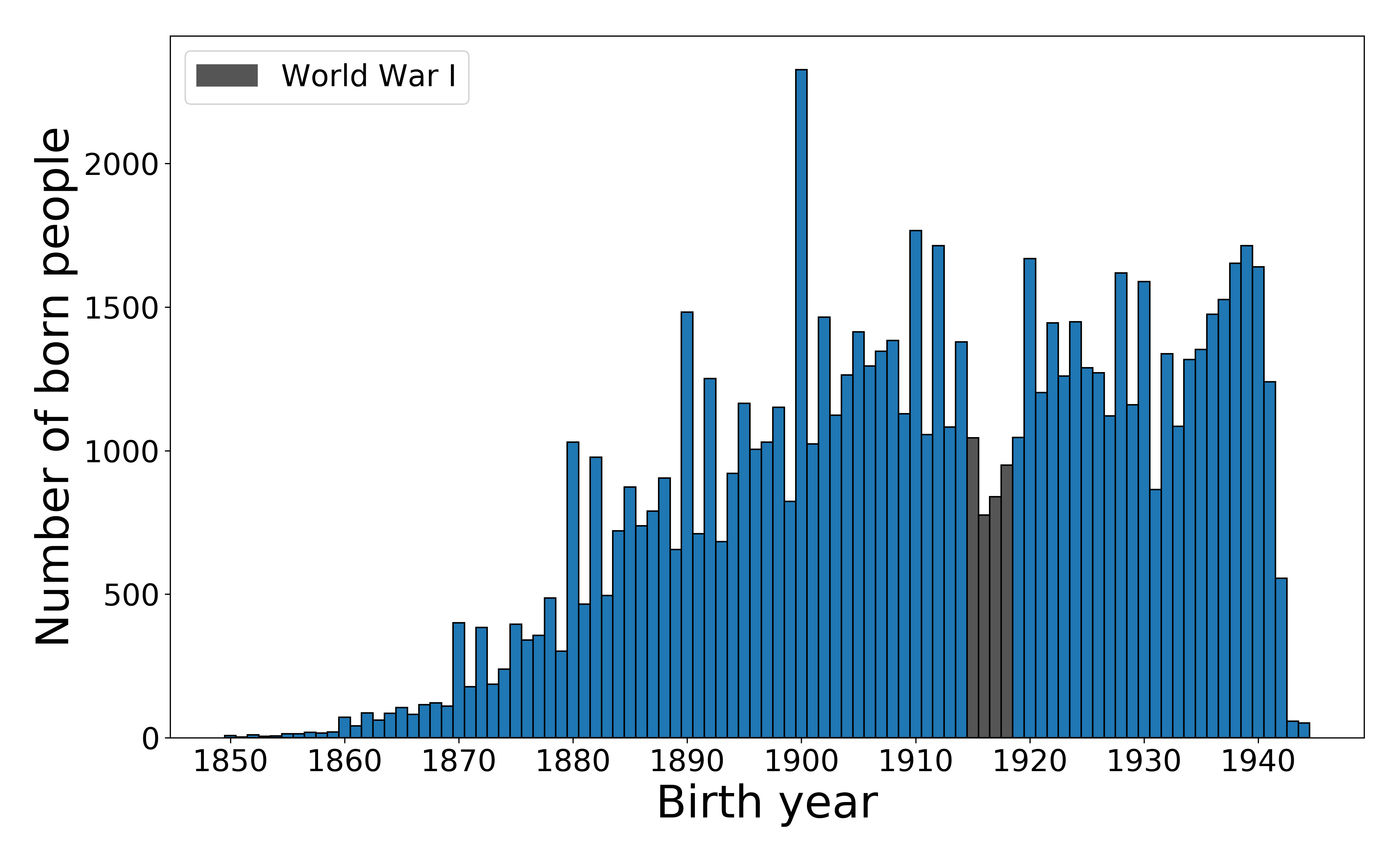}
	\label{fig:jms}
	}%
	\subfloat[]{
	\includegraphics[width=0.48\linewidth]{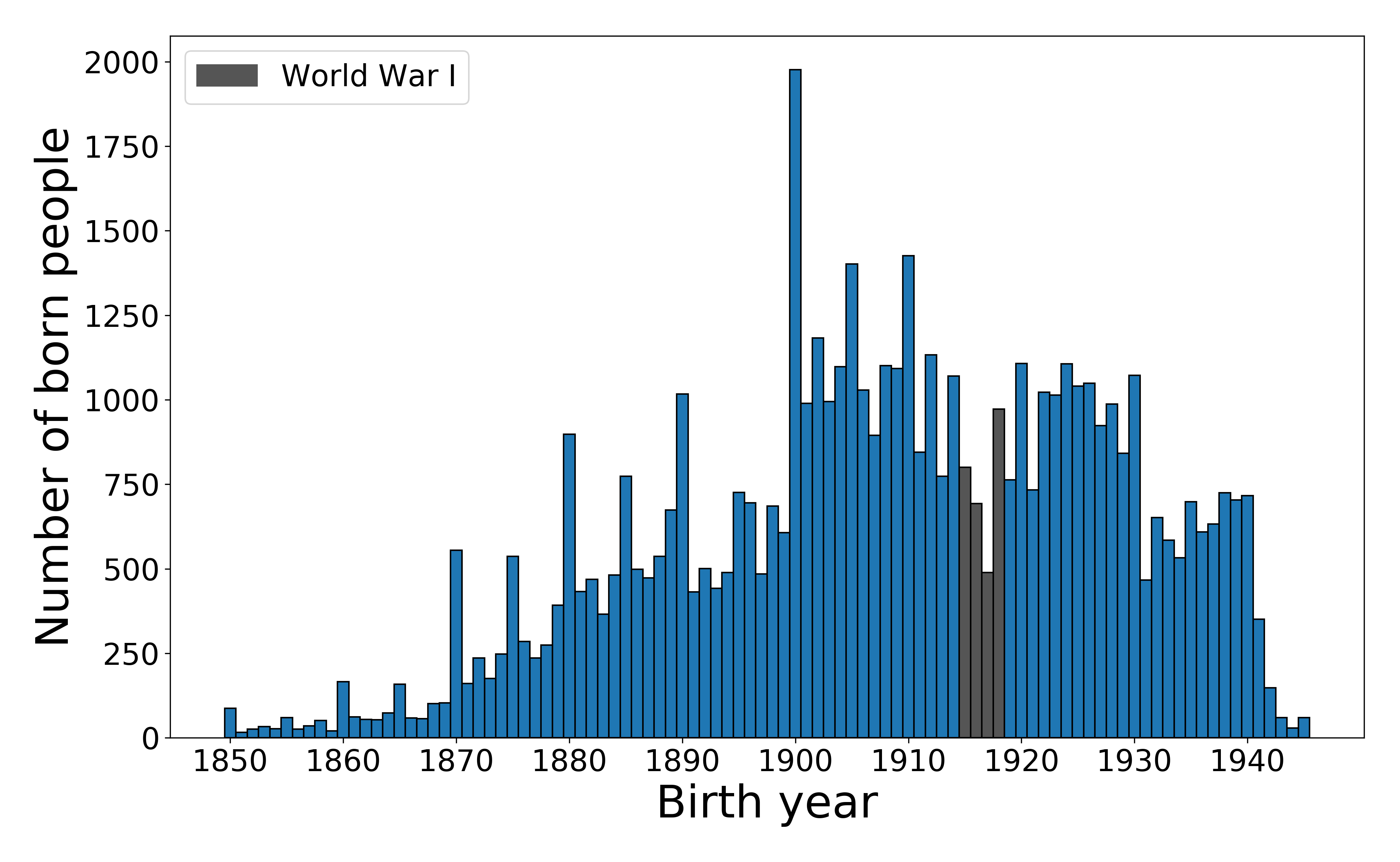}
	\label{fig:20781}
	}\\
	\subfloat[]{
	\includegraphics[width=0.48\linewidth]{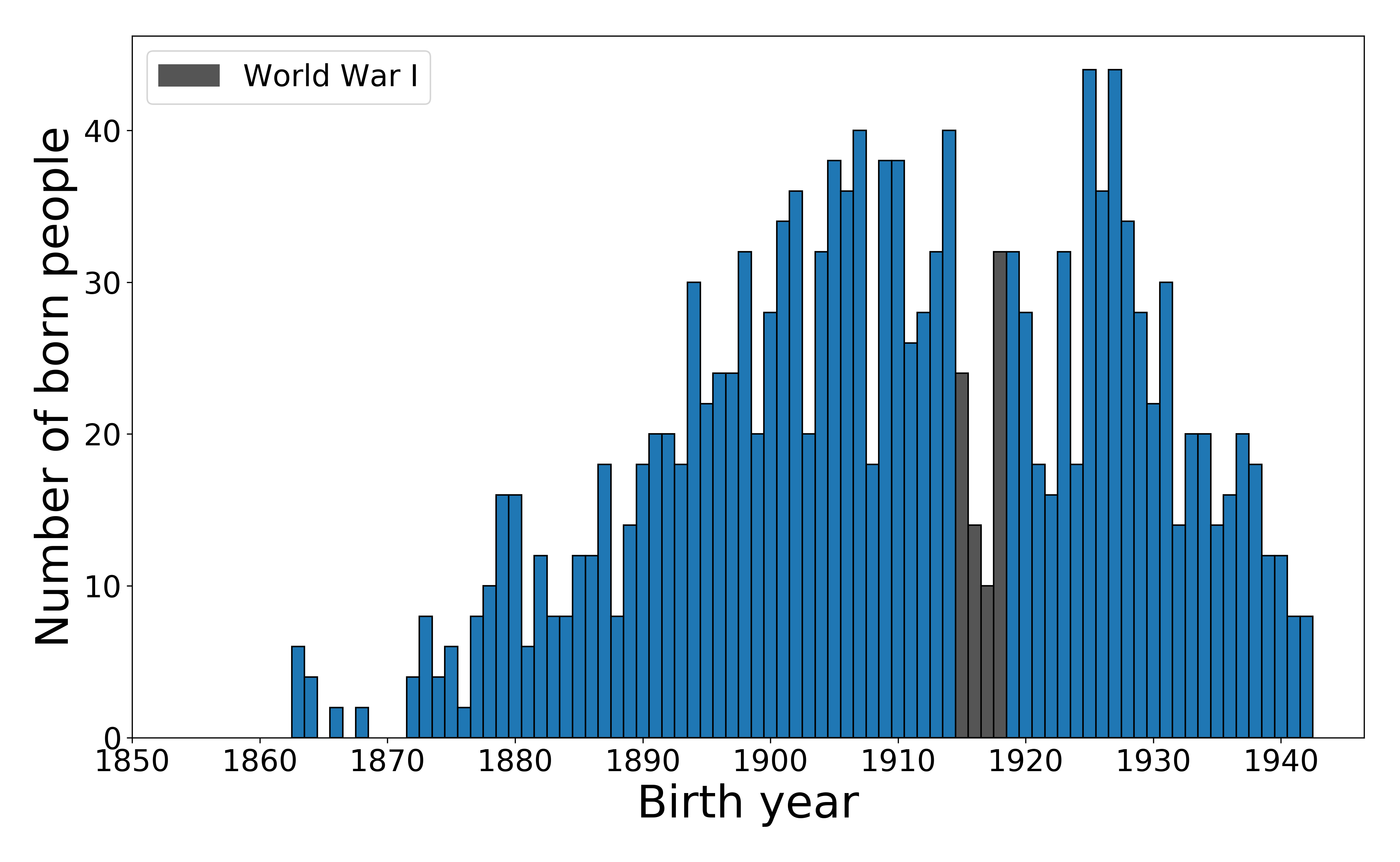}
	\label{fig:38665}
	}%
	\subfloat[]{
	\includegraphics[width=0.48\linewidth]{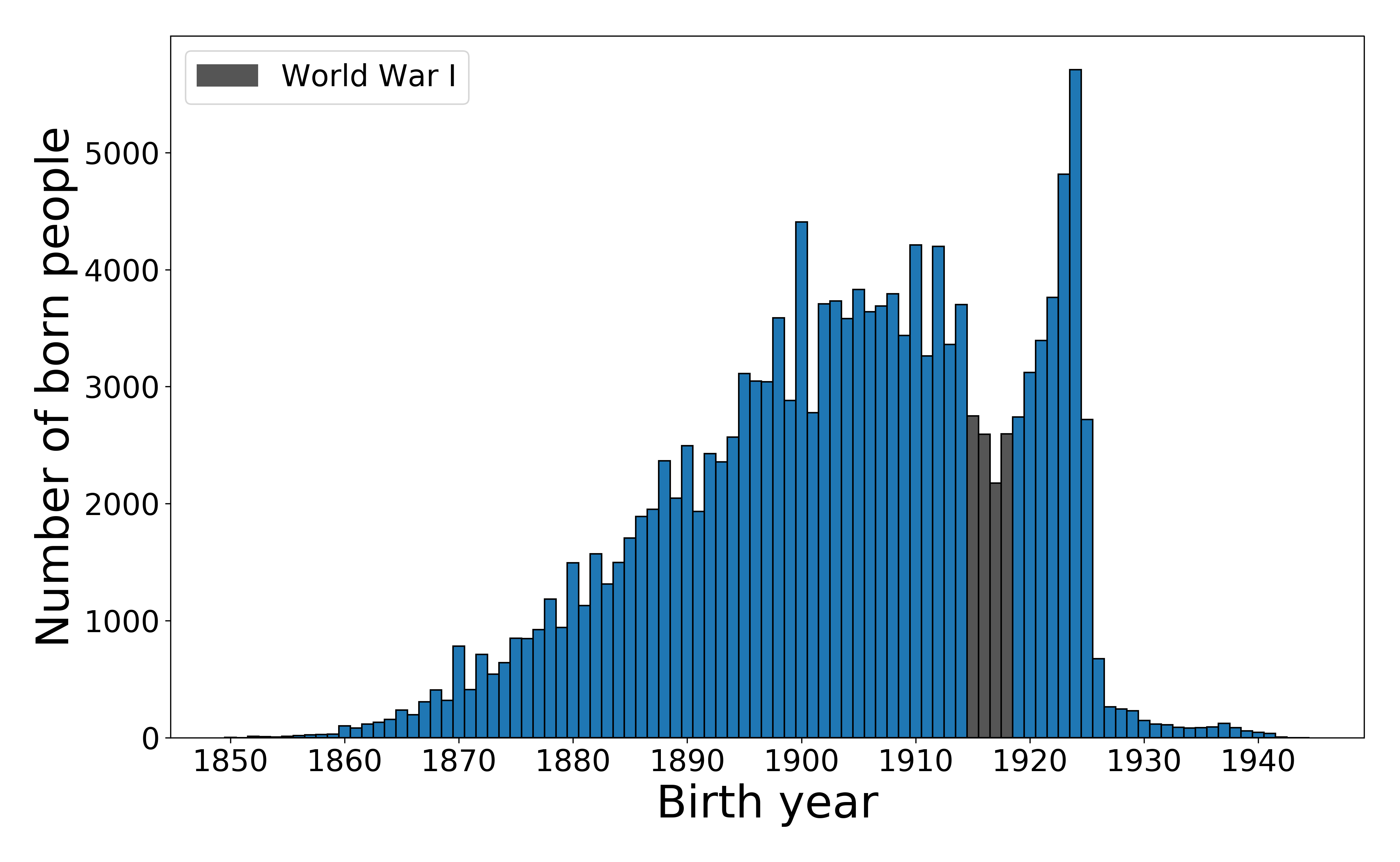}
	\label{fig:20492}
	}
    
	\caption{Top-scoring birth year lists out of $7106$ checked lists: a)~the Jasenovac camp inmates available at USHMM's webpage~\cite{ushmm2020jusp}, b)~the victims from the Soviet Extraordinary Commission~\cite{ushmm2020sec}, c)~the victims from the Franz Street Number 38~\cite{ushmm2020street2}, and d)~persons from the Registration cards of Jewish refugees in Tashkent, Uzbekistan during WWII~\cite{ushmm2020tashkent}.}
	
	\label{fig:histograms}
	
\end{figure*}

\begin{figure}[htb]
    \centering
    
	\includegraphics[width=\linewidth]{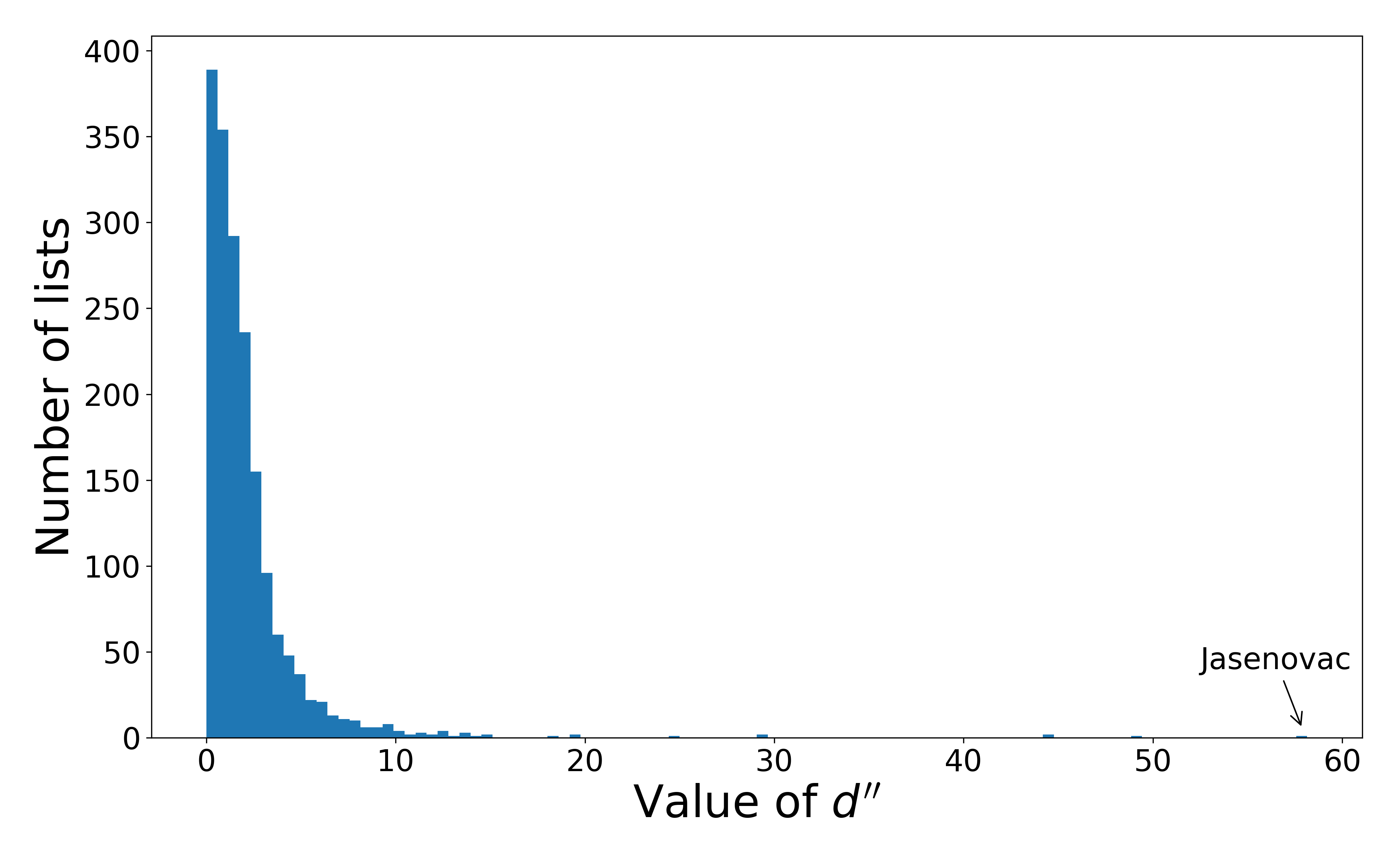}
	
    \caption{The distribution of values $d''$ calculated by the proposed method for birth years from the USHMM lists.}
	\label{fig:mc_scores}
    
\end{figure}

\subsection{Census data}
\label{subsec:census}

\subsubsection{The goal}
\label{subsubsec:census_goal}

The goal of this subsection is to test the proposed method on an example of real-life census data with sample sizes spanning several orders of magnitude and being drawn from slightly different, but similar distributions. Here a closer look is taken at the samples of the top-scoring histograms. This can show the robustness of the proposed method in noisy conditions and its usefulness for real-life data applications.

\subsubsection{Experimental setup}
\label{subsubsec:census_setup}

Several census data sources have been used for the experimental setup. The largest of them is the German census of 1939~\cite{population1939germany} with the corresponding birth year histogram being shown in Fig.~\ref{fig:germany1939}. Since the significant gap for the years of World War I can be traced in age composition of other similar lists and censuses of other countries as well~\cite{shryock1980methods,siegel2004methods}, this census data is used here as a gold standard for the discrete total variation of the population histograms for that time.

\begin{figure*}[htb]
    \centering
    
	\subfloat[]{
	\includegraphics[width=0.48\linewidth]{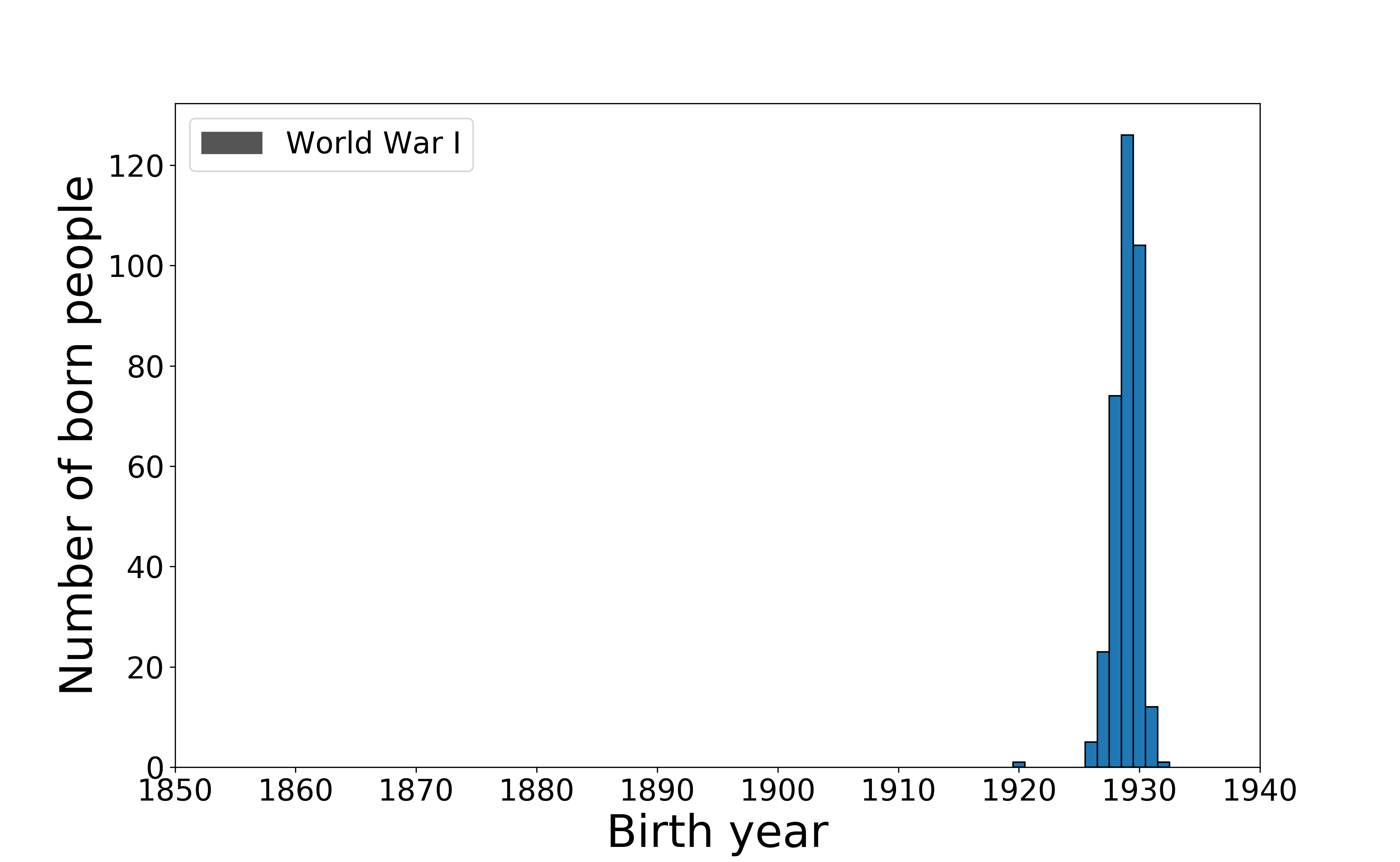}
	\label{fig:whipple}
	}%
	\subfloat[]{
	\includegraphics[width=0.48\linewidth]{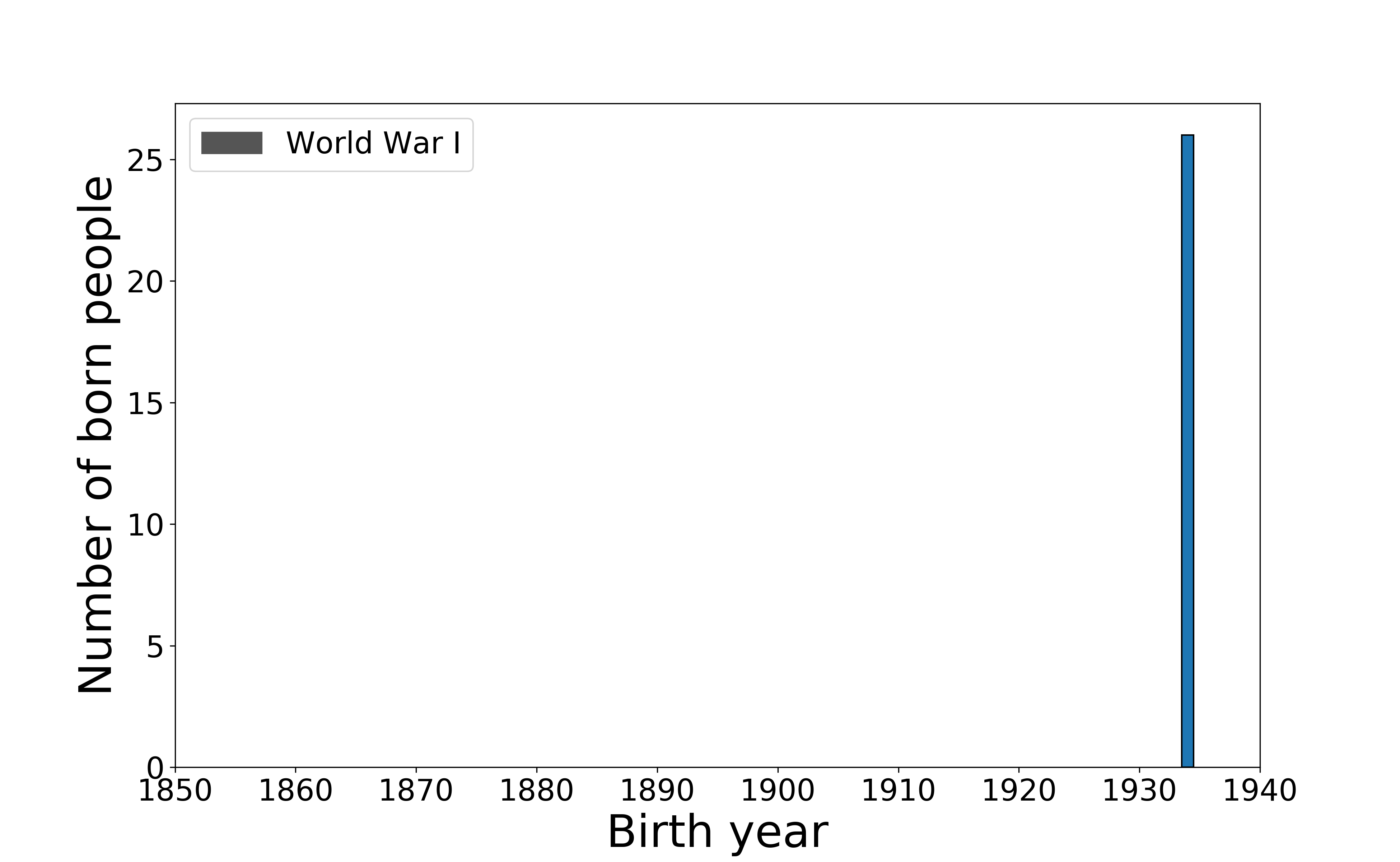}
	\label{fig:myers}
	}
    
	\caption{Top-scoring birth year lists out of $7106$ checked lists for a)~the Whipple's index and for b)~the Myers' index.}
	
	\label{fig:histograms2}
	
\end{figure*}

In addition to that, $7106$ variously sized censuses, i.e. lists of people with birth years available at the website of the United States Holocaust Memorial Museum~(USHMM)~\cite{ushmm2020search} are used since they were composed for the populations from roughly the same time frame. The distribution of the majority of the sizes with the largest ones being excluded for practical purposes is shown in Fig.~\ref{fig:ushmm_sizes}. The geographical locations of these populations differ, but they still mostly cover the populations whose birth year histograms should have similar discrete total variation properties. To make it clear immediately, this does not necessarily mean that the age distributions are similar as well. Namely, one census can have a significantly higher amount of e.g. young people in comparison to other censuses, but as it will be shown later on concrete examples, this should not necessarily affect the discrete total variation of the birth year histograms too significantly. Therefore, these lists available at USHMM constitute an interesting dataset in which to look for outliers in terms of discrete total variation.

\subsubsection{Results}
\label{subsubsec:census_results}

The first experiments that were carried out consisted of simply taking many variously sized subsamples of the birth years from the German census of 1939, calculating the discrete total variations of their birth year histograms, and fitting the proposed method's model in~\eqref{eq:non-uniform_model} to the data obtained in this way. Fig.~\ref{fig:simulation} shows the result of this experiment. The proposed model fits well to all data. This also holds for smaller subsamples where the influences of the two terms in~\eqref{eq:non-uniform_model} are still on par. It can also be seen how the discrete total variations get more dispersed as the sample size grows. While this may hint at heteroscedasticity, applying weighted regression or variance-stabilizing data transformations did not significantly change the results that are described here.

The relation between the sample size and the standard deviation of the discrete total variation is shown in Fig.~\ref{fig:sd}. Very similar results are obtained for other distributions as well. It can be seen that the relation is very similar to the square root function, which effectively justifies the use of~\eqref{eq:d2} for practical purposes. The distribution of the discrete total variations for the subsamples of the same size closely resembles the normal one as shown in Fig.~\ref{fig:sd_distribution} with the remark that the discrete total variations there are integers.

After conducting the relatively simple mentioned experiments in order to get a better insight into the inner workings of the proposed method, the next step was to apply the method to all USHMM lists whose data includes birth years. The distribution of the values of $d'$ described in~\eqref{eq:d2} and obtained by the proposed method in this way is shown in Fig.~\ref{fig:tvor_scores}, while the relation between the calculated discrete total variations and the predicted ones are shown in Fig.~\ref{fig:all}.

It can be seen that the majority of the values $d'$ in~Fig.~\ref{fig:tvor_scores} are not spread too widely with the exception of several outliers. Before analyzing these outliers in more detail and commenting on Fig.~\ref{fig:all}, it must be stressed that in Fig.~\ref{fig:all} the plot axes use the logarithmic scale to better accommodate the presentation to the list' size distribution. Therefore, the apparent misfit for the smallest lists can deceive into believing that the proposed model failed to fit properly, while it is actually only a misfit on a small scale. For similar reasons many of the differences between the calculated and the predicted discrete total variations for the larger lists are higher than they may appear to be on the plot. In addition to showing the proposed method's model, the Monte Carlo model based on the average discrete total variations of the variously sized subsamples of the German census of 1939 is shown in Fig.~\ref{fig:all} for comparison. It can be seen that on several places its predictions are not quite aligned with the ones of the proposed model, which can be attributed to the distribution shown in Fig.~\ref{fig:ushmm_sizes}, i.e. to the significant influence of samples of certain sizes during the model fitting. This can be alleviated by using techniques such as taking only samples of evenly spaced sizes, but as shown later in this subsection, the top results for the two models do not differ significantly even without applying such techniques. Therefore, the application of such techniques was omitted.

Out of the $7106$ lists that were analyzed, the top three outlier lists in terms of $d'$ were the Jasenovac camp inmates list~\cite{ushmm2020jusp} with $d'\approx 43.13$, the list of the Soviet Extraordinary Commission~\cite{ushmm2020sec} with $d'\approx 36.5$, and the list for the Franz Street Number 38~\cite{ushmm2020street2} with $d'\approx 31.29$. The histograms for these lists are shown in Figs.~\ref{fig:jms},~\ref{fig:20781}, and~\ref{fig:38665}, respectively. A more detailed analysis of the top-scoring histogram that provides additional insights and explanations of the behavior of the proposed method's scoring is available in Appendix.

\begin{figure*}[htb]
    \centering
    
	\subfloat[]{
	\includegraphics[width=0.48\linewidth]{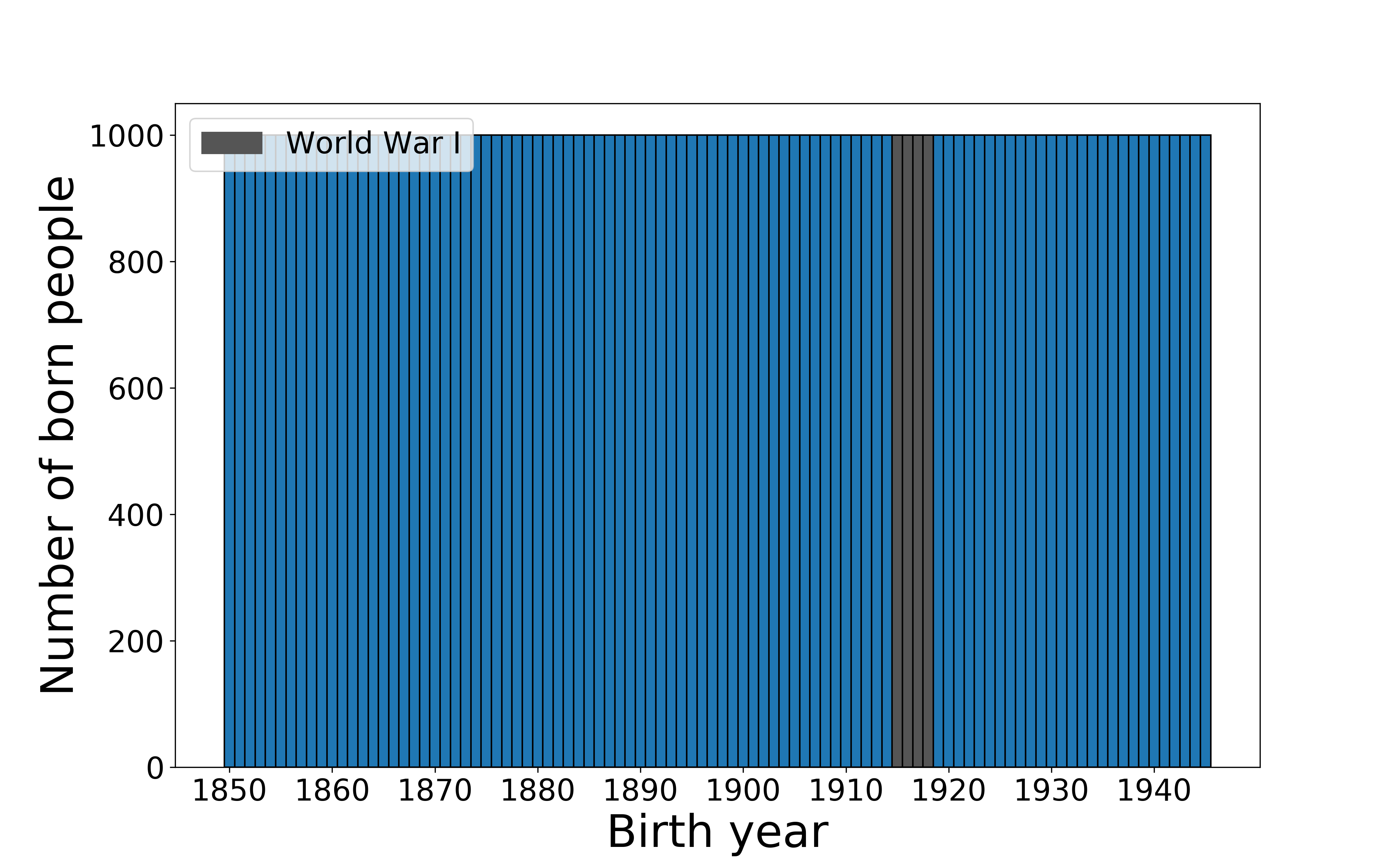}
	\label{fig:same_1}
	}%
	\subfloat[]{
	\includegraphics[width=0.48\linewidth]{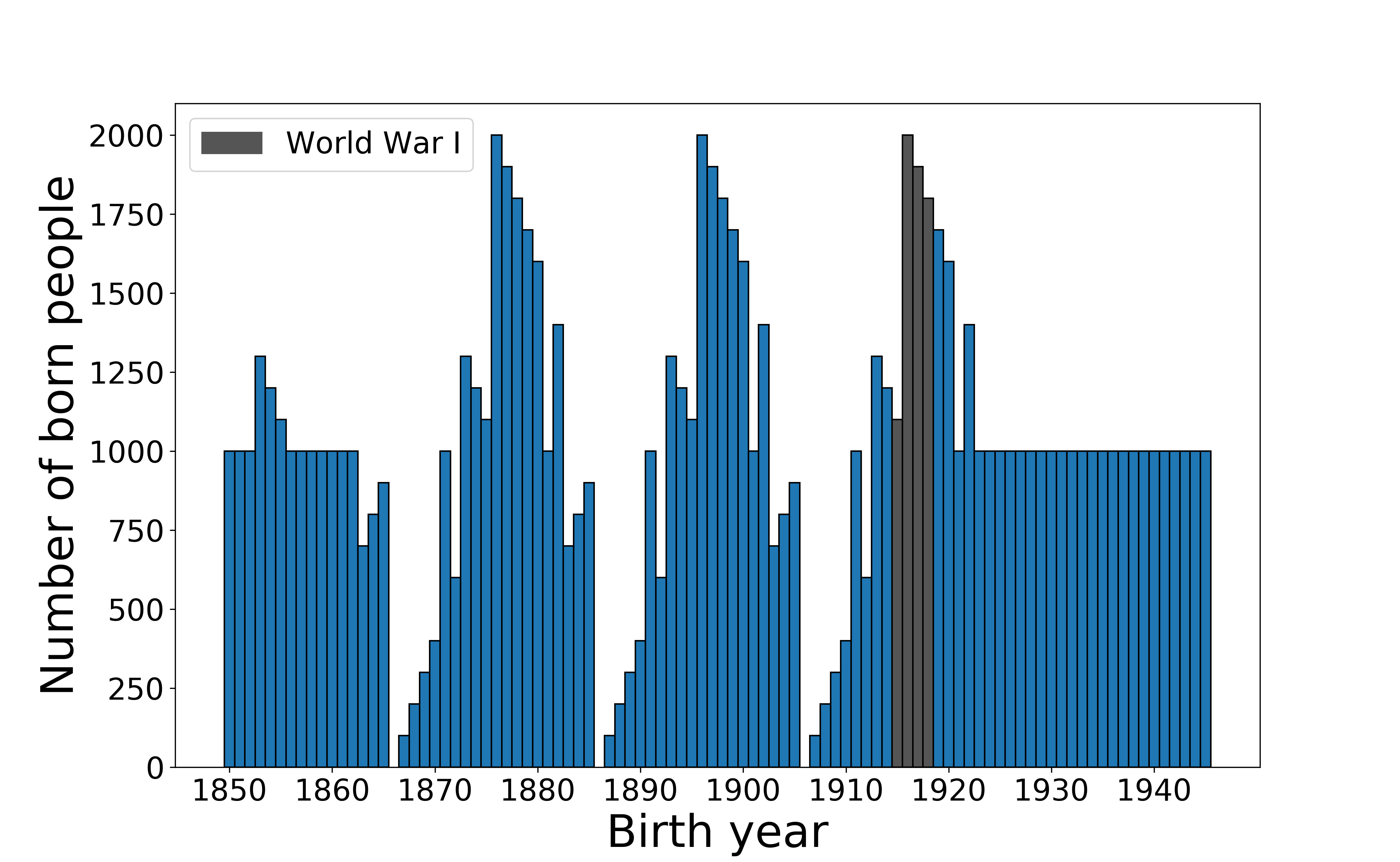}
	\label{fig:same_2}
	}
    
	\caption{Examples of histograms for all of which the Whipple's and the Myers' indices have exactly the same values.}
	
	\label{fig:histograms3}
	
\end{figure*}

In the case of Monte Carlo the score $d''$ was calculated as
\begin{equation}
\label{eq:d3}
\begin{gathered}
d''=\frac{\left|\V{\mathbf{x}_n}-\hat{\mu}_N\right|}{\hat{\sigma}_N}
\end{gathered}
\end{equation}
where $\hat{\mu}_N$ and $\hat{\sigma}_N$ are the mean and the standard deviation, respectively, of the discrete total variation obtained for a large number of subsamples of size $N$ of the German census of 1939. The distribution of the values of $d''$ obtained for all lists from the USHMM is shown in Fig.~\ref{fig:mc_scores}. The lists with the first and second highest value of $d''$ were the same as for $d'$ with $d''\approx 58.14$ and $d''\approx 49.04$, while the list with the third highest value of $d''$ was the list of Jewish refugees in Tashkent~\cite{ushmm2020tashkent} with $d''\approx 44.51$ and with the corresponding histogram shown in Fig.~\ref{fig:20492}. Already by looking at the mentioned figures for the top-scoring lists it can be seen that their corresponding histograms indeed have high values of discrete total variation with spikes, i.e. individual bins that significantly differ from their neighbors, which contrasts the smoothness of the histogram for the German census of 1939.

\begin{figure*}[htb]
    \centering
    
	\includegraphics[width=\linewidth]{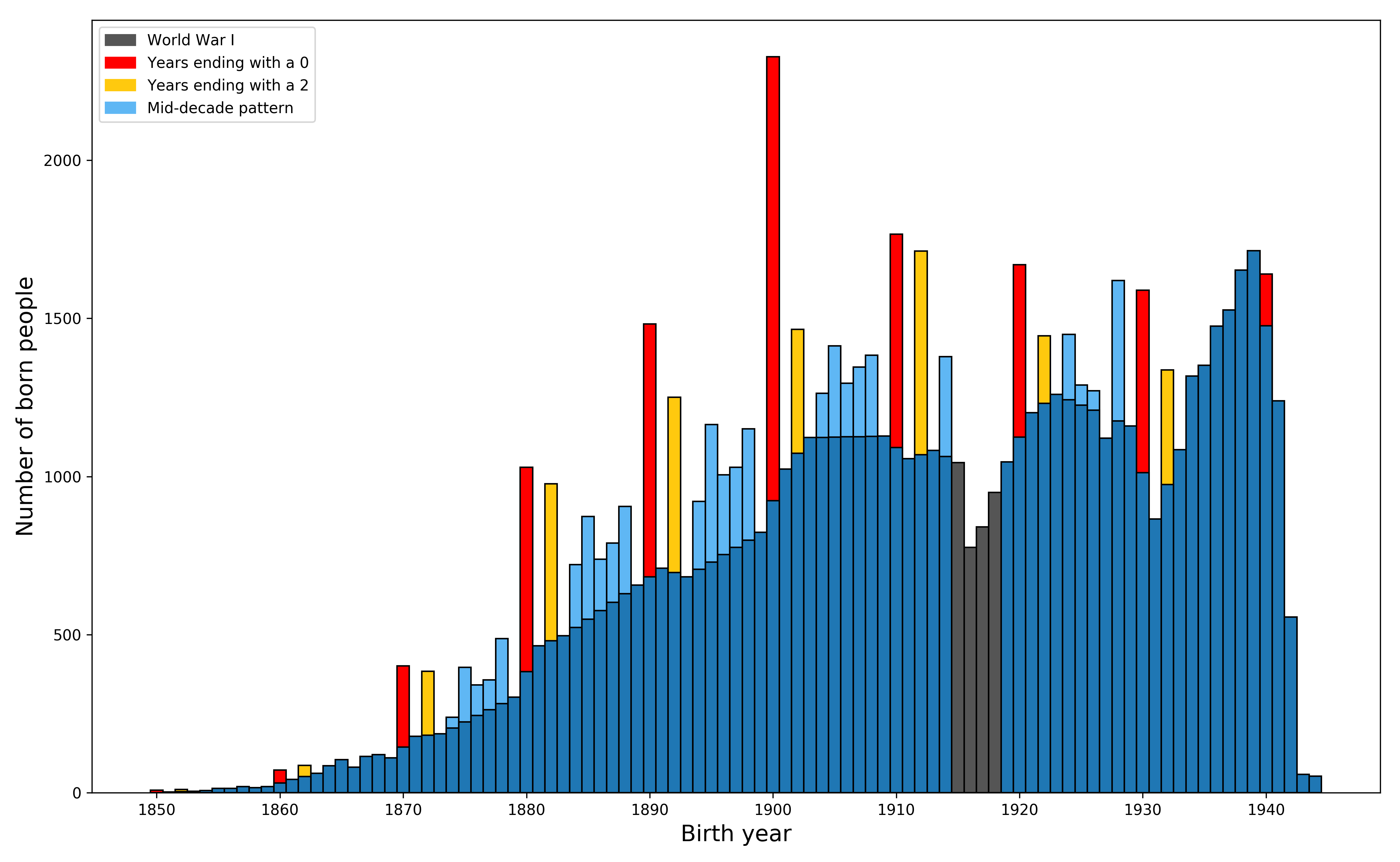}
	
    \caption{Same data for Jasenovac inmates as in Fig.~\ref{fig:jms}, but with additional markings for the age heaping~\cite{caselli2005demography} artifacts.}
	\label{fig:jms_marked}
    
\end{figure*}

\subsection{Advantages over existing metrics}
\label{subsec:advantages}

Like for many other groups of population histograms, there is no ground-truth ordering for USHMM lists in terms of their histograms' smoothness or accordance with historical populations. Because of that, the quality of ordering obtained by the proposed method and by existing metrics such as Whipple's and Myers' indices can not be compared directly. However, it is possible to show cases that are problematic for both of these indices, but not for the proposed method.

The first example is the histograms shown by Figs.~\ref{fig:whipple} and~\ref{fig:myers}, which represent the top-scoring histograms among the USHMM lists' histograms for the Whipple's and Myers' indices, respectively. It can be seen that these histograms are actually relatively smooth, but they also contain only a few non-zero values: the first one $8$ and the second one only a single. These histograms can hardly be considered outliers in terms of smoothness when compared to the histograms in Fig.~\ref{fig:histograms}, but rather outliers in terms of covered years span, which is different and also detectable by much simpler techniques. Additionally, the lists that produced these histograms have only a relatively small number of birth years and since the mentioned indices, unlike the proposed method, do not take into account the sample size, they are also more prone to anomalies that arise in smaller samples due to randomness.

Another problem with metrics such as Whipple's and Myers' indices is that they are mainly concerned with frequencies and do not take into account other properties such as shape or smoothness. Because of that, for different samples that have the same frequencies of last digits of their numbers, it is still possible to obtain the same values of the mentioned indices even if the samples' histograms differ significantly. An example of this is given in Fig.~\ref{fig:histograms3} with a fully smooth histogram that has the same indices values as a histogram that can hardly be considered smooth. While numerous similar examples exist, the ones presented are enough to show the frequency-based weakness of the Whipple's and Myers' indices. On the other hand, the proposed method has no such problems and its values for the histograms in Fig.~\ref{fig:histograms3} differ significantly with one being zero and the other one non-zero.

In short, while being widely used and useful in certain cases, metrics such as the Whipple's and the Myers' indices are too simple to properly handle properties such as smoothness. Therefore, the proposed method's ability to specifically target smoothness is its main advantage over other metrics.

\subsection{Discussion}
\label{subsec:discussion}

Looking at the distributions shown in Figs.~\ref{fig:tvor_scores} and~\ref{fig:mc_scores} and observing the significant difference between the majority of the scores and the highest scores, it can be concluded that the histograms of the used USHMM lists that obtained the highest scores are indeed outliers in terms of the discrete total variation. Since the analyzed data consisted of birth years, it may seem that an appropriate tool for identifying outliers such as the ones in Fig.~\ref{fig:histograms} could be the Whipple's index~\cite{shryock1980methods}, but due to its fixed nature of checking only specific kinds of data, it is often inappropriate~\cite{wang1998age,spoorenberg2009whipple}. This also holds in the case of the histogram of the Jasenovac inmates shown in~\ref{fig:jms} whose artifacts are marked more closely in Fig.~\ref{fig:jms_marked}. It can be seen that age heaping occurs in several forms that the Whipple's index not only cannot pick up, but it also gets hampered by them. Namely, in its slightly changed form the Whipple index checks for a surplus of years ending in $0$ or $5$ when compared to other years, but in the case of Jasenovac there is also a surplus of years ending in $2$, which is not checked by the Whipple's index and it actually reduces the overall surplus of years ending in $0$ or $5$, thus hampering the Whipple's index in detecting the unusual data patterns. Since the proposed method has no such problems, it may be more appropriate in situations similar to the one in this experiment.

Besides all these histogram artifacts, there are other peculiarities with the Jasenovac list. Namely, if it is compared to other USHMM lists used here, it directly contradicts some of them. For example, the list available at~\cite{ushmm2020oral} states that a certain Stanko Nick survived the war~\cite{ushmm2020nickoral}, while the Jasenovac list claims that he was killed~\cite{ushmm2020nickjusp}, which is known to be wrong~\cite{enwiki2020nick}. In another example, the list available at~\cite{ushmm2020auschwitz} states that a certain Josip Stern arrived at Auschwitz in 1942~\cite{ushmm2020sternau}, while the Jasenovac list claims that he was killed in 1941~\cite{ushmm2020sternjusp}.
This means that the proposed method can also be used to detect samples that contain potentially problematic data with properties not always shared with the usual outliers.

\subsection{Source code and data repository}
\label{subsec:repository}

The source code written in the Python programming language and the data required to recreate the results described in this section are publicly available in a dedicated GitHub repository.\footnote{{https://github.com/DiscreteTotalVariation/TVOR}} At the time of writing this paper, the census data used in this section was publicly available at the USHMM website, but for the sake of simplicity of recreating the results, it is also available in the repository. While the census data also contains other information alongside the birth years, only the birth years were copied to the repository in order to avoid data privacy violation for potentially still living persons. For example, according to the Jasenovac camp inmates list~\cite{ushmm2020jusp}, which was already shown to be problematic, a certain Stojan Ra{\v{z}}okrak~\cite{ushmm2020razokrak} was allegedly killed in 1942, but a publicly available video of him\footnote{{https://www.youtube.com/watch?v=S5lRwT63as0}}\textsuperscript{,}\footnote{{https://archive.is/48sKw}} from 2012 and its transcript\footnote{https://archive.is/RtnsJ} clearly show the opposite. Because of that, it seemed reasonable to copy only the birth years, while any interested reader can check the rest of the data at the USHMM website by using the appropriate list identifier given in the repository.

\begin{figure*}[htbp]
    \centering
    
  \subfloat[]{
  \includegraphics[width=0.48\linewidth]{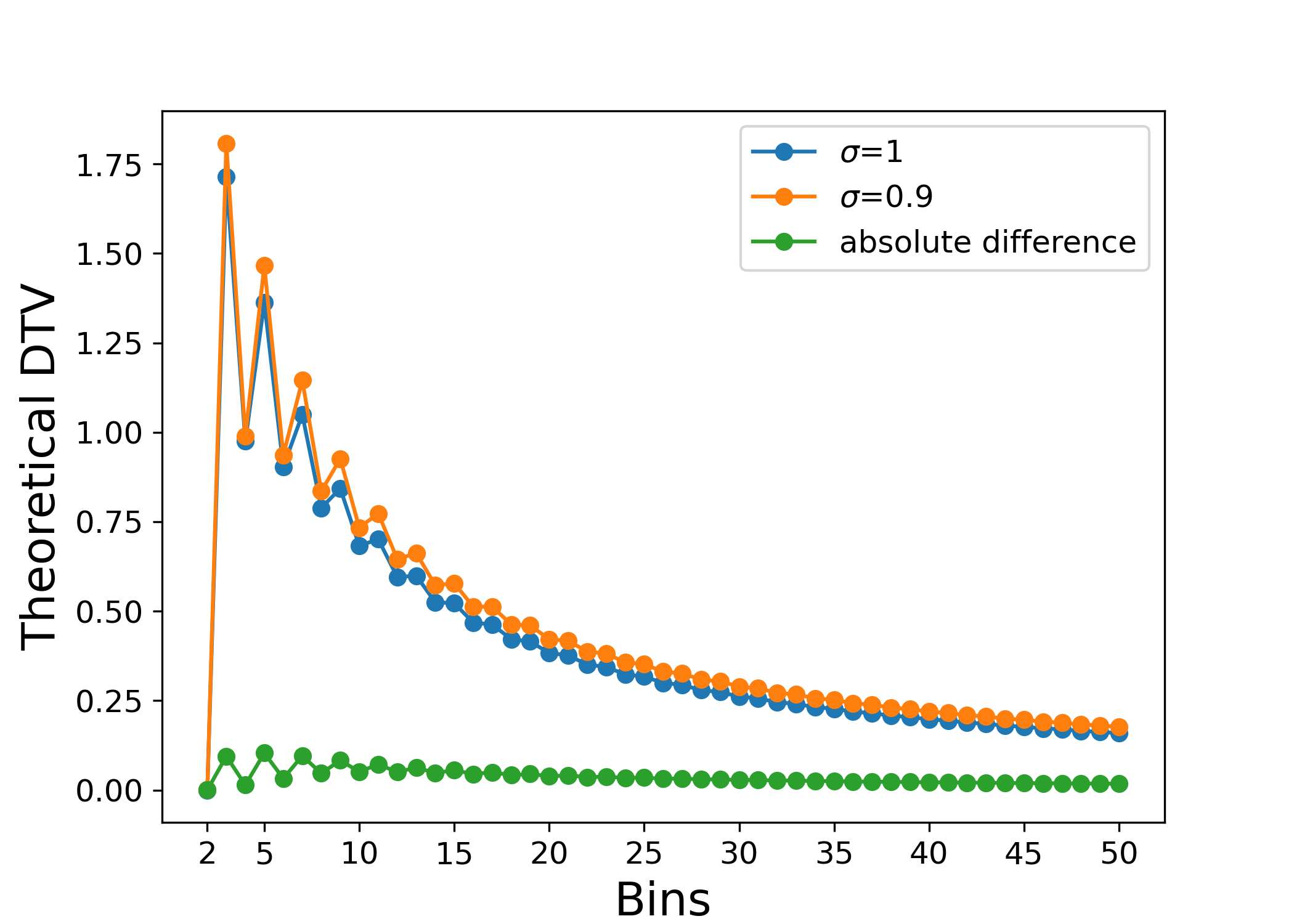}
  \label{fig:n_b_5_s_0.9}
  }%
  \subfloat[]{
  \includegraphics[width=0.48\linewidth]{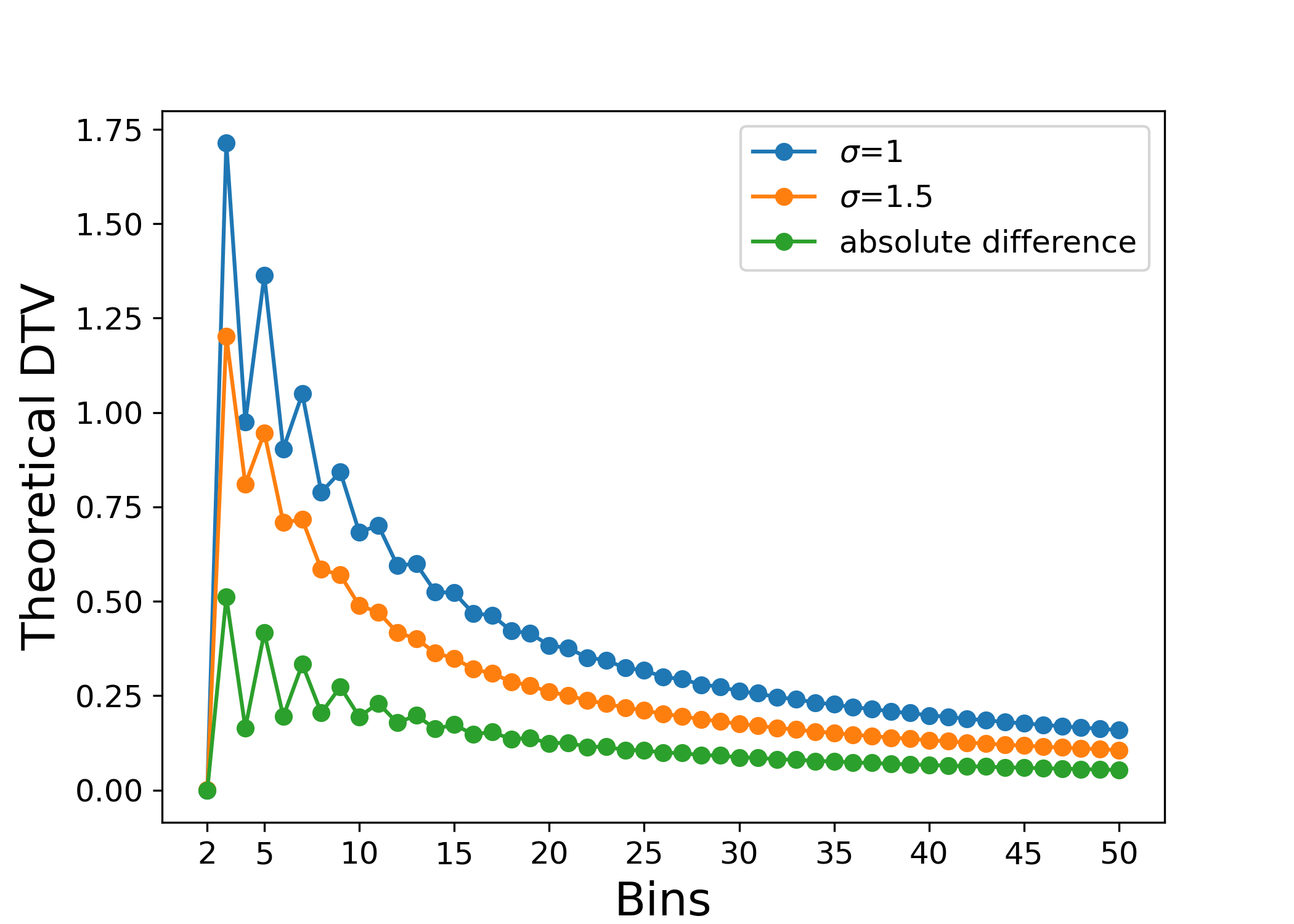}
  \label{fig:n_b_5_s_1.5}
  }
  \\
  \subfloat[]{
  \includegraphics[width=0.48\linewidth]{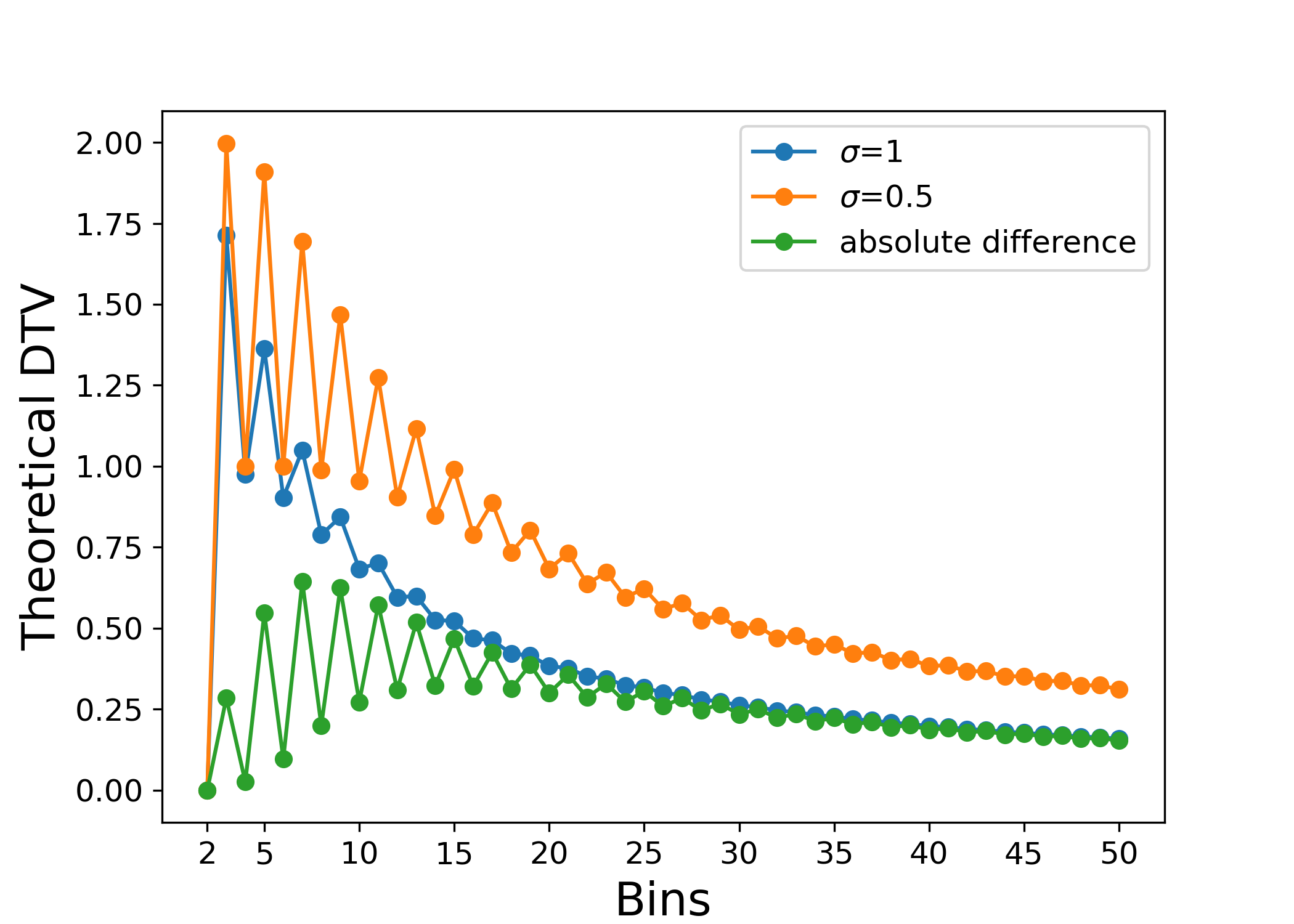}
  \label{fig:n_b_5_s_0.5}
  }%
  \subfloat[]{
  \includegraphics[width=0.48\linewidth]{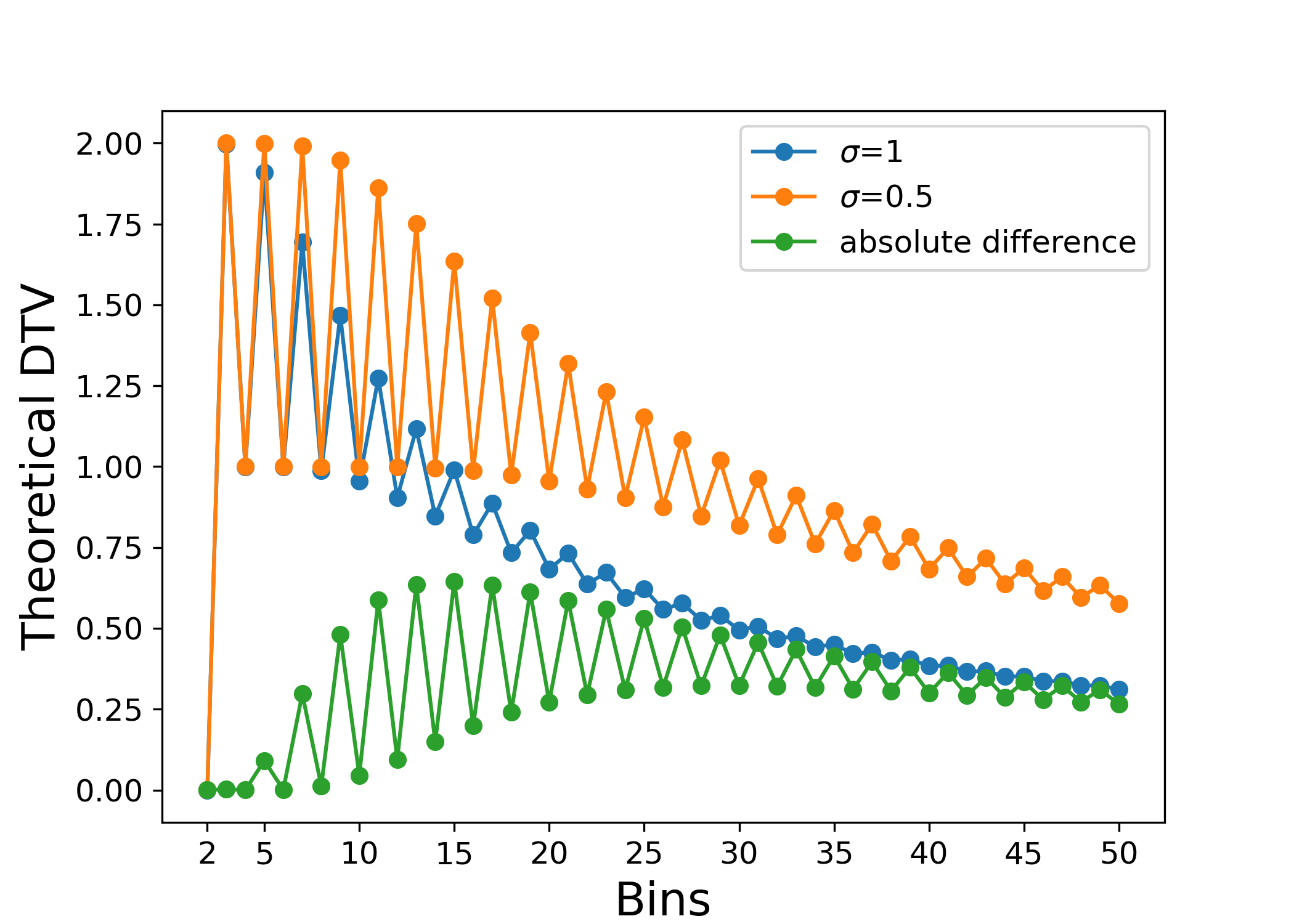}
  \label{fig:n_b_10_s_0.5}
  }
  
    \caption{The comparison of the values of the theoretical discrete total variation $\V{{\cal D}}$ of the histograms of normal distribution ${\cal N}(0,\sigma^2)$ for the values in the interval $\left[-b, b\right]$ for various number of histogram bins used to obtain the experimental results that were shown earlier in Fig.~\ref{fig:tvor_vs_baseline}: a)~$c=5, \sigma=0.9$, b)~$c=5, \sigma=1.5$, c)~$c=5, \sigma=0.5$, and d)~$c=10, \sigma=0.5$.}
  \label{fig:tvor_vs_baseline_dtvs}
    
\end{figure*}

\begin{figure*}[htbp]
    \centering
    
  \subfloat[]{
  \includegraphics[width=0.48\linewidth]{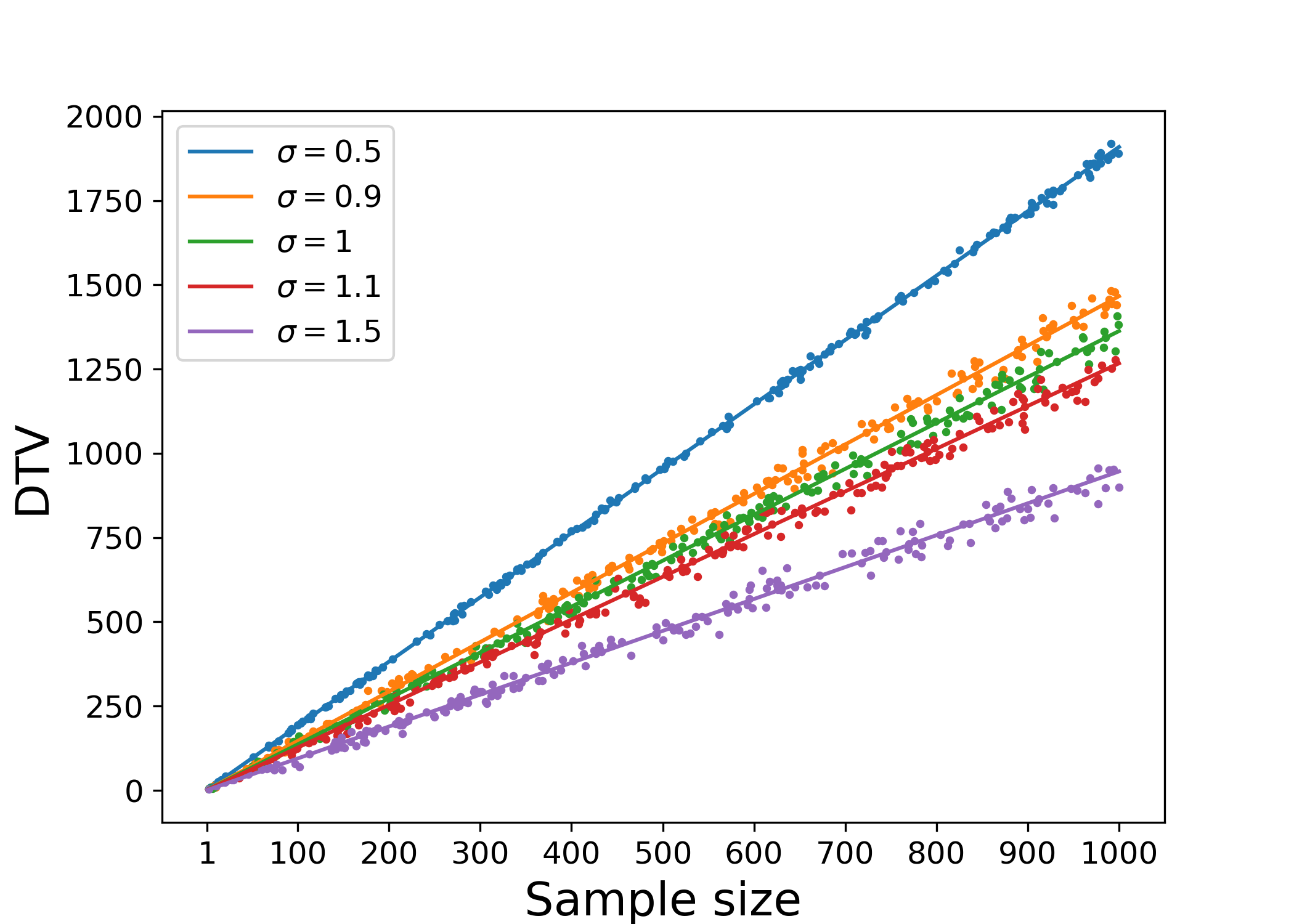}
  \label{fig:n_5_up_1000}
  }%
  \subfloat[]{
  \includegraphics[width=0.48\linewidth]{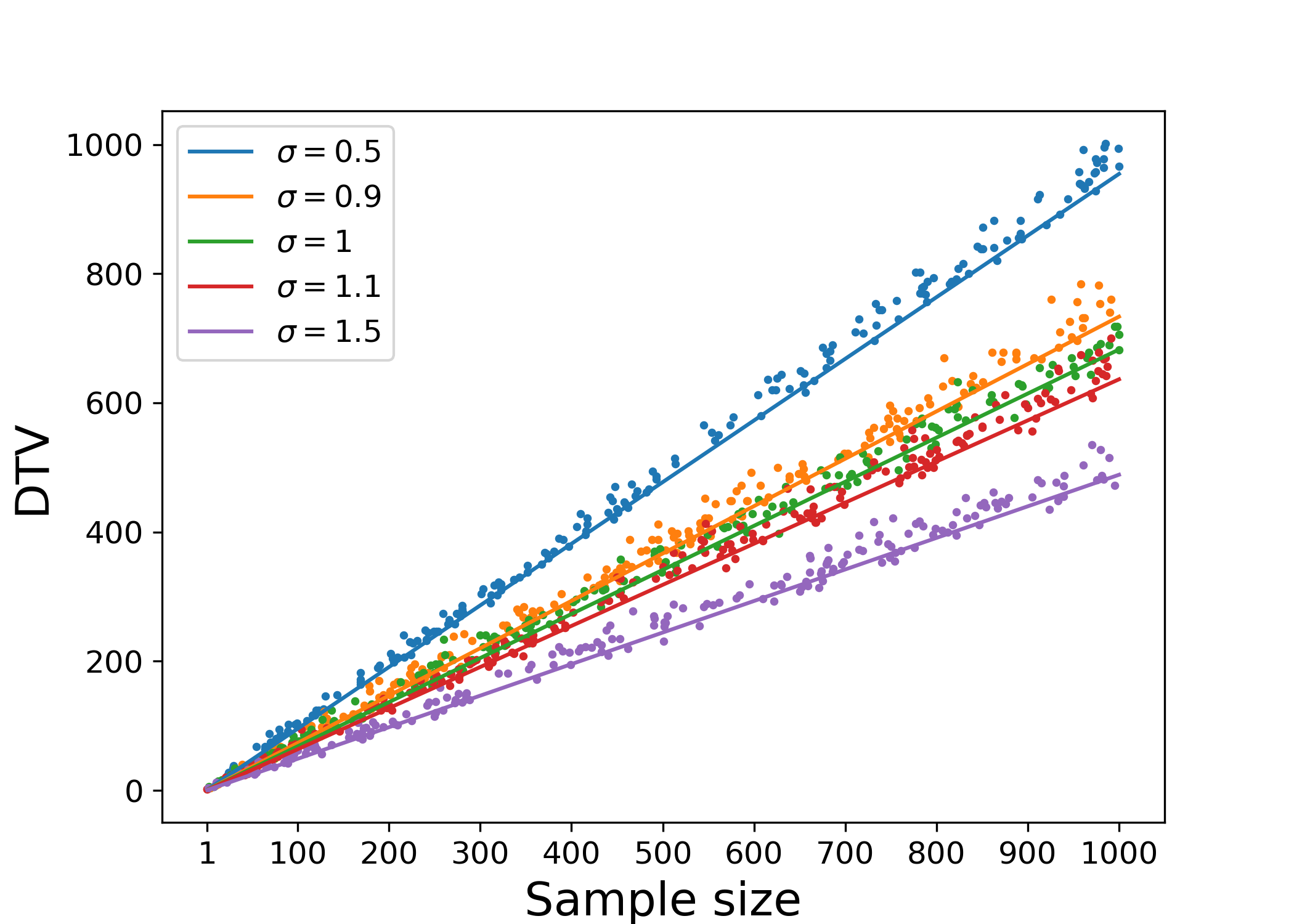}
  \label{fig:n_10_up_1000}
  }
  \\
  \subfloat[]{
  \includegraphics[width=0.48\linewidth]{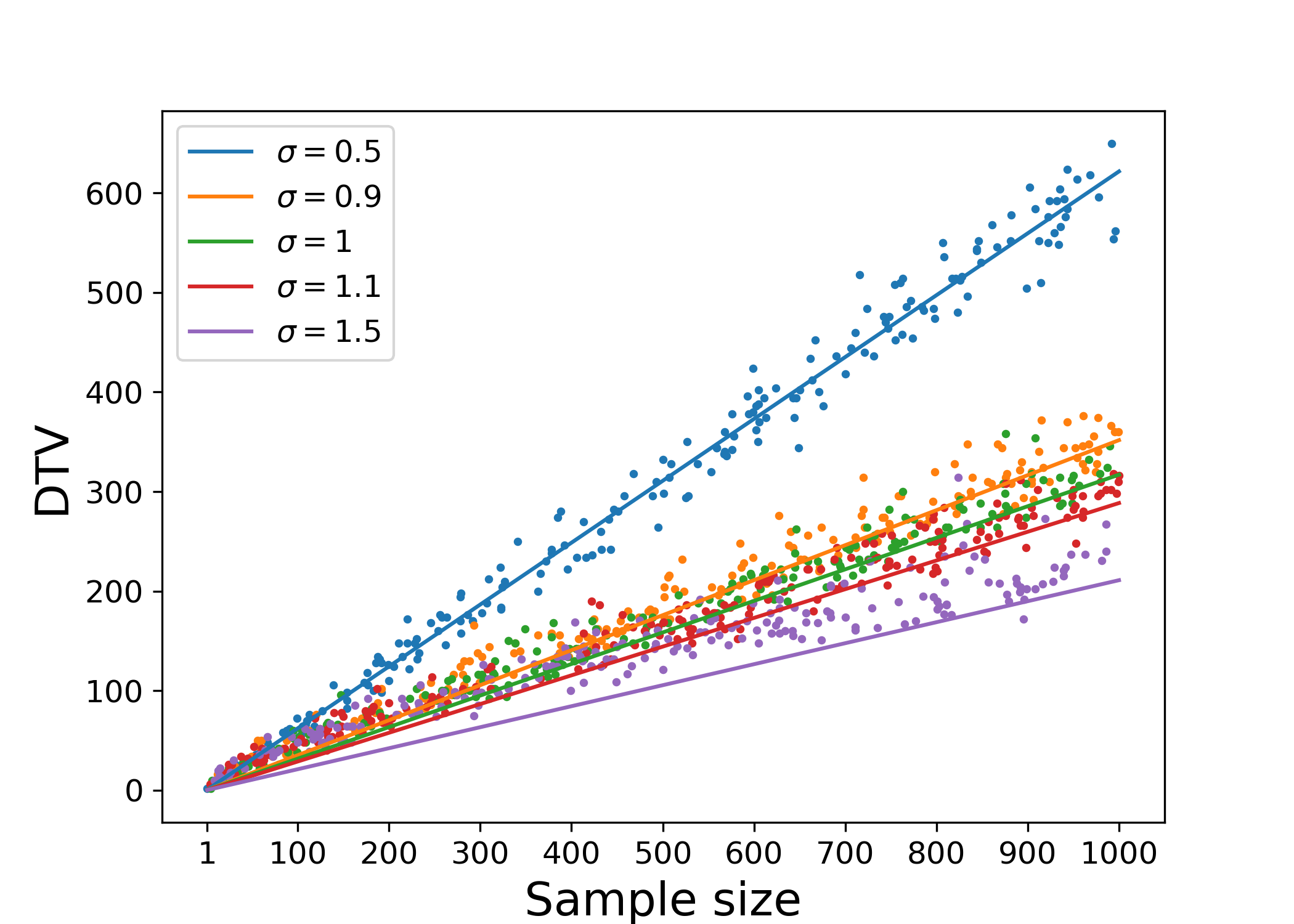}
  \label{fig:n_25_up_1000}
  }%
  \subfloat[]{
  \includegraphics[width=0.48\linewidth]{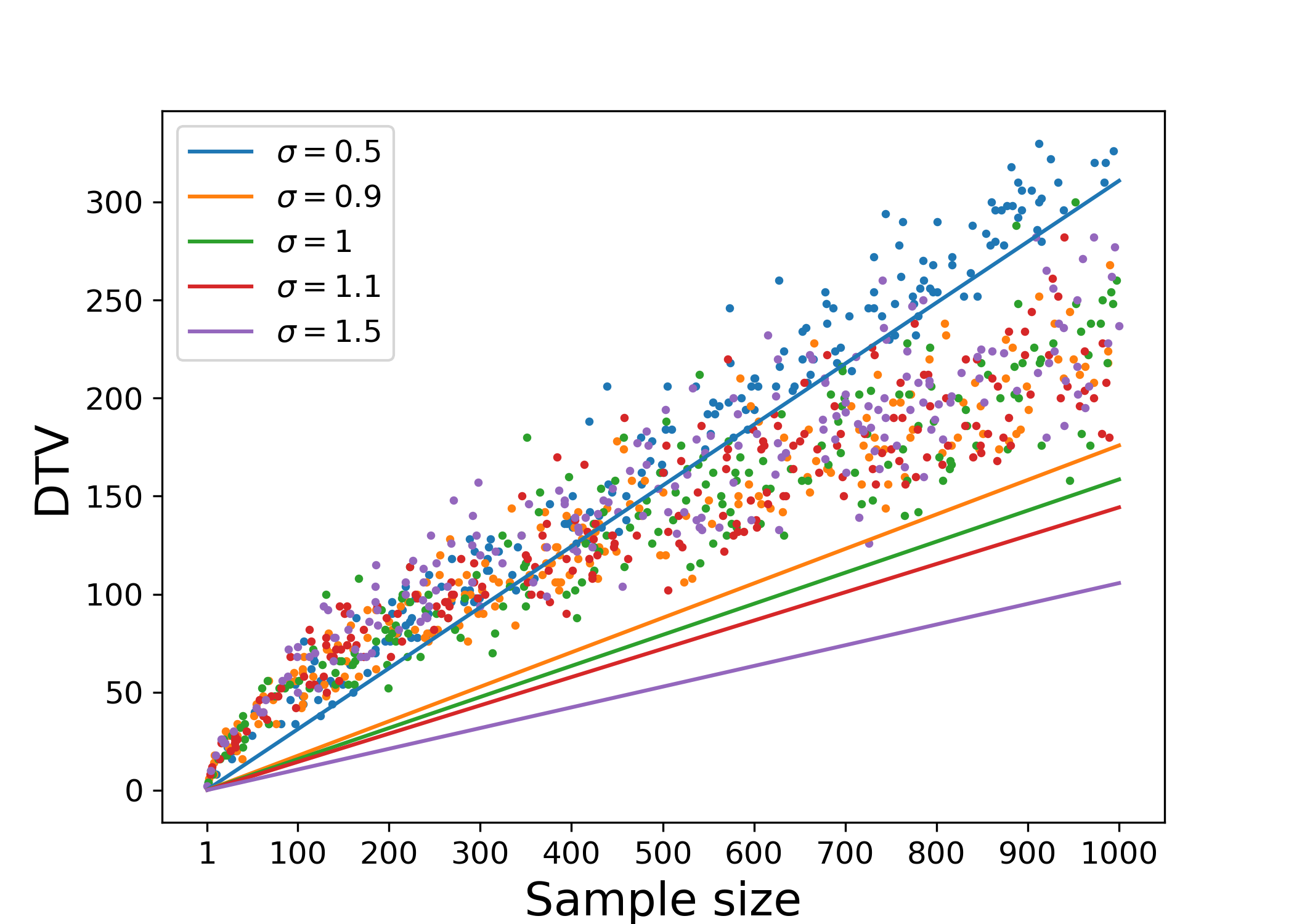}
  \label{fig:n_50_up_1000}
  }
  \\
  \subfloat[]{
  \includegraphics[width=0.48\linewidth]{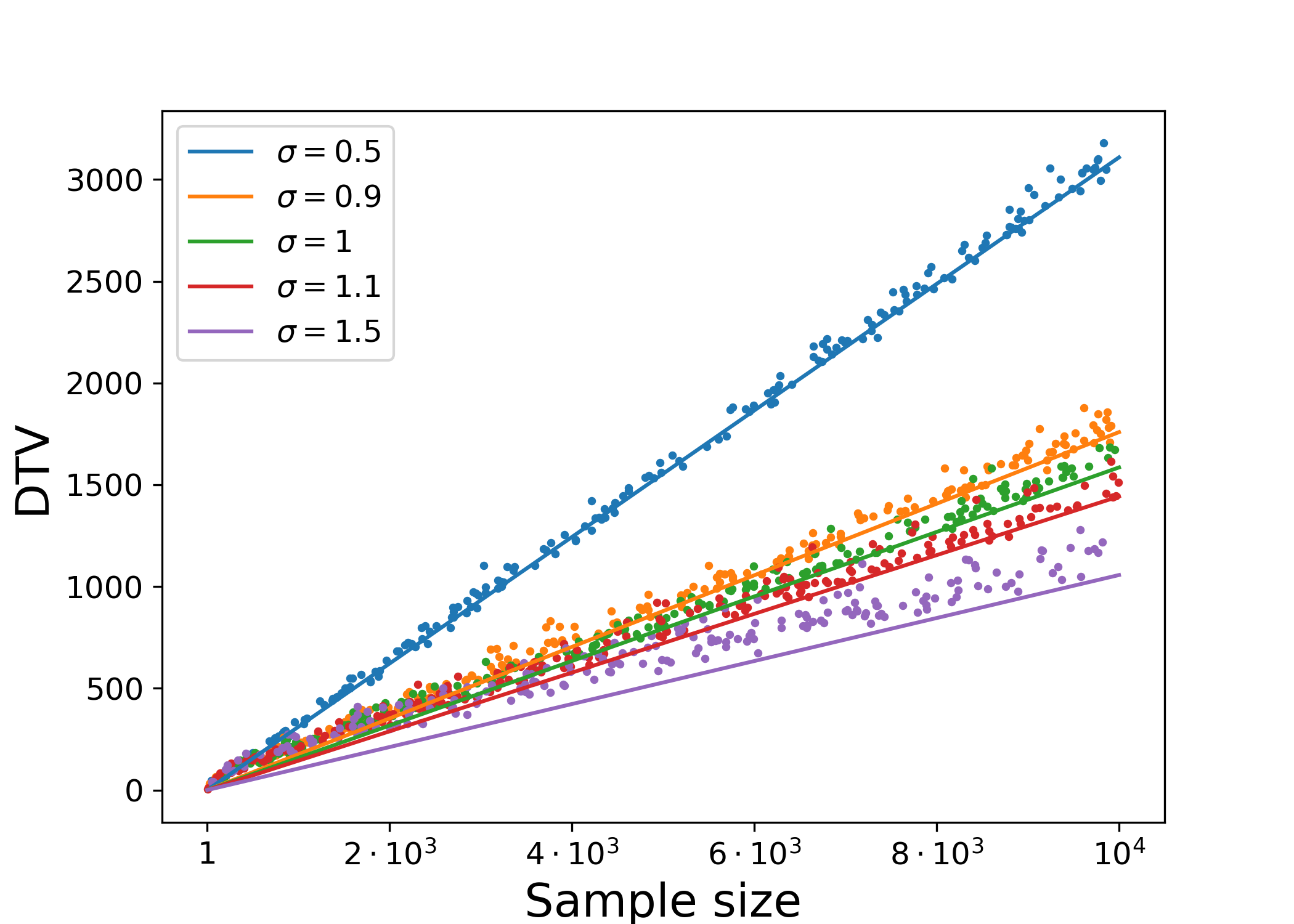}
  \label{fig:n_50_up_10000}
  }%
  \subfloat[]{
  \includegraphics[width=0.48\linewidth]{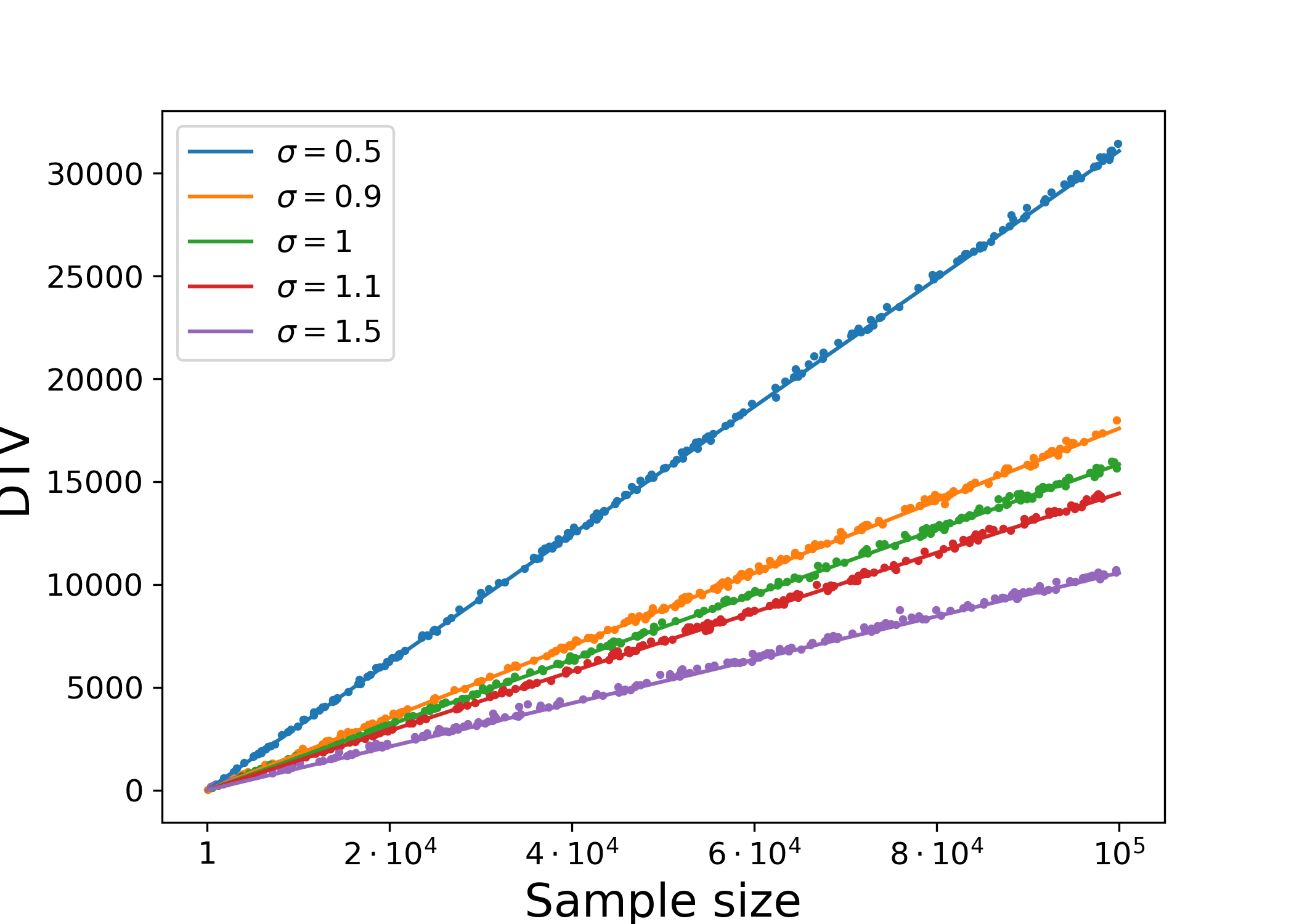}
  \label{fig:n_50_up_100000}
  }
  
    \caption{The DTVs of histograms of random samples drawn from $\N{0}{\sigma^2}$ and of sizes randomly chosen to be between $1$ and $U$. The number of bins $n$ and the upper size bound $U$ are set to a)~$n=5$ and $U=1000$, b)~$n=10$ and $U=1000$, c)~$n=25$ and $U=1000$, d)~$n=50$ and $U=1000$, e)~$n=50$ and $U=10^4$, and f)~$n=50$ and $U=10^5$. The lines represent the value of $\V{{\cal D}}$ described in \eqref{eq:theoretical_variation} and multiplied by the sample size, while the dots represent the random samples.}
  \label{fig:sizes}
    
\end{figure*}

\begin{figure}[htb]
    \centering
    
	\includegraphics[width=\linewidth]{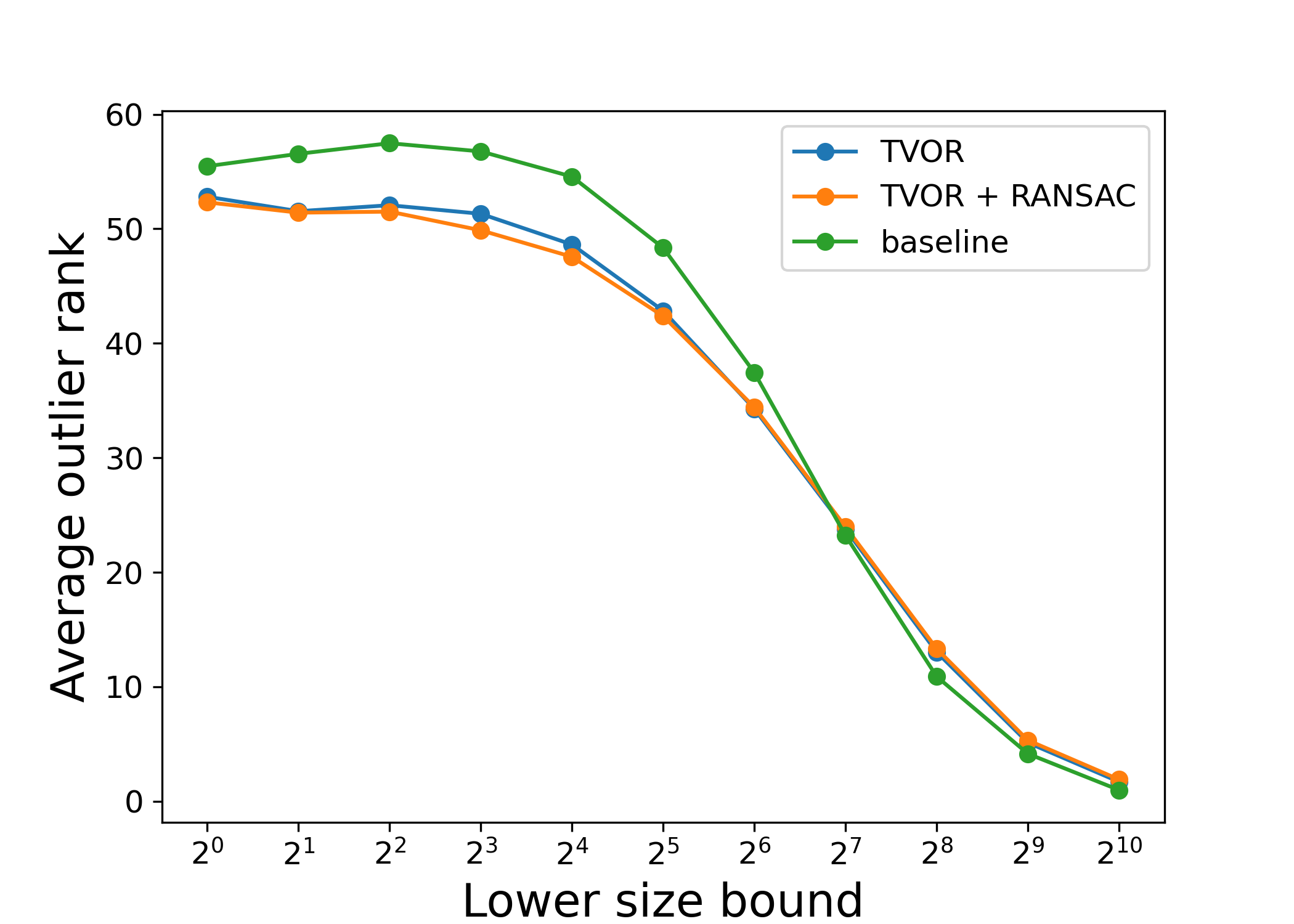}
	
    \caption{The dependence of the performance of the baseline and the proposed method on the random samples' size range when $100$ inlier samples are drawn from $\N{0}{1}$, a single outlier sample is drawn from $\N{0}{0.9^2}$, the number of bins $n$ is $15$, $c=5$, and the size of the inlier and the outlier samples is randomly chosen to be between $L$ and $10\cdot L$ where $L$ is the lower size bound that is shown on the $x$-axis.}
	\label{fig:sizes2}
    
\end{figure}

\section{Conclusions and future work}
\label{sec:conclusions}

In this paper, a method for finding discrete total variation outliers among histograms has been proposed. It scores histograms based on the deviation of their discrete total variation from its expected value. To carry out this scoring, a statistical framework has been proposed. One of the method's main advantages is that in order to work it requires no information about the distribution of the samples that are being described by histograms. In some special cases the proposed method even outperforms the Pearson's chi-squared test when looking for the outlier histograms in terms of the sample distribution despite the fact that is was not designed for this task. On the other hand, the proposed method clearly outperforms the Pearson's chi-squared test when looking for discrete total variation outliers, especially in cases of a huge amount of outliers. Overall, the proposed method represents a successful proof-of-concept of how discrete total variation that is used in the method's modelling can be applied to histogram outlier detection in terms of discrete total variation, which has been experimentally confirmed on synthetic and real-life data. Future work may include looking for some other histogram properties that can also be used for histogram outlier detection in terms of their smoothness in the cases where the distribution of the histogram samples is unknown. As for improving the proposed method, future work will include at least two things. The first of them is the analysis of variance for the discrete total variation to potentially improve the scoring criteria. The second of them comprises other aspects of the histogram's discrete total variation that could decrease the scores obtained for the inlier samples, but simultaneously keep the scores obtained for the outliers high.

\begin{figure*}[htb]
    \centering
    
  \subfloat[]{
  \includegraphics[width=0.48\linewidth]{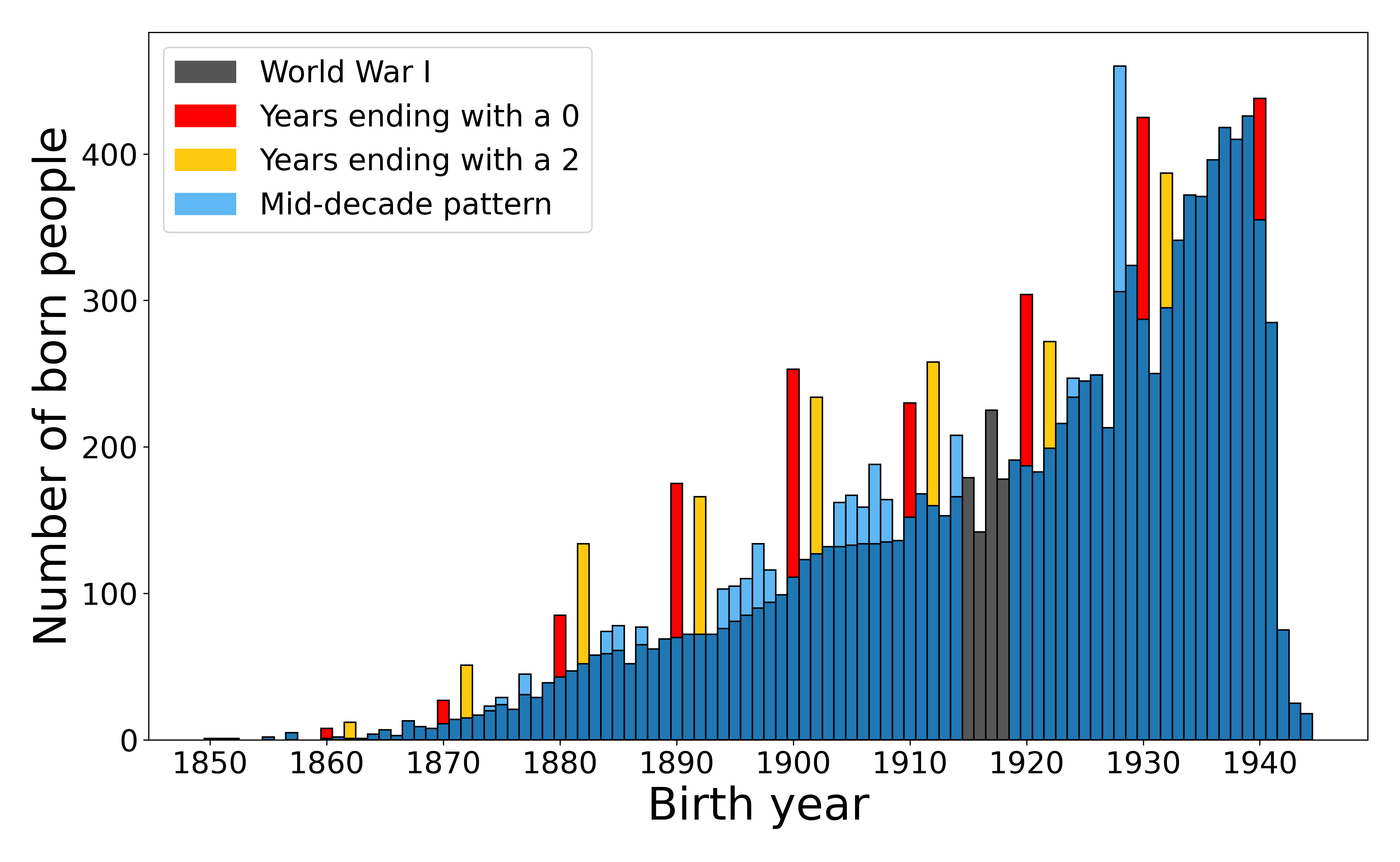}
  \label{fig:romani_marked}
  }%
  \subfloat[]{
  \includegraphics[width=0.48\linewidth]{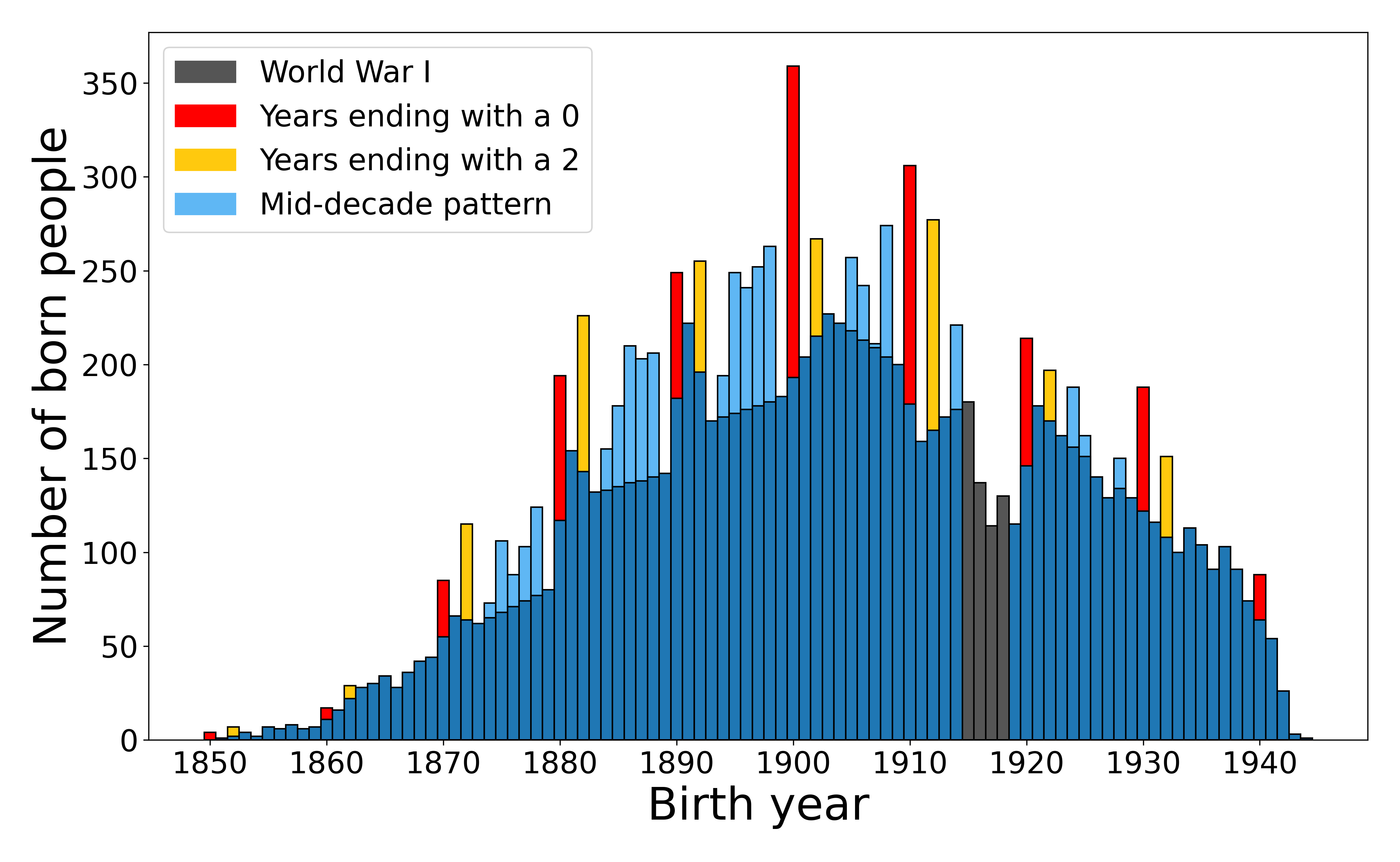}
  \label{fig:jews_marked}
  }
  \\
  \subfloat[]{
  \includegraphics[width=0.48\linewidth]{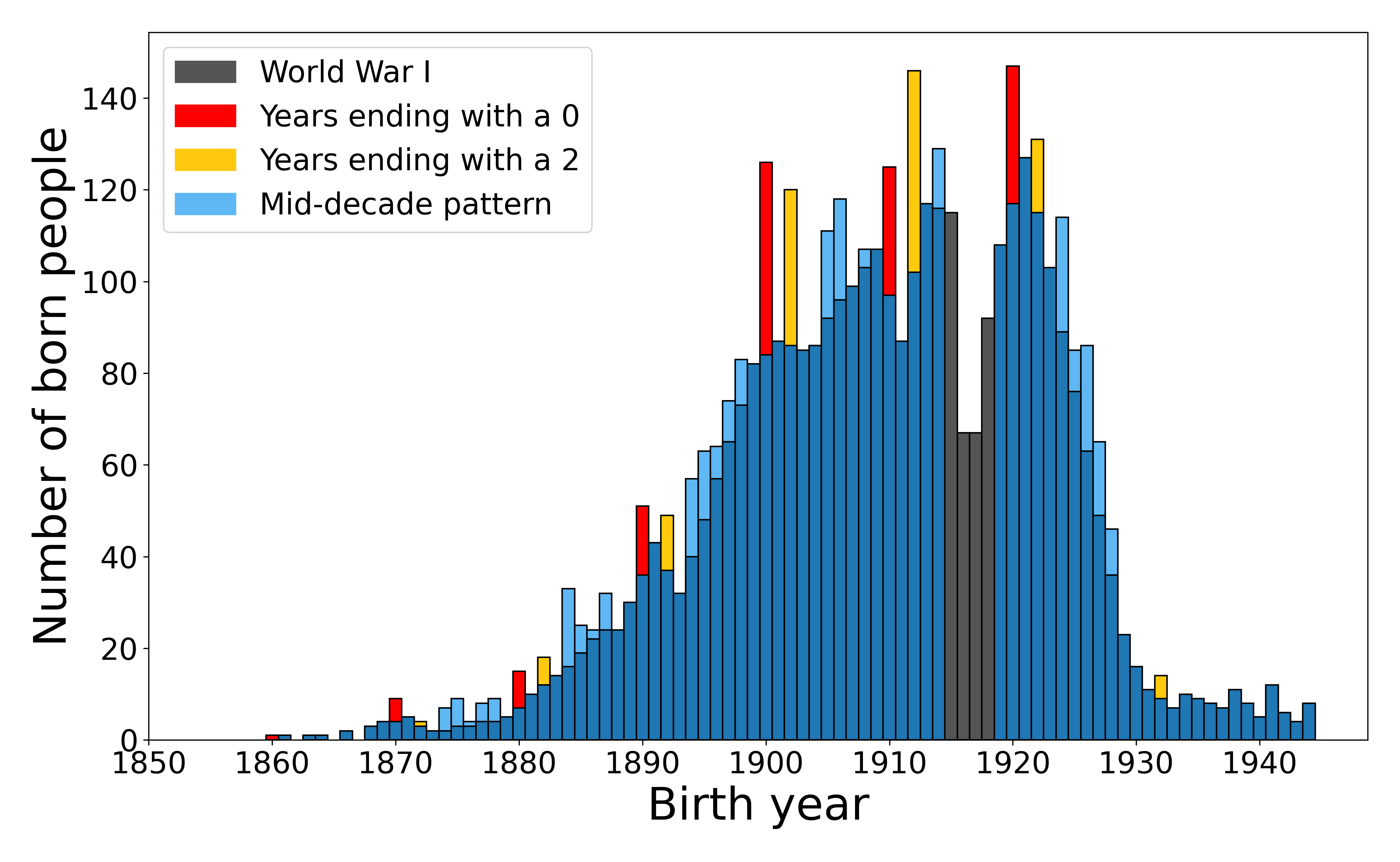}
  \label{fig:croats_marked}
  }%
  \subfloat[]{
  \includegraphics[width=0.48\linewidth]{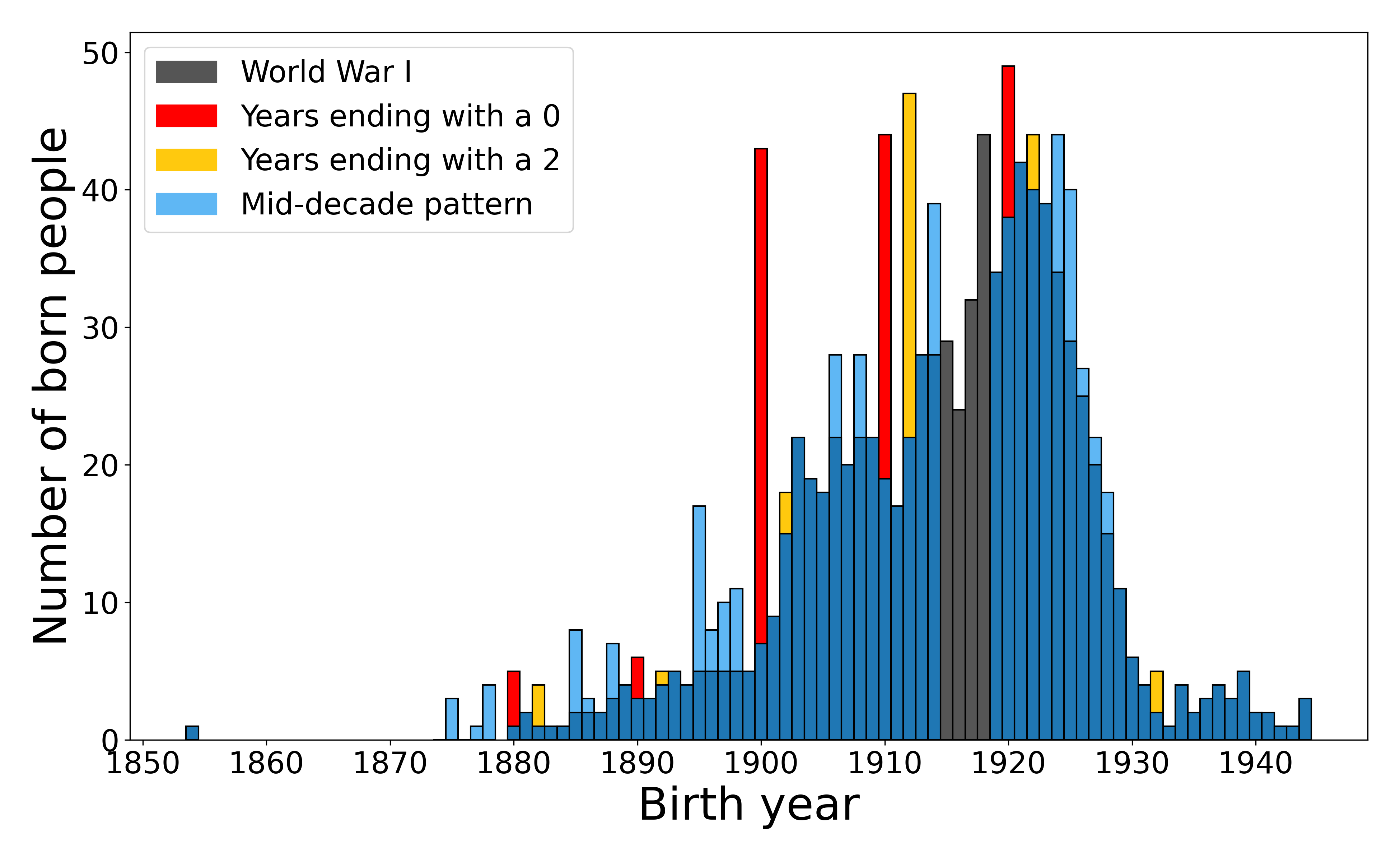}
  \label{fig:muslims_marked}
  }
  
    \caption{Birth year histograms of Jasenovac camp inmates~\cite{ushmm2020jusp} by nationality with markings for age heaping: a)~Roma inmates, b)~Jewish inmates, c)~Croatian inmates, and d)~Muslim inmates. Only the histograms for nationalities for which there are more than $1000$ listed inmates are shown here, while the histogram for the Serbian inmates is given separately in Fig.~\ref{fig:jms_serbian}.}
  \label{fig:jms_parts}
    
\end{figure*}

\section*{Appendix}
\label{sec:appendix}

\subsection{Proofs of the theorems}
\label{subsec:theorems}

\subsubsection{Proof of Theorem~\ref{th:2-dimensional}}
\label{subsubsec:theorem-2}

By the definition of $F(2,N)$, it can be developed as follows:
\begin{align}
    \label{eq:edtvb2}
    F(2,N)&=\E{\V{\mathbf{x}_2}}\nonumber\\
    &=2^{-N+1}\sum\limits_{k=0}^{\lfloor\frac{N-1}{2}\rfloor}\binom{N}{k}\left(N-2k\right).
\end{align}
	For an even $N=2r$, the equality $\sum_{k=0}^{N}{N \choose k}=2^N$ leads to
\begin{equation}
    \label{eq:ncks}
    \begin{gathered}
    \sum_{k=0}^{r-1}{2r \choose k}\left(2r-2k\right)\\
    =2r\sum_{k=0}^{r-1}{2r \choose k}-4r\sum_{k=1}^{r-1}{2r-1 \choose k-1}\\
    =r\left(2^{2r}-{2r \choose r}\right)-2r\left(2^{2r-1}-2{2r-1 \choose r-1}\right)\\
    =-r{2r \choose r}+2r{2r \choose r}=r{2r \choose r}.
    \end{gathered}
\end{equation}
Since here $\lfloor(N+1)/2\rfloor=\lfloor(2r+1)/2\rfloor=r$ and $\lfloor N/2\rfloor=\lfloor(2r+1)/2\rfloor=r$, it follows that \eqref{eq:ncks} matches \eqref{eq:F(2,N)}.
For an odd $N=2r+1$, a similar calculation as before gives
\begin{equation}
    \label{eq:ncks2}
    \begin{gathered}
    \sum_{k=0}^{r}{2r+1 \choose k}\left(2r+1-2k\right)=\left(r+1\right){2r+1 \choose r}.
    \end{gathered}
\end{equation}

\medskip

To avoid any possible confusion, it has to be mentioned that the lower index of the binomial coefficient in \eqref{eq:ncks2} can also be set to $r+1$ because $N$ is supposed to be odd there. Furthermore, like in the previous case, it can be seen that \eqref{eq:ncks2} also matches \eqref{eq:F(2,N)}, which proves Theorem~\ref{th:2-dimensional}.

\begin{figure*}[htb]
    \centering
    
	\includegraphics[width=\linewidth]{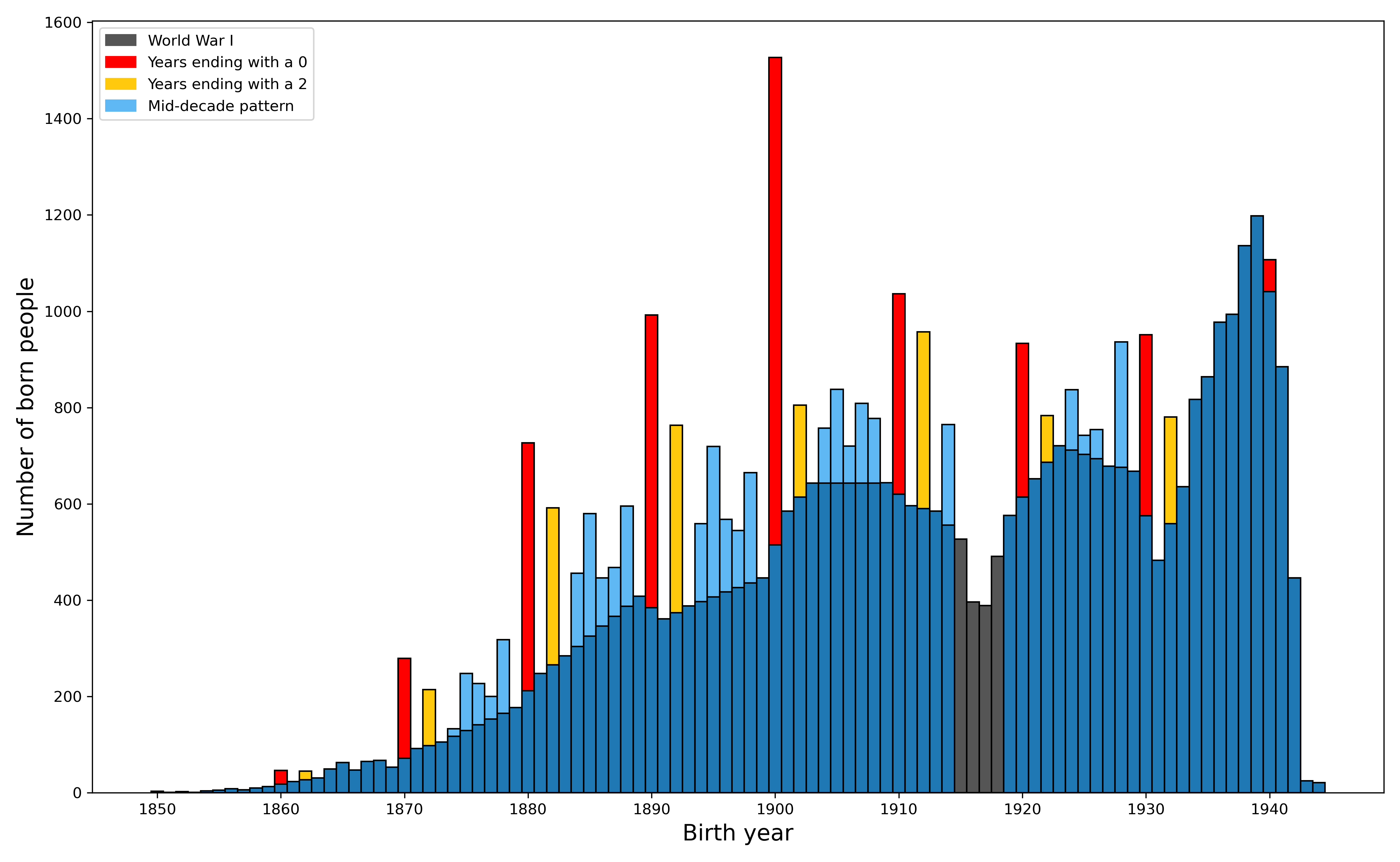}
	
    \caption{Birth year histogram of Jasenovac inmates of Serbian nationality with same markings for age heaping as in Fig.~\ref{fig:jms_marked}.}
	\label{fig:jms_serbian}
    
\end{figure*}

\subsubsection{Proof of Theorem~\ref{th:n-dimensional}}
\label{subsubsec:theorem-n}

	The expectation $\E{|x_2-x_1|}$ can be written as follows:
	\begin{equation}
	\label{eq:multinomial}
	\E{|x_2-x_1|}=n^{-N}\sum_{k_1+\dots+k_n=N}{N\choose k_1,\dots,k_n}|k_2-k_1|.
	\end{equation}
	The right-hand side of \eqref{eq:multinomial} can further be written as
	\begin{equation}
	\label{eq:rhs}
	\begin{gathered}	
	n^{-N}\sum_{k_1+k_2\le N}{N\choose k_1,k_2,N-k_1-k_2}\times\\
	\times|k_2-k_1|
	\sum_{k_3+\dots+k_n=N-k_1-k_2}
	{N-k_1-k_2\choose k_3,\dots,k_n}\\
	=n^{-N}\sum_{k_1+k_2\le N}{N\choose k_1,k_2,N-k_1-k_2}\times\\
	\times|k_2-k_1|
	(n-2)^{N-k_1-k_2}\\
	=\left(\frac{n{-}2}n\right)^N\kern-8pt
	\sum_{k_1+k_2\le N}\kern-6pt
	{N\choose k_1,k_2,N{-}k_1{-}k_2}(n{-}2)^{-(k_1+k_2)}|k_2-k_1|.
	\end{gathered}
	\end{equation}
	To obtain \eqref{eq:n-dimensional} from here, it is sufficient to note that
	\begin{align}
		\label{eq:e2e}
		\begin{split}
		\E{|x_2-x_1|}
		&=\E{x_2-x_1\mid x_2>x_1}\\
		&\qquad+\E{x_1-x_2\mid x_2<x_1}\\
		&=\E{x_2-x_1\mid x_2>x_1}\\
		&\qquad+\E{x_2-x_1\mid x_1<x_2}\\
		&=2\E{x_2-x_1\mid x_2>x_1}.
		\end{split}
	\end{align}
	\eqref{eq:e2e} can be applied to \eqref{eq:rhs}, which can then be applied to \eqref{eq:x_2-x_1}. This results in \eqref{eq:n-dimensional}, which proves Theorem~\ref{th:n-dimensional}.

\subsection{Theoretical discrete total variations}
\label{subsec:theoretical_dtvs}

To facilitate a better understanding of the experimental results that were discussed in Section~\ref{subsec:synthetic_distribution} and shown in Fig.~\ref{fig:tvor_vs_baseline}, the comparison of the values of the theoretical discrete total variation $\V{{\cal D}}$ of the histograms of normal distribution with parameters used to obtain these results are given in Fig.~\ref{fig:tvor_vs_baseline_dtvs}. By comparing Figs.~\ref{fig:tvor_vs_baseline} and~\ref{fig:tvor_vs_baseline_dtvs}, it is relatively easy to explain phenomena such as the sudden drops in the proposed method's performance that can be seen in Fig.~\ref{fig:b_b} when $10$ bins are used. Namely, Fig.~\ref{fig:n_b_10_s_0.5} clearly shows that for $10$ bins the difference between the theoretical DTVs of the distributions used there is very small, which renders the proposed method inadequate for recognizing outlier samples for that specific case. Similar reasoning can also be applied to successful cases where this difference is sufficiently large.

\subsection{Dependence of variation on the sample size}
\label{subsec:variation_size}

Fig.~\ref{fig:beta_triangular_dtv} clearly shows how randomness can have a significant impact on the performance of the proposed method. Nevertheless, as described by~\eqref{eq:non-uniform2}, when the samples' sizes grow, this impact becomes ever smaller. However, in order to decrease this impact in cases of e.g. larger values of $n$, the samples' sizes have to grow significantly more than in the cases of smaller values of $n$. This is illustrated on several examples shown in Fig.~\ref{fig:sizes}. There it can be seen that for $n=5$ the samples with random sizes up to $1000$ are clearly separated, while for the same sizes and $n=50$ the samples can hardly be separated. However, as shown in Fig.~\ref{fig:n_50_up_10000} and Fig.~\ref{fig:n_50_up_100000}, if the upper bound for the sizes of random samples gets increased even further, the separation again becomes clear. As shown in Fig.~\ref{fig:sizes2}, this has a direct influence on the performance of both the baseline and the proposed methods.

In short, a successful application of the proposed method assumes a reasonably high ratio between the number of bins $n$ and the sizes of samples. How high this ratio should be, however, depends on the specific distributions of the samples.

\subsection{Partitioning the top-scoring histogram}
\label{subsec:partial_jms}

In order to describe the behavior of the proposed method in more detail, it may be useful to additionally analyze the top-scoring histogram shown in Figs.~\ref{fig:jms} and~\ref{fig:jms_marked}. By partitioning the initial birth year sample into more smaller samples, it is possible to examine the behavior of the proposed method when the sample size is changing. One way of partitioning the sample is by nationality of the inmates. The nationalities for which there are more than $1000$ listed inmates are, as specified in the Jasenovac inmates list, the following ones: Serbian, Roma, Jewish, Croatian, and Muslim. While the histograms of the Roma, Jewish, Croatian, and Muslim nationalities shown in Fig.~\ref{fig:jms_parts} all exhibit signs of age heaping similar to the ones in Fig.~\ref{fig:jms_marked}, by far more prominent signs are exhibited by the Serbian nationality as shown in Fig.~\ref{fig:jms_serbian}.

If the histograms for separate nationalities are also added to the set of USHMM lists and the proposed method is applied to this extended set, then the histogram for the Serbian nationality ends up being the second most likely outlier just after the whole Jasenovac list with $d'=40.82$. The Romani nationality histogram ends up on the 21st place with $d'=15.12$, the Jewish nationality histogram ends up on the 66th place with $d'=9.08$, while other histograms are not inside the $100$ most likely outliers. This shows how the proposed method can also be used to detect the potentially problematic parts of a sample, which in the case of the Jasenovac list lies in the birth years of Serbian inmates.

Additionally, there is another thing to be observed here. Namely, while Figs.~\ref{fig:jms_marked} and~\ref{fig:jms_serbian} seem to be very similar, the score $d'$ for the histogram of the birth years of the Serbian inmates was nevertheless smaller than the one for the whole Jasenovac list. This has to do with the fact that the sample with birth years of Serbian inmates has fewer values than the whole Jasenovac list, i.e. it makes up roughly 57\% of the Jasenovac list. Because of that, such similar deviations are considered to be less likely on a larger sample and thus the whole Jasenovac list has a slightly larger value of score $d'$.

\section*{Acknowledgments}
\label{sec:acknowledgements}

\textit{(Nikola Bani{\'{c}} and Neven Elezovi{\'{c}} contributed equally to this work.)} The authors would like to thank the reviewers for their constructive and useful feedback on the research. Additionally, the authors would like to thank Prof.~Branko Jeren for the discussions, advice for a clearer presentation, and his networking efforts, Dr. Juraj Radi{\'{c}} for the significant help on the initial theoretical derivation of the used statistical model, Dr.~Mladen Koi{\'{c}} for motivating to provide better visualization and explanations, Dr.~Josip Stjepandi{\'{c}} for his role in motivating the research, Dr.~Tomislav Petkovi{\'{c}} for his useful advice, which have made the paper more readable, Dr. Vuko Brigljevi{\'{c}} for his advice on language improvement, Dr. Stjepan {\v{S}terc} for his advice on some demographic topics, Dr. Viktoria Oliver for her proof-reading of the paper and suggestions on English style improvement as a native speaker, Dr. Ivan Hrvoi{\'{c}} and Ivana Kova{\v{c}}evi{\'{c}} Mandac for connecting the authors with Dr. Oliver. Finally, the authors would also like to thank Dr. Julio Guijarro Garcia, Dr. Josep Peguera Poch, Dr. Jorge Ramos, and Dr. Jure Bogdan for their kind support.

\balance
\bibliographystyle{IEEEtran}
\bibliography{literature}

\begin{IEEEbiography}[{\includegraphics[width=1in,height=1.25in,clip,keepaspectratio]{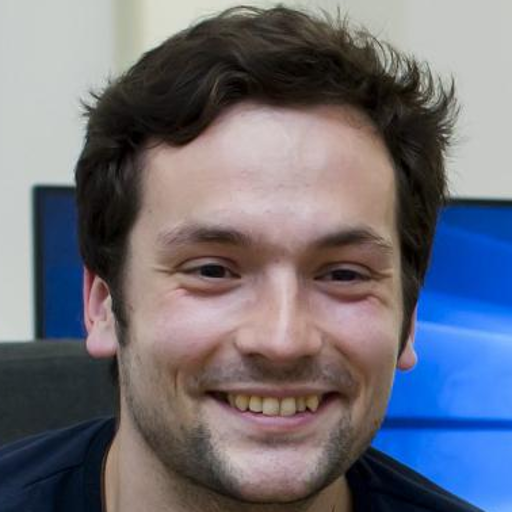}}]{Nikola Bani{\'{c}}} received B.Sc., M.Sc., and Ph.D. degrees in computer science in 2011, 2013, and 2016, respectively. He is currently working as a senior computer vision engineer at Gideon Brothers, Croatia. He has worked in real-time image enhancement for embedded systems, digital signature recognition, people tracking and counting, and image processing for stereo vision. His research interests include image enhancement, color constancy, image processing for stereo vision, and tone mapping.
\end{IEEEbiography}

\begin{IEEEbiography}[{\includegraphics[width=1in,height=1.25in,clip,keepaspectratio]{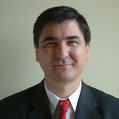}}]{Neven Elezovi{\'{c}}} 
received B.Sc., M.Sc., and Ph.D. degrees in mathematics in 1977, 1981, and
1985, respectively. He published numerous publications in high ranking journals and books for college mathematics. Currently he is a full professor at the University of Zagreb Faculty of Electrical Engineering and Computing and the founder of the publishing house Element. His research interests include mathematical analysis and probability theory.
\end{IEEEbiography}

\EOD

\end{document}